\pdfoutput=1
\documentclass[submission, Phys]{SciPost}

\usepackage{amsmath}
\usepackage{amssymb}
\usepackage{amsfonts}
\usepackage{mathrsfs}
\usepackage{mathtools}
\usepackage{cite}
\usepackage{xcolor}
\usepackage{graphicx}
\usepackage{tikz}
\usetikzlibrary{decorations.pathmorphing,backgrounds,calc,shapes,arrows,shapes.misc,shadows,intersections}
\tikzset{cross/.style={cross out, draw=black, minimum size=2*(#1-\pgflinewidth), inner sep=0pt, outer sep=0pt},
cross/.default={1pt}}
\usetikzlibrary{decorations.pathreplacing}
\usepackage{epiolmec}

 \hypersetup{
     colorlinks,
     linkcolor={red!70!black},
     citecolor={green!70!black},
     urlcolor={blue!70!black}
 }

\usepackage{silence}
\usepackage{longtable}
\usepackage{booktabs}
\usepackage{siunitx}
\usepackage{etoolbox}

 \definecolor{gammagray}{gray}{0.6}
 \newcommand{\Gnum}[2]{\ifstrequal{#2}{F}{(\num{#1})}{\num{#1}}}

\newcommand{\eq}[2]{\begin{equation} #1 \label{#2} \end{equation}}

\DeclareMathOperator{\extdm}{d}
\newcommand{\extd}{\extdm \!}
\newcommand{\dd}{\extdm \!}
\newcommand{\xl}{x_{\textrm{\tiny L}}}
\newcommand{\xr}{x_{\textrm{\tiny R}}}
\newcommand{\xm}{x_{\textrm{\tiny match }}}

\newcommand{\win}{n} 

\newcommand{\scri}{{\mathscr{I}}}

\definecolor{tabutter}{rgb}{0.98824, 0.91373, 0.30980}		
\definecolor{ta2butter}{rgb}{0.92941, 0.83137, 0}		
\definecolor{ta3butter}{rgb}{0.76863, 0.62745, 0}		

\definecolor{taorange}{rgb}{0.98824, 0.68627, 0.24314}		
\definecolor{ta2orange}{rgb}{0.96078, 0.47451, 0}		
\definecolor{ta3orange}{rgb}{0.80784, 0.36078, 0}		

\definecolor{tachocolate}{rgb}{0.91373, 0.72549, 0.43137}	
\definecolor{ta2chocolate}{rgb}{0.75686, 0.49020, 0.066667}	
\definecolor{ta3chocolate}{rgb}{0.56078, 0.34902, 0.0078431}	

\definecolor{tachameleon}{rgb}{0.54118, 0.88627, 0.20392}	
\definecolor{ta2chameleon}{rgb}{0.45098, 0.82353, 0.086275}	
\definecolor{ta3chameleon}{rgb}{0.30588, 0.60392, 0.023529}	

\definecolor{taskyblue}{rgb}{0.44706, 0.56078, 0.81176}		
\definecolor{ta2skyblue}{rgb}{0.20392, 0.39608, 0.64314}	
\definecolor{ta3skyblue}{rgb}{0.12549, 0.29020, 0.52941}	

\definecolor{taplum}{rgb}{0.67843, 0.49804, 0.65882}		
\definecolor{ta2plum}{rgb}{0.45882, 0.31373, 0.48235}		
\definecolor{ta3plum}{rgb}{0.36078, 0.20784, 0.4}		

\definecolor{tascarletred}{rgb}{0.93725, 0.16078, 0.16078}	
\definecolor{ta2scarletred}{rgb}{0.8, 0, 0}			
\definecolor{ta3scarletred}{rgb}{0.64314, 0, 0}			

\definecolor{taaluminium}{rgb}{0.93333, 0.93333, 0.92549}	
\definecolor{ta2aluminium}{rgb}{0.82745, 0.84314, 0.81176}	
\definecolor{ta3aluminium}{rgb}{0.72941, 0.74118, 0.71373}	

\definecolor{tagray}{rgb}{0.53333, 0.54118, 0.52157}		
\definecolor{ta2gray}{rgb}{0.33333, 0.34118, 0.32549}		
\definecolor{ta3gray}{rgb}{0.18039, 0.20392, 0.21176}		

\makeatletter
\def\@hex@@Hex#1%
 {\if a#1A\else \if b#1B\else \if c#1C\else \if d#1D\else
  \if e#1E\else \if f#1F\else #1\fi\fi\fi\fi\fi\fi \@hex@Hex}
\makeatother

\newcommand{\intfct}{\beta} 

\begin{document}


\newcommand{\mytitle}{Critical spacetime crystals in continuous dimensions}

\begin{center}{\Large \textbf{\mytitle
}}\end{center}

\begin{center}
Christian Ecker\textsuperscript{1},
Florian Ecker\textsuperscript{2}, Daniel Grumiller\textsuperscript{2}, and Tobias Jechtl\textsuperscript{2}
\end{center}

\begin{center}
{\bf 1} Institute for Theoretical Physics, Goethe University\\ 60438, Frankfurt am Main, Germany\\ \smallskip
{\bf 2} Institute for Theoretical Physics, TU Wien\\ Wiedner Hauptstrasse~8-10, A-1040 Vienna, Austria, Europe
\\ \smallskip
\href{mailto:ecker@itp.uni-frankfurt.de}{ecker@itp.uni-frankfurt.de},
\href{mailto:fecker@hep.itp.tuwien.ac.at}{fecker@hep.itp.tuwien.ac.at},
\href{mailto:grumil@hep.itp.tuwien.ac.at}{grumil@hep.itp.tuwien.ac.at},
\href{mailto:tobias.jechtl@tuwien.ac.at}{tobias.jechtl@tuwien.ac.at}

\end{center}

\section*{Abstract}{%
We numerically construct a one-parameter family of critical spacetimes in arbitrary continuous dimensions $D>3$. This generalizes Choptuik's $D=4$ solution to spherically symmetric massless scalar-field collapse at the threshold of $D$-dimensional Schwarzschild--Tangherlini black hole formation. We refer to these solutions, which share the discrete self-similarity of their four-dimensional counterpart, as \textit{critical spacetime crystals}. Our main results are the echoing period and Choptuik exponent of the crystals as continuous functions of $D$, with detailed data for the interval $3.05\leq D\leq 5.5$. Notably, the echoing period has a maximum near $D\approx 3.76$. As a by-product, we recover the echoing periods and Choptuik exponents in $D=4\,(5)$: $\Delta = 3.445453\,(3.22176)$ and $\gamma = 0.373961\,(0.41322)$. We support these numerical results with analytical expansions in $1/D$ and $D-3$. They suggest that both the echoing period and Choptuik exponent vanish as $D\to 3^+$. This paves the way for a small-$(D-3)$ expansion, paralleling the large-$D$ expansion of general relativity. We also extend our results to two-dimensional dilaton gravity.
}


\vspace{10pt}
\noindent\rule{\textwidth}{1pt}
\tableofcontents\thispagestyle{fancy}
\noindent\rule{\textwidth}{1pt}
\vspace{10pt}


\section{Introduction}
\label{se:1}
\begin{figure}[!htb]
    \centering
    \includegraphics[width=0.85\linewidth]{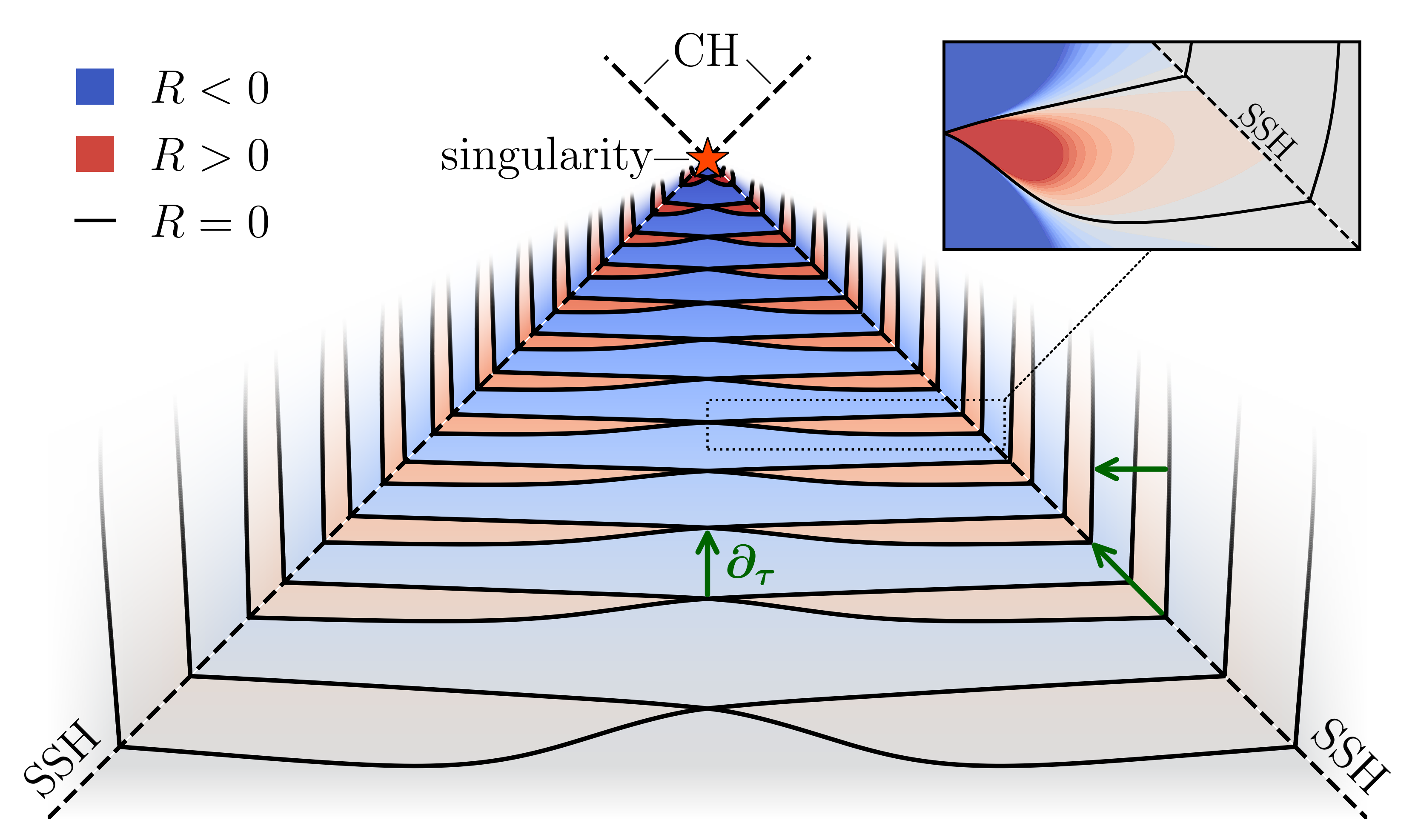}
    \caption[Illustration of critical spacetime crystal]{Illustration of spacetime crystal for critical solutions in spherically symmetric collapse. The past lightcone of the naked singularity is the patch we solve, but the spacetime may be extended beyond the selfsimilar horizon (SSH) until the Cauchy horizon (CH). The crystal vector $\partial _\tau $ (in green) is timelike, null, or spacelike depending on whether one is inside, on, or outside the SSH.}
    \label{fig:illustration_crystal}
\end{figure}

At the threshold of black hole formation, general relativity can admit discretely self-similar (DSS) solutions. In adapted coordinates $(\tau,x^i)$, this means the metric $g$ is related to a periodic metric $\tilde g$ by a Weyl factor $e^{-2\tau}$,
\eq{
g_{\mu\nu}(\tau,\,x^i) = e^{-2\tau}\,\tilde g_{\mu\nu}(\tau,\,x^i)\qquad\qquad\tilde g_{\mu\nu}(\tau+\Delta,\,x^i)=\tilde g_{\mu\nu}(\tau,\,x^i)
}{eq:1}
where $\Delta$ is called ``echoing period''. We refer to the metric $\tilde g$ (and to ease the notation, also to $g$ when there can be no confusion) as ``critical spacetime crystal'' (CSC), since the crystal vector $\partial_\tau$ can be spacelike or timelike in different regions of spacetime, as illustrated in Fig.~\ref{fig:illustration_crystal}. The null separatrix between such regions is known as ``selfsimilar horizon'' (SSH). This notion of a spacetime crystal generalizes the one introduced in a condensed matter context, where the crystal vector can also have any signature but is fixed \cite{Xu_2018,Zhang_2023,Zhao:2025}. For our purposes, it is crucial that the crystal vector can change its signature, since we are exclusively interested in spacetime crystals with an SSH.

The first explicit example of such a solution, i.e., a DSS spacetime Weyl-related to a spacetime crystal with an SSH, was numerically constructed by Choptuik \cite{Choptuik:1993jv} in the massless Einstein--Klein--Gordon model in four spacetime dimensions \cite{Christodoulou:1986zr}. Besides evaluating the echoing period $\Delta\approx 3.4$, in the language of the renormalization-group flow interpretation of \cite{Koike:1995jm}, Choptuik discovered that the spectrum of linearized perturbations around the crystal features a single unstable mode and determined its Lyapunov exponent $\lambda=1/\gamma\approx2.7$, where $\gamma\approx0.37$ is known as ``Choptuik exponent''. In this work, we generalize Choptuik's construction to arbitrary spacetime dimension $D>3$, treating the dimension as a continuous parameter. 

Our main aim is to determine the dependence of the echoing period $\Delta$ and the Choptuik exponent $\gamma$ on $D$, and to push as close as possible to the limit $D\to 3^+$, aided by analytical methods where we use $(D-3)$ as a small parameter. The opposite limit, $D\to\infty$, was recently studied by the first three authors using paper-and-pencil methods, providing the first analytical results for DSS spacetimes \cite{Ecker:2026akf}. A secondary aim is to elaborate on the large-$D$ limit, with particular focus on the large-$D$ scaling of the echoing period $\Delta$.

\newcommand{\coupling}{\kappa}

In the following, we present our motivations and goals from the perspective of different communities; the items are largely redundant and rephrase the same goals in different terminologies (we postpone the superstring perspective to the concluding Section):
\begin{itemize}
    \item \textbf{General relativity perspective.} Critical collapse is a core feature of general relativity, see \cite{Gundlach:1999cu,Gundlach:2002sx,Gundlach:2007gc} for reviews, but so far it is accessible mostly through numerical methods, in part because there is no small parameter within general relativity that could be used for an expansion. By introducing the spacetime dimension $D$ as a free parameter, we may overcome this, for instance, along the lines of the large-$D$ expansion \cite{Emparan:2013moa,Emparan:2020inr}. To achieve this, we first deal with general relativity in any spacetime dimension $D>3$ and then analytically continue in the dimension so that $D\in(3,\infty)$. Our goal is to study critical collapse of the spherically symmetric massless Einstein--Klein--Gordon model for arbitrary (real) spacetime dimensions $D>3$, particularly in the limit $D\to 3^+$. Three dimensions are special insofar as there are no black hole solutions without a negative cosmological constant \cite{Ida:2000jh}. By contrast, asymptotically flat black holes exist in $3+\epsilon$ dimensions. As we shall see, gravity in $3+\epsilon$ dimensions has complementary features to the large $D$ limit and it seems promising to pursue this line of research to see whether some of the successes of the large $D$ expansion \cite{Emparan:2020inr} carry over to this limit. Our work is a first step in this direction.
    \item \textbf{Two-dimensional dilaton gravity perspective.} By integrating out the angular part, we reduce the $D$-dimensional massless Einstein--Klein--Gordon model to a 1-parameter family of two-dimensional (2d) dilaton gravity models with matter that can be further generalized by dilaton-dependent Weyl rescalings. The family parameter $\coupling=D-3$ is the original spacetime dimension, and analytic continuation to real values is trivial in this formulation. We aim to study critical collapse of 2d dilaton gravity non-minimally coupled to a massless scalar field and to derive critical parameters, such as the echoing period $\Delta$, the scaling exponent $\gamma$, and the opening angle between lines saturating the null energy condition (NEC) $\alpha$, depending on the parameter $\coupling$, especially for $\coupling\to0^+$. 
    \item \textbf{Quantum field theory perspective.} We have at our disposal a 2d effective field theory that depends on one effective coupling constant, $\coupling=D-3\in(0,\infty)$, and are interested in constructing (unstable) solitonic solutions of the classical field equations that exhibit universal features, such as spontaneous symmetry breaking of spacetime translations to a discrete subgroup and universal scaling exponents of linearized perturbations around these solitons. Some ${\cal O}(1)$ values of $\coupling$ were studied in the literature using numerical methods. We plan to consider specifically the strong ($\coupling\to\infty$) and weak ($\coupling\to 0$) coupling limits to check whether there are any simplifications or asymptotic trends for our key observables and the equations of motion, and intend to study how the physical observables depend on the effective coupling constant.
    \item \textbf{Mathematics perspective.} We are interested in solving a specific 1-parameter set of coupled nonlinear first-order PDEs for which none of the known existence and uniqueness theorems apply. Apart from providing numerical evidence concerning existence and uniqueness, we find that there are two singular limits of the parameter where the set of PDEs simplifies, in one limit to an exactly soluble system (though in this case we lose uniqueness) and in another limit to a Fuchsian system (though in this case we lose existence of regular solutions). A crucial property of the solutions is that they are necessarily periodic, and the period, $\Delta$, is related in a non-obvious way to the family parameter $\coupling$. Since the solution associated with $\coupling=1$ exists \cite{Reiterer:2012hnr}, it is plausible that it exists also for other values of $\coupling$.
    \item \textbf{Condensed matter perspective.} Spacetime crystals are generalizations of crystals, where the broken translation symmetries may involve space and time \cite{Xu_2018,Zhang_2023,Zhao:2025}. Our notion of CSC is a further generalization where the lattice vector can change its signature from spacelike to timelike, thus allowing a codimension-1 region of the spacetime crystal where the lattice vector is null (this region corresponds to the SSH defined above). We intend to numerically (and in certain limits also analytically) construct a family of $(1+1)$-dimensional spacetime crystals (with SSH), where the lattice length $\Delta$ can be calculated from first principles. These CSC are codimension-1 attractors, so they have a single unstable linearized mode, the Lyapunov exponent of which is the inverse of the Choptuik exponent $\gamma$. They resemble prethermalized states \cite{Berges:2004ce,Langen:2016vdb}, in the sense that they are not thermal states but decay into them when excited and serve as an intermediate attractor for generic initial states. The CSCs we intend to construct play the role of the non-thermal fixed point in this jargon, compare, e.g., the respective Figs.~1 in \cite{Gundlach:2007gc} and \cite{Langen:2016vdb}.
\end{itemize}

In the body of the paper, we combine some of these perspectives when technically advantageous. For example, from a general relativity perspective, it may seem non-obvious why an analytic continuation to arbitrary real spacetime dimensions makes sense. However, from a 2d dilaton gravity perspective, this analytic continuation just means that a certain coupling constant $\coupling$ is continued from integer to real values.

Our primary motive is to gain a better understanding of the limit $D\to3^+$ of general relativity, so we aim to push the dimension as close to three from above as possible. Approaching this limit is numerically costly, but this challenge motivates our analytic treatment with $D-3$ as a small parameter, leading to considerable simplifications of the non-linear equations of motion. This is in a similar spirit as the $1/D$-expansion of general relativity, but technically, it has complementary features to such an expansion. To make this explicit, we also reconsider the expansion where $1/D$ is treated as a small parameter, extending the analysis of \cite{Ecker:2026akf}.

Notable earlier work on critical collapse in non-integer dimensions below $D=4$ is the research paper \cite{Bland:2005kk}, which provides as a single data point the Choptuik exponent for $D=3.5$ (although without determining the echoing period) and the PhD thesis by Bland \cite{blandthesis}, which provides nine data points from $D=3.02$ to $D=3.9$, with error bars in the percent range. Our work improves on these results in terms of precision and number of data points by several orders of magnitude. Moreover, we complement our numerical results with analytical discussions, as outlined above. A final difference between our work and their paper/PhD thesis is the methodology: while they implemented critical collapse \'a la Choptuik, i.e., by considering 1-parameter families of initial data such that on one end of the family they always get black hole formation and on the other hand always dispersion, we follow Gundlach's method \cite{Gundlach:1995kd} of determining directly the CSC that represents the critical solution at the threshold of black hole formation. This method is particularly efficient in determining the echoing period $\Delta$ and permits using a CSC in any given dimension as an approximate ansatz for a CSC in a nearby dimension. Since we intend to provide a fine scan through the dimensions (with neighboring data points differing by a dimension of 0.01 or less), this method is especially suitable for our purposes. We review its essence in the next Section.
Readers who want to get a quick overview of our main results can skip ahead to the conclusions in Section \ref{sec:8} and backtrack from there to check the details of the aspects they care about most.

This paper is organized as follows. In Section \ref{sec:2}, we review essential aspects of critical collapse and critical solutions. In Section \ref{sec:3}, we discuss analytically the massless Einstein--Klein--Gordon model in arbitrary spacetime dimensions $D>3$ through the lens of 2d dilaton gravity, such that analytic continuation in $D$ is straightforward. In Section \ref{sec:4}, we describe our numerical setup to obtain the critical solution in arbitrary spacetime dimensions $D>3$. In Section \ref{sec:5}, we present our results for the echoing period $\Delta$ and the Choptuik exponent $\gamma$, for dimensions $D\in[3.05,5.5]$. In Section \ref{sec:6}, we discuss the large-$D$ and the small-$(D-3)$ expansions and use them to formulate conjectures about the echoing period in these limits. In Section \ref{sec:7}, we apply our results to intrinsic 2d dilaton gravity models, such as the Jackiw--Teitelboim model or the Witten black hole. In Section \ref{sec:8}, we conclude and give an outlook on generalizations and open issues, especially regarding the $D\to3^+$ limit. Appendix \ref{app:Taylor} displays details on the Taylor expansions near the center and the SSH. Appendix \ref{app:SingularBranch} analyzes numerically the singular solution branch near the SSH. Appendix \ref{app:Error} contains numerical convergence tests, consistency checks, and error estimates. Appendix \ref{app:numbers} lists our data for echoing period $\Delta$ and Choptuik exponent $\gamma$.

Below is a list of the figures presented in this work, with the four most important ones highlighted in bold faced letters.

\vspace{10pt}
\noindent\rule{\textwidth}{1pt}
\listoffigures
\noindent\rule{\textwidth}{1pt}
\vspace{10pt}


\section{Constructing critical spacetime crystals}\label{sec:2}

In this Section, we outline the conceptual framework for constructing CSCs, emphasizing the determination of key observables such as the echoing period $\Delta$ and the Choptuik exponent $\gamma$.

The perspective reviewed in this Section follows Gundlach's insightful papers \cite{Gundlach:1995kd,Gundlach:1996eg}, which in turn were inspired by \cite{Evans:1994pj,Koike:1995jm}. The key idea is to avoid the method used by Choptuik in his original work \cite{Choptuik:1993jv} that iteratively approaches the CSC and instead directly construct the latter by imposing the crystal property \eqref{eq:1} as an input. In other words, instead of constructing a 1-parameter family of spacetimes (with family parameter $p$), some of which contain black holes as final states while others do not, and then iterating our way towards the critical parameter $p=p_\ast$, we directly construct the CSC.

The main disadvantage of this method is that it assumes such a critical solution exists (and is unique) in the first place. However, since we know this to be the case for the Choptuik system in various integer dimensions, we pragmatically assume this to be the case also in real dimensions $D>3$.

Its main advantage is that we need to construct only a single spacetime rather than a 1-parameter family of them, which allows a higher precision in determining the echoing period $\Delta$. Another considerable advantage for our purposes is that we can take a critical solution in a given dimension $D$ and use it as a reasonable ansatz in a nearby dimension $D-\epsilon$ as long as $\epsilon$ is sufficiently small. Since we intend to provide a fine scan through the dimension, having a small $\epsilon$ is desirable anyhow.

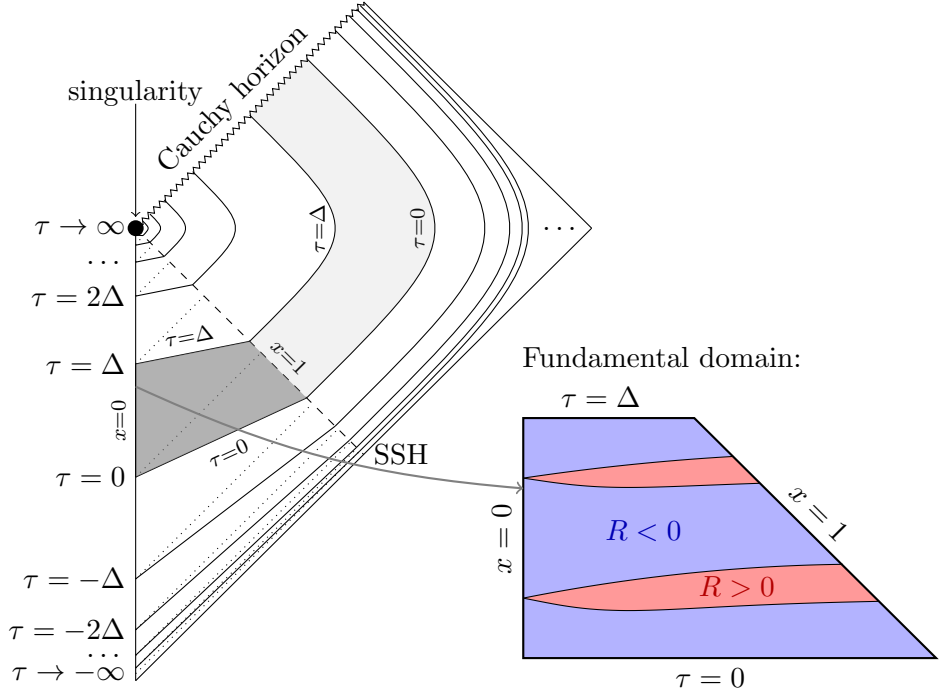
\begin{figure}[htb]
\centering
\begin{tikzpicture}[xscale=3.0,yscale=3.0]
\draw (0,0) coordinate (ori) -- (0,2) coordinate (sin) node[left, yshift=1mm] at (ori) {$\tau\to-\infty$};
\draw (ori) -- (2,2) coordinate (i0); 
\draw (i0) -- (1,3) coordinate (iplus) node[left] at (i0) {$\dots$};
\draw[thin,decorate,decoration={zigzag, amplitude=1pt, segment length=3pt}] (iplus) -- (sin) node[midway, above, sloped] {Cauchy horizon};
\fill (sin) circle (1pt);
\draw[thin,dashed] (sin) -- (1,1) coordinate (ssh) node[left] at (sin) {$\tau\to\infty$} node[right] at (ssh) {SSH} node[pos=0.62, above, sloped] {$\scriptstyle x=1$};
\draw[thin,dotted] (0,0.03125) coordinate (lo5) -- (0.983375,1.015625) coordinate (tm5);
\draw[thin,draw=none] (0,0.0625) coordinate (lo4x) -- (0.96875,1.03125) coordinate (tm4);
\draw[thin,dotted] (0,0.05625) coordinate (lo4) -- (0.971875,1.028125) coordinate (tm4x);
\draw[thin,draw=none] (0,0.125) coordinate (lo3x) -- (0.9375,1.0625) coordinate (tm3);
\draw[thin,dotted] (0,0.1125) coordinate (lo3) -- (0.94375,1.05625) coordinate (tm3x) node[left] at (lo3) {$\dots$};
\draw[thin,draw=none] (0,0.25) coordinate (lo2x) -- (0.875,1.125) coordinate (tm2);
\draw[thin,dotted] (0,0.225) coordinate (lo2) -- (0.8875,1.1125) coordinate (tm2x) node[left] at (lo2) {$\tau=-2\Delta$};
\draw[thin,draw=none] (0,0.5) coordinate (lo1x) -- (0.75,1.25) coordinate (tm1);
\draw[thin,dotted] (0,0.45) coordinate (lo1) -- (0.775,1.225) coordinate (tm1x) node[left] at (lo1) {$\tau=-\Delta$};
\draw[thin,draw=none] (0,1) coordinate (lox) -- (0.5,1.5) coordinate (t0);
\draw[thin,dotted] (0,0.9) coordinate (lo) -- (0.55,1.45) coordinate (t0x) node[left] at (lo) {$\tau=0$};
\draw[thin,draw=none] (0,1.5) coordinate (hix) -- (0.25,1.75) coordinate (t1);
\draw[thin,dotted] (0,1.4) coordinate (hi) -- (0.3,1.7) coordinate (t1x) node[left] at (hi) {$\tau=\Delta$};
\draw[thin,draw=none] (0,1.75) coordinate (hi2x) -- (0.125,1.875) coordinate (t2);
\draw[thin,dotted] (0,1.7) coordinate (hi2) -- (0.15,1.85) coordinate (t2x) node[left] at (hi2) {$\tau=2\Delta$};
\draw[thin,draw=none] (0,1.875) coordinate (hi3x) -- (0.0625,1.9375) coordinate (t3);
\draw[thin,dotted] (0,1.85) coordinate (hi3) -- (0.075,1.925) coordinate (t3x) node[left] at (hi3) {$\dots$};
\draw[thin,draw=none] (0,1.9375) coordinate (hi4x) -- (0.03125,1.96875) coordinate (t4);
\draw[thin,dotted] (0,1.925) coordinate (hi4) -- (0.0325,1.9675) coordinate (t4);
\draw[thin,dotted] (0,1.96875) coordinate (hi5) -- (0.015625,1.984375) coordinate (t5);
\draw[thin] (lo) -- (tm1) node[below, midway, sloped] {$\scriptstyle\tau=0$};
\draw[thin] (hi) -- (t0)  node[above, midway, sloped] {$\scriptstyle\tau=\Delta$};
\fill[gray, opacity=0.6] (lo) -- (tm1) -- (t0) -- (hi) -- cycle;
\draw[thin] (hi5) -- (t4);
\draw[thin] (hi4) -- (t3);
\draw[thin] (hi3) -- (t2);
\draw[thin] (hi2) -- (t1);
\draw[thin] (lo1) -- (tm2);
\draw[thin] (lo2) -- (tm3);
\draw[thin] (lo3) -- (tm4);
\draw[thin] (lo4) -- (tm5);
\draw[thin] (t4) .. controls (0.0625,2.0) .. (0.03125,2.03125);
\draw[thin] (t3) .. controls (0.125,2.0) .. (0.0625,2.0625);
\draw[thin] (t2) .. controls (0.25,2.0) .. (0.125,2.125);
\draw[thin] (t1) .. controls (0.5,2.0) .. (0.25,2.25);
\draw[thin] (t0) .. controls (1.0,2.0) .. (0.5,2.5) coordinate (end0) node[above, midway, sloped] {$\scriptstyle \tau=\Delta$};
\draw[thin] (tm1) .. controls (1.5,2.0) .. (0.75,2.75) coordinate (endm1)  node[above, midway, sloped] {$\scriptstyle \tau=0$};
\draw[thin] (tm2) .. controls (1.75,2.0) .. (0.875,2.875);
\draw[thin] (tm3) .. controls (1.875,2.0) .. (0.9375,2.9375);
\draw[thin] (tm4) .. controls (1.9375,2.0) .. (0.96875,2.96875);
\draw[thin] (tm5) .. controls (1.96875,2.0) .. (0.984375,2.984375);
\begin{scope}
    \fill[gray, opacity=0.1] 
        (tm1) .. controls (1.5,2.0) .. (endm1) -- 
        (end0) .. controls (1.0,2.0) .. (t0) -- cycle;
\end{scope}
\draw[thin] (lo) -- (hi) node[above, midway, sloped] {$\scriptstyle x=0$};
\node at (0,2.6) {singularity};
\draw[->, thin] (0,2.55) -- (0,2.05);
\begin{scope}[shift={(1.7,0.1)},scale=0.75]
\def\H{1.4142136}          
\def\B{2.4142136}         
\def\T{1}   
\coordinate (A) at (0,0);
\coordinate (B) at (\B,0);
\coordinate (C) at (\T,\H);
\coordinate (D) at (0,\H);
\def\s{0.75} 
\def\lowUp{0.35355339 + 0.20*\x - 0.05*\x*\x}
\def\lowDown{0.35355339 - 0.20*\x/(1+\x*\x)^2 - 0.05*\x*\x/(1+\x*\x)^2 + 0.01*\x*\x*\x/(1+\x*\x)^2}
\def\upUp{1.0606602 + \s*(0.20*\x - 0.05*\x*\x)}
\def\upDown{1.0606602 + \s*(-0.20*\x/(1+\x*\x)^2 - 0.05*\x*\x/(1+\x*\x)^2 + 0.01*\x*\x*\x/(1+\x*\x)^2)}
\begin{scope}
  \clip (A)--(B)--(C)--(D)--cycle;
  \fill[red!40] plot[smooth,domain=0:2.0] (\x,{\upUp}) -- plot[smooth,domain=2.0:0] (\x,{\upDown}) -- cycle;
  \fill[red!40] plot[smooth,domain=0:2.757] (\x,{\lowUp}) -- plot[smooth,domain=2.757:0] (\x,{\lowDown}) -- cycle;
\end{scope}
\begin{scope}
  \clip (A)--(B)--(C)--(D)--cycle;
  \fill[blue!30] plot[smooth,domain=0:3] (\x,{\lowDown}) -- plot[smooth,domain=3:0](\x,{0}) -- cycle;
  \fill[blue!30] plot[smooth,domain=0:3] (\x,{\lowUp}) -- plot[smooth,domain=3:0](\x,{\upDown}) -- cycle;
  \fill[blue!30] plot[smooth,domain=0:3] (\x,{1.4142136}) -- plot[smooth,domain=3:0](\x,{\upUp}) -- cycle;
\end{scope}    
\draw[name path=l1] plot[smooth,domain=0:1.87] (\x,{\lowUp});
\draw[name path=l2] plot[smooth,domain=0:2.08] (\x,{\lowDown});
\draw[name path=u1] plot[smooth,domain=0:1.23] (\x,{\upUp});
\draw[name path=u2] plot[smooth,domain=0:1.38] (\x,{\upDown});
\draw[thick] (A) 
node[rotate=90,xshift=45,yshift=8] at (A) {$x=0$}
node[xshift=80,yshift=27,text=red!70!black] at (A) {$R>0$} 
node[xshift=45,yshift=48, text=blue!70!black] at (A) {$R<0$}  
-- (B) node[rotate=-45,xshift=-70,yshift=8] at (B) {$x=1$} 
to[thick,dashed] (C) node[xshift=-35,yshift=8]  at (C) {$\tau=\Delta$}
-- (D) node[xshift=70,yshift=-98] at (D) {$\tau=0$} node[xshift=52,yshift=23] at (D) {Fundamental domain:} 
-- cycle;
\end{scope}
\draw[gray,thick,->] (0,1.3) to[bend left=-10] (1.7,0.85);
\end{tikzpicture}
\caption[Penrose diagram of critical spacetime crystal]{Penrose diagram of a CSC taken from \cite{Ecker:2024haw}. Dark gray shaded: time crystal region. Light gray shaded: (space) crystal region. Dashed line: SSH. Black circle: Naked singularity. Zig-zag line: Cauchy horizon. Vertical line: Center. Solid 45-degree lines: $\scri^\pm$. Dotted 45-degree lines: Null lines in fundamental domain. CSC metric identified along bold $\tau=\rm const.$ lines. Inset: Fundamental domain with NEC lines (black), positive (red) and negative (blue) Ricci regions.
}
\label{fig:1}
\end{figure}

Due to spherical symmetry, the relevant information about our CSCs is conveyed by 2d metrics. Their Penrose diagram is depicted in Fig.~\ref{fig:1}. These spacetimes have a naked curvature singularity, located at the black circle in the diagram, corresponding to $\tau\to\infty$ in the adapted coordinates \eqref{eq:1}. There is a Cauchy horizon emanating from the singularity, which, however, is of no concern for us.

We only need to solve the field equations in the dark gray shaded region between the center (without loss of generality at $x=0$) and the SSH (without loss of generality at $x=1$), i.e., the fundamental domain in the inset, since this turns out to be sufficient to determine the echoing period $\Delta$, other descendant observables, like the NEC angle $\alpha$ \cite{Ecker:2024haw} or, using perturbation theory, the Choptuik exponent $\gamma$  (see below).

Regularity at both the center and the SSH reduces the continuum of solutions to a discrete set of allowed echoing periods, in complete analogy with quantized eigenvalues in quantum mechanics. In practice, only a single such solution exists, so the CSC and its echoing period $\Delta$ are uniquely determined.

For the numerical algorithms described in the following Sections, the setup reviewed above requires that we begin with boundary data at the center and at the SSH, together with an estimate of the echoing period $\Delta$, that are sufficiently close to the true solution. Then we evolve radially from both sides to a suitably placed matching surface, determine the inevitable mismatch between the left and the right side, and use a Newton algorithm to optimize the initial data and $\Delta$, until the mismatch is minimized to the desired accuracy.

To initiate the algorithm, we start with the well-studied case of $D=4$ and increase/decrease the dimension in sufficiently small steps to numerically build a CSC in nearby dimensions. If the dimension is sufficiently close, the final solutions for the boundary data and echoing period of the previous dimension provide a reasonable ansatz for the neighboring dimension, which then is again optimized using the same Newton algorithm.

While in principle (and also in practice \cite{Martin-Garcia:2003xgm}), we could continue the CSC through the SSH all the way to the Cauchy horizon (and with additional assumptions even beyond), this is neither necessary for the determination of the echoing period $\Delta$ nor for the Choptuik exponent $\gamma$ or the NEC angle $\alpha$. We shall refrain from doing so, since we have a lot of crystals to build and want to remain as efficient as possible.

Given a numerically accurate CSC, we proceed to determine its observables besides the echoing period $\Delta$. Of particular interest is the Ricci scalar, not only because it becomes singular at the black circle in Fig.~\ref{fig:1}, but also because lines of vanishing Ricci scalar turn out to intersect in the center at a fixed angle that depends only on the spacetime dimension. The lines happen to coincide with lines where one of the NECs saturates, so the determination of the zero-Ricci-lines and intersection angle also determines the NEC-saturation lines and the NEC angle \cite{Ecker:2024haw}, which provides a purely geometric observable that depends only on the spacetime dimension.

To get the Choptuik exponent, we linearize around the CSC and solve the corresponding (linear) eigenvalue problem that determines the spectrum of the perturbations. We assume there is a single unstable mode and thus focus on determining the unique (in our conventions positive) eigenvalue $\lambda$ that constitutes the Lyapunov exponent of the unstable mode. The Choptuik exponent is the inverse of this Lyapunov exponent, $\gamma=1/\lambda$. Since non-spherical perturbations always decay \cite{Martin-Garcia:1998zqj}, it is sufficient to perform this analysis in a strictly spherically symmetric setting.

For completeness of this brief summary, we recall the significance of the Choptuik exponent $\gamma$. Physical observables sufficiently close to criticality scale with a certain power (provided by simple dimensional analysis) of $\gamma$. For example, above the threshold of black hole formation, $p>p_\ast$, the mass of the final black hole scales like
\eq{
M = M_0\,(p-p_\ast)^{(D-3)\gamma}
}{eq:2}
where $M_0$ is a non-universal mass scale, $p>p_\ast$ is the value of the parameter of the black hole solution under consideration, $p_\ast$ is the critical value of this parameter where the spacetime crystal solves the field equations, and $\gamma$ is the universal Choptuik exponent. Similar considerations apply to subcritical solutions; e.g., the maximal value of the Kretschmann scalar scales as $R_{\mu\nu\lambda\sigma}R^{\mu\nu\lambda\sigma}|_{\textrm{\tiny max}}\propto(p_\ast-p)^{-4\gamma}$.

In the following Sections, we implement the procedure outlined above for the massless Einstein--Klein--Gordon model in any dimension. We start in the next Section with $D$-dimensional Einstein gravity minimally coupled to a massless scalar field. Like in Choptuik's original work, we impose strict spherical symmetry, but unlike his work, we do so already at the level of the action.\footnote{%
Imposing spherical symmetry commutes with varying the action \cite{Palais:1979} so classically, our procedure does not lose any information.
}
This allows a reformulation of the theory as a specific 2d dilaton gravity model, where the original spacetime dimension $D$ appears merely as a parameter in the action, which permits a straightforward analytic continuation to arbitrary (real) values of $D$. Since our main goal is to construct CSCs, we adopt a convenient gauge fixing inspired by work in $D=4$. 


\section{Massless Einstein--Klein--Gordon model in any dimension}\label{sec:3}

The main results of this Section are the equations of motion for arbitrary values of $D$ (displayed already in \cite{Ecker:2026akf}), the boundary/regularity conditions imposed on the fields, and analytic expressions for key observables that we shall exploit in later Sections.

This Section is organized as follows. Section \ref{sec:3.1} reviews the spherical reduction of $D$-dimensional Einstein gravity with a minimally coupled massless scalar field to 2d dilaton gravity with a non-minimally coupled massless scalar field. Section \ref{sec:3.2} addresses gauge-fixing. Section \ref{sec:3.3} displays our first key result, the gauge-fixed first-order equations of motion. Section \ref{sec:3.4} discusses parity, regularity, and boundary conditions needed to construct the CSC and to determine the echoing period $\Delta$. Section \ref{sec:3.5} derives the linearized equations of motion needed to determine the Choptuik exponent $\gamma$. Section \ref{sec:3.6} summarizes analytic expressions for key observables, including scalar invariants, null projections of the energy-momentum tensor, NEC inequalities, and their saturations. 

\subsection{Spherical reduction}\label{sec:3.1}

The bulk action of $D$-dimensional Einstein gravity (without cosmological constant) minimally coupled to a massless scalar field $\psi $ is given by
\eq{
    S\big[g^{(D)}_{\mu\nu},\,\psi\big]=\frac{1}{16\pi G^{(D)}}\int \extd^Dx\sqrt{-g^{(D)}}\, R^{(D)}
    -\frac{1}{2}\int \extd^Dx\sqrt{-g^{(D)}}\, g_{(D)}^{\mu \nu}(\nabla_\mu \psi)(\nabla_\nu \psi) ~.
}{eq:EMKG_ddim}
The value of the $D$-dimensional Newton constant $G^{(D)}$ is irrelevant in the classical theory and can be absorbed by rescaling the scalar field. We are going to exploit this property below to absorb some $D$-dependent factors.

Since we impose spherical symmetry, it is convenient to partially fix the coordinates such that we can split off a 2d part $g_{\alpha \beta}$ in the metric,
\eq{
    \extd s_{(D)}^2=g_{\alpha \beta}(x^\gamma)\extd x^\alpha \extd x^\beta +\Phi ^2(x^\gamma)\extd \Omega ^2_{S^{D-2}}
}{eq_spher_ansatz}
where $\alpha,\beta,\gamma=0,1$ and the scalar field $\Phi (x^\gamma)$ is the area radius of the transverse spheres. This allows writing the $D$-dimensional Ricci scalar in terms of $g_{\alpha \beta}$ and $\Phi$ as
\eq{
    R^{(D)}=R-2(D-2)\frac{\square \Phi}{\Phi}-(D-3)(D-2)\frac{(\nabla \Phi)^2}{\Phi^2}+\frac{(D-3)(D-2)}{\Phi^2} 
}{eq:RicciD}
with the Ricci scalar associated to $g_{\alpha \beta}$ just written as $R$. Up to boundary terms, the action \eqref{eq:EMKG_ddim} can be recast in a 2d, spherically reduced form
\eq{
    S_{\textrm{\tiny 2d}}[X,g_{\alpha\beta},\psi ]=\frac{1}{2}\int\!\extd^2x\sqrt{-g}\big(XR-U(X)(\partial X)^2-2V(X)\big)-\frac{1}{2} \int\!\extd^2 x\sqrt{-g}X(\partial \psi)^2
}{eq:2d_SRG}
where we have absorbed the gravitational coupling constant by a suitable rescaling of $\psi$. We define the dilaton $X$ by
\eq{
\Phi =X^{\frac{1}{D-2}}
}{eq_dildef}
such that the now non-minimal coupling to the scalar field takes a canonical form. The two potential functions are given by
\eq{
    U(X)=-\frac{D-3}{D-2}\frac{1}{X} \qquad\qquad V(X)=-\frac{1}{2}(D-2)(D-3)X^{\frac{D-4}{D-2}} 
}{eq:UV_srg}
and are just one choice from a large class of other 2d dilaton gravity models, see \cite{Grumiller:2002nm} for a review. Here, one can see explicitly that the dimension of the higher-dimensional spacetime only enters the action as a real parameter. Clearly, $D=3$ is special since both potentials vanish. So from now on, we assume $D>3$ but do not restrict $D$ to be integer.

The equations of motion descending from this action are
\begin{align}
    \mathcal{E}_{\alpha \beta}&= T_{\alpha \beta} \label{eq:eomg}\\ 
    \mathcal{E}_X&= T_X \label{eq:eomX}\\
    \nabla _\alpha \big(g^{\alpha \beta}X\partial_\beta \psi \big)&=0 \label{eq:eomKG}
\end{align}
where
\begin{align}
    \mathcal{E}_{\alpha \beta}&=g_{\alpha \beta}\square X-\nabla_\alpha \partial_\beta X+\frac{1}{2}g_{\alpha \beta}U(X)(\partial X)^2-U(X)(\partial_\alpha X)(\partial_\beta X)+g_{\alpha \beta}V(X)\\
    \mathcal{E}_X&=R+2U(X)\square X+\partial_XU(X)(\partial X)^2-2\partial_XV(X)
\end{align}
and the stress tensor components are given by
\begin{align}
    T_{\alpha \beta}&=-\frac{2}{\sqrt{-g}}\frac{\delta S_m}{\delta g^{\alpha \beta}}= X\big((\partial_\alpha \psi) (\partial_\beta \psi) -\frac{1}{2}g_{\alpha \beta}(\partial\psi)^2\big)\label{eq:stresstensor1}\\
    T_X&=-\frac{2}{\sqrt{-g}}\frac{\delta S_m}{\delta X}= (\partial\psi)^2 ~.\label{eq:stresstensor2}
\end{align}

While the system of equations \eqref{eq:eomg}-\eqref{eq:stresstensor2} is complete, it is convenient to present specific redundant equations. The first one is the trace of the tensor equations \eqref{eq:eomg}, 
\eq{
\square X = -2V(X)
}{eq:traceeq}
which is very simple as a consequence of 2d tracelessness, $g^{\alpha\beta}T_{\alpha\beta} = 0$. Finally, the contracted Bianchi identities
\begin{align}
    \nabla_\alpha \mathcal{E}^\alpha{}_\beta +\frac{1}{2}(\partial_\beta X)\mathcal{E}_X\equiv 0
\end{align}
imply that the stress tensor components satisfy a generalized conservation law
\begin{align}
    \nabla_\alpha T^\alpha {}_\beta +\frac{1}{2}(\partial_\beta X)T_X= 0 \label{eq:cons}
\end{align}
not only if \eqref{eq:eomg}, \eqref{eq:eomX} hold, but also if the Klein--Gordon equation \eqref{eq:eomKG} alone holds.

The usual gauge redundancy from diffeomorphisms prevents one from directly solving the second-order field equations \eqref{eq:eomg}-\eqref{eq:stresstensor2}  numerically. Thus, in the next Subsection, we completely fix the gauge to eliminate this redundancy, and in the Subsection thereafter, we convert the system of second-order PDEs into an equivalent system of first-order PDEs that is suitable for numerical purposes.

\subsection{Gauge fixing}\label{sec:3.2}

A characteristic feature of the critical collapse for a spherically symmetric scalar field is the emergence of a DSS solution.

Since our goal is to compute the critical solution directly, i.e., with DSS imposed from the start, it is convenient to introduce coordinates that make this symmetry manifest, like the adapted coordinates in \eqref{eq:1}. For further reduction of the redundancies, we adhere to the gauge choices made in \cite{Martin-Garcia:2003xgm}. Together with the gauge fixing adapted to spherical symmetry \eqref{eq_spher_ansatz}, this leads to the DSS metric
\begin{equation}
     \extd s^2=e^{-2\tau}\big(\tilde g_{\alpha\beta}(\tau,x)\,\extd x^\alpha\extd x^\beta+x^2\extd \Omega^2_{S^{D-2}}\big)
     \label{eq:whynolabel}
\end{equation}
and the 2d CSC metric
\begin{equation}
\tilde g_{\alpha\beta}\extd x^\alpha\extd x^\beta = e^{\omega }\,\big((x^2-f^2)\extd\tau^2-2x\extd\tau \extd x+\extd x^2\big)
\label{eq:lalapetz}
\end{equation}
with $(\tau, x)\in \mathbb{R}\times \mathbb{R}^+_0$ defining the coordinates in the spherically reduced, 2d theory. The two functions $\omega(\tau,x)$ and $f(\tau,x)$ in the metric $\tilde g_{\alpha \beta}$ are both periodic in $\tau $ with period $\Delta$ such that $g_{\mu \nu}(\tau +\Delta ,x)=e^{-2\Delta }g_{\mu \nu}(\tau,x)$, as required by \eqref{eq:1}. We refer to $\omega$ as ``Weyl factor'' and to $f$ as ``SSH function''.

This gauge choice also implies a condition on the dilaton and area radius $\Phi(\tau,x)$ of the $S^{D-2}$ given by
\eq{
    \Phi(\tau,x) = e^{-\tau}\,x \qquad \qquad X(\tau,x)=e^{-\tau (D-2)}\,x^{D-2}
}{eq:dilaton}
ensuring that the center of the spheres coincides with the $x=0$ line. 

One can see in \eqref{eq:lalapetz} that $x =\text{const.}$ lines are timelike as long as $0\leq x<|f|$ but become null at $x=|f|$ in which case $\tau $ parametrizes a radial null geodesic. This line represents the SSH, and like the authors of \cite{Martin-Garcia:2003xgm}, we fix $x=1$ at this locus, such that $f(\tau,1)=1$. Once this is fixed, the function $f(\tau,x)$ is strictly positive and is related to the volume form by $\sqrt{-g}=e^{-2\tau }e^\omega |f|$. 

As shown in the dark gray region and the inset of the Penrose diagram Fig.~\ref{fig:1}, we intend to solve the equations of motion for one period in $\tau$ and only between the center and the SSH. The boundary conditions chosen will be explained in the next Subsection and are partly fixed by demanding regularity at the respective boundaries.

There is still a residual redundancy in the equations of motion associated with a generalized symmetry of the action \eqref{eq:2d_SRG} under constant Weyl rescalings ($\tau_0\in\mathbb{R}$),
\eq{
g_{\mu\nu}\to e^{-2\tau_0}\,g_{\mu\nu}\qquad\qquad X\to e^{-(D-2)\tau_0}\,X\qquad\qquad\Rightarrow\qquad\qquad S\to e^{-(D-2)\tau_0}\,S
}{eq:gen_sym}
that rescales the action by an overall constant and hence leaves invariant the equations of motion. In our coordinates with the gauge-fixing condition $f(\tau,1)=1$, this residual redundancy amounts to arbitrary shifts of time,
\eq{
    \tau \to \tau +\tau_0\,.
}{eq:res_gauge}
Following \cite{Martin-Garcia:2003xgm}, we fix this redundancy by requiring that in a Fourier expansion, the $k=2$ mode of the function $f$ is purely imaginary at $x=0$. That this is always possible can be seen by expanding in Fourier modes
\begin{align}
    f(\tau ,0)=\sum_{k=0}^\infty \text{Re}(\hat{f}_{k})\cos \Big(\frac{2\pi k}{\Delta }\tau \Big)+\text{Im}(\hat{f}_k)\sin \Big(\frac{2\pi k}{\Delta }\tau \Big)
\end{align}
with $\hat{f}_{k}=\frac{1}{\Delta }\int _0^\Delta \extd \tau e^{2\pi ik\tau /\Delta }f(\tau,0)$. Under the transformation \eqref{eq:res_gauge} the coefficients change as
\begin{align}
    \text{Re}(\hat{f}_k)&\to \text{Re}(\hat{f}_k)\cos\Big(\frac{2\pi k}{\Delta }\tau _0\Big)+\text{Im}(\hat{f}_k)\sin\Big(\frac{2\pi k}{\Delta }\tau _0\Big)\\
     \text{Im}(\hat{f}_k)&\to \text{Im}(\hat{f}_k)\cos\Big(\frac{2\pi k}{\Delta }\tau _0\Big)-\text{Re}(\hat{f}_k)\sin\Big(\frac{2\pi k}{\Delta }\tau _0\Big)
\end{align}
such that choosing
\begin{align}
    \tau_0=-\frac{\Delta}{2\pi k}\tan^{-1}\Big(\frac{\text{Re}(\hat{f}_k)}{\text{Im}(\hat{f}_k)}\Big)
\end{align}
for any fixed $k>0$ yields $\text{Re}(\hat{f}_k)=0$. As stated, at $x=0$ we demand
\begin{align}\label{eq:gaugef2}
    \text{Re}(\hat{f}_2)=0 ~.
\end{align}
This last condition removes all redundancies from the equations of motion, which we present in the next Subsection.

\subsection{Gauge-fixed first-order equations of motion}\label{sec:3.3}

To obtain first-order equations of motion, we introduce null projections of derivatives of the scalar field, 
\begin{equation}
\psi_+:=\sqrt{\frac{1}{D-2}}\frac{x}{f}\,v^\mu\partial_\mu \psi\qquad\qquad \psi _-:=\sqrt{\frac{1}{D-2}}\frac{x}{f}\,u^\mu\partial_\mu \psi
\label{eq:psipm}
\end{equation}
with the future-pointing radial null vectors
\begin{equation}
v^\mu\partial_\mu=\partial_\tau+(f+x)\partial_x \qquad\qquad u^\mu\partial_\mu=\partial_\tau-(f-x)\partial_x\,. 
\label{eq:nullvectors}
\end{equation}
As a consequence of the equations of motion \eqref{eq:eomg}-\eqref{eq:eomX} and the assumption of DSS the stress energy tensor components have to satisfy $T_{\alpha \beta}(\tau,x)=T_{\alpha \beta}(\tau +\Delta,x)$ and $T_{X}(\tau,x)=T_{X}(\tau +\Delta,x)$. This in turn implies that $\psi _\pm $ are periodic functions themselves. For the massless scalar field this leaves the possibility for quasi-periodicity, i.e.
\eq{
    \psi(\tau,x) =\win\,\tau +\psi_{\textrm{\tiny periodic}}(\tau,x)
}{eq:winding}
with some $\win\in\mathbb{R}$ that can be interpreted as a winding number if the $\tau$-direction is compactified and $\win$ takes integer values. While there could be scenarios where this coefficient plays a role, for the present case there is numerical evidence that it vanishes (see, e.g., \cite{Gundlach:2002sx}). We therefore fix $\win=0$.

Once these variables are adopted, one can write out the equations of motion \eqref{eq:eomg}-\eqref{eq:eomKG} explicitly. As it will be practical for us to think of $x$ as an evolution direction, they are split into four dynamical equations
\begin{subequations}
    \label{eq:eom}
\begin{align}
    x\partial _x\omega &=(D-3)(1-e^\omega )+\frac{1}{2}\big(\psi _+^2+\psi _-^2\big) \label{eq:eom1}\\
    x\partial _xf&=(D-3)(e^\omega -1)f
    \label{eq:eom2} \\
    \frac{2x}{f}v^\mu \partial _\mu \psi _-&=\big(D-2-2(D-3)e^\omega \big)\psi _-+(D-2)\psi _+\label{eq:eom3}\\
    \frac{2x}{f}u^\mu \partial _\mu \psi _+&=\big(2-D+2(D-3)e^\omega \big)\psi _++(2-D)\psi _-\label{eq:eom4}
\end{align}
and one constraint that is redundant with the equations above
\begin{align}
    \partial_\tau \omega &=\frac{(f-x)\psi _+^2-(f+x)\psi _-^2}{2x}+(D-3)(e^\omega -1) ~. \label{eq:constr}
\end{align}
\end{subequations}

The main task of the remainder of our paper is to (numerically) solve the system of coupled nonlinear first-order PDEs \eqref{eq:eom}, which is our first key result. We postpone a thorough discussion of the limits $D\to\infty$ and $D\to3^+$ until Section \ref{sec:6}, but just by looking at the equations above, it is clear that one might expect dramatic changes in the behavior of the CSC in these limits, since various numerical factors either diverge or become zero in such limits.

\subsection{Parity, regularity, convexity and boundary conditions}\label{sec:3.4}

We impose additional parity conditions on the Weyl factor $\omega$ and the SSH function $f$ which is likewise motivated by empirical evidence \cite{Martin-Garcia:2003xgm}. Requiring them to be even functions, $\omega(\tau+\Delta/2,x)=\omega(\tau,x)$ and similarly for $f$ one can infer that the matter variables $\psi_\pm$ have to satisfy $\psi_\pm (\tau +\Delta/2,x)= -\psi_\pm(\tau,x)$. This is so, because they cannot be also even functions since otherwise we would just have half the echoing period, $\Delta\to\Delta/2$. The oddness of $\psi_\pm$ is compatible with the evenness of the Weyl factor and the SSH function since the matter variables only couple quadratically to the geometric variables.

For solving the system we restrict to one fundamental domain $0\leq \tau \leq \Delta $ and impose boundary conditions at the center $x=0$ and at the SSH $x=1$. In both cases, we require that all fields are Taylor expandable such that the solution is analytic in the whole domain. While one can, in principle, continue the solution to larger values of $x$ (see \cite{Martin-Garcia:2003xgm}), this is not necessary in our case since we are mainly interested in the echoing period and the critical exponent, which can already be extracted by solving the system for $x\in [0,1]$. From the expansion around the center of the $D$-dimensional Ricci scalar
\begin{multline}
     R^{(D)} \big|_{x\ll 1}=\big(1-e^{-\omega}\big)\Big \vert _{x=0}\frac{(D-2)(D-3)e^{2\tau}}{x^2} \\
     +\big[e^{-\omega }\big((D-3)\partial_x\omega-2\partial_x\ln{f}\big)\big]\Big \vert_{x=0} \frac{(D-2)e^{2\tau}}{x} + \mathcal{O}(1)
\label{eq:riccireg}
\end{multline}
one finds the regularity conditions $\omega(\tau ,0)=0$ and $(D-3)\partial_x\omega (\tau ,0)=2\partial_x\ln f(\tau,0)$ that cancel the second and first-order poles in $x$.

Solving the Taylor-expanded equations of motion \eqref{eq:eom} order by order in $x$ we find that regularity at $x=0$ imposes additional conditions,
\eq{
\omega=\psi^\pm=\partial_x\omega=\partial_x f=\partial_x(\psi^+-\psi^-)=\partial_x^2(\psi^++\psi^-)=0
}{eq:regularityorigin}
evaluated in the center $x=0$ for all values of time $\tau$. In principle, one can push the Taylor expansion for $x\to 0^+$ to arbitrarily high order, but this perturbative approach is insufficient to construct the CSC. Nevertheless, such a Taylor expansion is useful to numerically provide boundary data at a cutoff surface close to $x=0$ and then use that expansion to extrapolate all the way to the center. We discuss details of this numerical scheme in Section \ref{sec:4} and provide formulas for the Taylor expansions around $x=0^+$ and $x=1^-$ in Appendix \ref{app:Taylor}.

Using the near-center expansion $\omega = \omega_2(\tau)\,x^2 + \mathcal{O}(x^4)$, we prove that $\omega$ is non-negative in the whole fundamental domain for $D>3$. First, as shown in Appendix~\ref{app:Taylor}, the coefficient $\omega_2 = \psi_{1+}(\tau)^2/(D-1)$ is non-negative, which implies that $\omega$ is non-negative close to the center. Second, suppose there exist values $x\in(0,1)$ for which $\omega<0$, and define $x_\ast$ as the infimum of such values. Since we know $\omega$ is positive for small positive $x$ this infimum must be zero because the positive part of $\omega$ must connect smoothly to the negative part, i.e., $\omega|_{x=x_\ast}=0$. This implies $\partial_x\omega|_{x=x_\ast}<0$ since the values of $\omega$ are negative by assumption to the immediate right of $x_\ast$. However, evaluating \eqref{eq:eom1} at this infimum yields a contradiction: $x_\ast\partial_x\omega|_{x=x_\ast}=\frac12(\psi_+^2+\psi_-^2)\geq 0$, i.e., the $x$-derivative of $\omega$ is non-negative at the infimum. Therefore, our assumption that $\omega$ becomes negative somewhere is incorrect, and hence $\omega$ must remain non-negative.

Using the convexity condition that we just proved, $\omega\geq 0$, we prove an additional chain of inequalities, 
\eq{
0\leq x\leq f\leq 1\,. 
}{eq:inequalities}
The inequalities $0\leq x\leq 1$ and $x\leq f$ are trivial and simply follow from our choice of domain and the fact that $x=1$ was designed to be the SSH. While the remaining inequality $f\leq 1$ is non-trivial, it follows from \eqref{eq:eom2} whose right-hand side must be non-negative since $0\leq x\leq f$ implies $f$ is non-negative and we have shown above that $e^\omega-1$ is non-negative, too. Therefore, $f$ is monotonically increasing with $x$. Since by construction $f=1$ at $x=1$ this implies $f\leq 1$ in the whole interval $x\in[0,1]$, which is what we wanted to prove. The inequalities above played an important role in checking the consistency of subleading corrections in the large-$D$ expansion \cite{Ecker:2026akf}.

There is also a regularity condition implied by \eqref{eq:eom4} when evaluated at the SSH,
\eq{
\partial_\tau\psi_+\big|_{x\to 1^-}=\frac D2\,\big(\psi_+-\psi_-\big) + \psi_- - 2\psi_+ + (D-3)\,\big(e^\omega-1\big)\,\psi_+\Big|_{x\to 1^-} 
}{eq:SSH}
which linearly relates $\psi_\pm$ with each other for given $\omega$. Taylor expanding near $x=1$ does not yield additional regularity conditions at the SSH. It is technically a bit more involved than the expansion near the center since it requires at each order in $(1-x)$ integrating some function in $\tau$ with the integration constant fixed by periodicity, see Appendix \ref{app:Taylor}.

Having imposed the parity and regularity conditions, we still need to impose boundary conditions on the free functions to uniquely solve the first-order equations of motion \eqref{eq:eom}. If we shoot from the center, due to the regularity condition of finite Ricci scalar \eqref{eq:riccireg}, we have only two free boundary functions available. At the SSH, due to the regularity condition \eqref{eq:SSH} and the choice $f(\tau,x=1)$, we only have one free boundary function. Their choices, discussed below, are compatible with the Taylor expansions in Appendix \ref{app:Taylor}. 

As boundary conditions in the center (or a cutoff surface nearby), we start from the two free functions
\eq{
    f_c(\tau ):= f(\tau,0) \qquad\qquad \Psi_c(\tau):=\lim_{x\to 0}\Big(\frac{\psi_+(\tau,\,x)-\psi_-(\tau,\,x)}{2x^2}\Big)
}{eq:inidata_center}
where $f(\tau,0)$ is subject to the residual gauge choice \eqref{eq:gaugef2}. Note that $\Psi_c(\tau)$ is finite since the regularity conditions \eqref{eq:regularityorigin} imply that the difference between $\psi_+$ and $\psi_-$ is only of order $\mathcal{O}(x^2)$. It is numerically advantageous to use this function instead of, say, $\psi_+(\tau,0)$ because $\Psi_c(\tau)$ has pronounced features that we will discuss in later Sections. At the SSH (or a cutoff surface nearby), we use the free function 
\eq{
    \psi_{-p}(\tau):=\psi _-(\tau,1)\,.  
}{eq:up_def}

The right definition \eqref{eq:inidata_center} and the Taylor expansions in Appendix \ref{app:Taylor} suggest defining the matter field combinations
\eq{
\Pi(\tau,\,x):=\frac{\psi_+(\tau,\,x) + \psi_-(\tau,\,x)}{2x}\qquad\qquad\Psi(\tau,\,x):=\frac{\psi_+(\tau,\,x) - \psi_-(\tau,\,x)}{2x^2}
}{eq:pipsi}
that are finite in the center and have only even Taylor coefficients $x^{2n}$ with $n\in\mathbb{Z}^\ast$.

\subsection{Linearized equations}\label{sec:3.5}

We linearize the equations of motion \eqref{eq:eom} by introducing perturbations $\{\delta f,\,\delta \omega,\,\delta \psi_+,\,\delta \psi_-\}$ collectively denoted by $\delta Z$, i.e., the linearized equations of motion are linear PDEs for $\delta Z$. While $\delta Z$ is not going to be periodic in $\tau$, we can still expand it in terms of periodic functions $\delta_i Z$,
\begin{align}\label{eq:linsol_general}
    \delta Z(\tau,x)&=\sum_{i=1}^\infty e^{\lambda _i\tau}\,\delta_iZ(\tau,x) & \delta_iZ(\tau+\Delta,x)=\delta_iZ(\tau,x)
\end{align}
where the exponents $\lambda_i$ are either real or come in complex conjugate pairs.

For the Choptuik system, it was found that there exists only a single $\lambda _i=:\lambda $ with $\text{Re}(\lambda )>0$ and $\text{Im}(\lambda )=0$ which corresponds to the single unstable mode of the critical solution. The critical exponent is then given by $\gamma =1/\lambda $ \cite{Gundlach:1995kd}. Following \cite{Gundlach:1996eg}, we formulate the problem of computing $\lambda $ as an eigenvalue problem. Inserting the ansatz \eqref{eq:linsol_general} into the linearized system one finds equations for the $\delta _iZ$ which read 
\begin{align}
x\partial _x\delta _i\omega &=-(D-3)e^\omega \delta _i\omega +\psi _+\delta _i\psi _++\psi _-\delta _i\psi _-\label{eq:lin1}\\[.5em]
    x\partial _x\delta _if&=(D-3)e^\omega f\delta _i\omega +(D-3)(e^\omega -1)\delta _if \label{eq:lin2}\\[.5em]
    \frac{2x}{f}v^\mu \partial _\mu \delta _i\psi _- &=\frac{2x}{f^2}\big((\partial_\tau+x\partial_x)\psi_-\big)\,\delta_if -2(D-3)e^\omega \psi _-\delta _i\omega \nonumber \\
     & \quad +\big(D-2-2(D-3)e^\omega \big)\delta _i\psi _-+(D-2)\delta _i\psi _+-\frac{2x}{f}\lambda _i\delta _i\psi _-\label{eq:lin3}\\
     \frac{2x}{f}u^\mu \partial _\mu \delta _i\psi _+ &=\frac{2x}{f^2}\big((\partial_\tau+x\partial_x)\psi_+\big)\,\delta _if +2(D-3)e^\omega \psi _+\delta _i\omega \nonumber \\
     & \quad +\big(2-D+2(D-3)e^\omega \big)\delta _i\psi _++(2-D)\delta _i\psi _--\frac{2x}{f}\lambda _i\delta _i\psi _+ \label{eq:lin4}\\
     \partial _\tau \delta _i\omega &=\frac{\psi _+^2-\psi _-^2}{2x}\delta _if-\psi _+\delta _i\psi _+\Big(1-\frac fx\Big)-\psi _-\delta _i\psi _-\Big(1+\frac fx\Big)\nonumber\\
     & \quad +(D-3)e^\omega \delta _i\omega - \lambda_i\delta_i\omega \label{eq:lin5}
\end{align}
and explicitly include the eigenvalue $\lambda_i$ corresponding to a specific solution. We also assume that the $\delta _iZ$ have the same parities as the corresponding background fields, i.e., $\delta_i\omega$, $\delta_if$ are even functions in $\tau$ while $\delta_i\psi_\pm$ are odd. In principle, there is also a sector with the opposite symmetry properties which, however, was not found to include unstable modes \cite{Gundlach:1996eg} and shall therefore be discarded hereafter. 

Formally, finding $\lambda$ at fixed $D$ amounts to solving $\det(L - \lambda) = 0$, where $L$ is the differential operator in the eigenvalue problem $L\,\delta_i Z = \lambda_i\,\delta_i Z$ obtained from the subset of equations \eqref{eq:lin3}--\eqref{eq:lin5} that involve $\lambda_i$. Imposing analyticity over the entire integration domain then yields a unique real value of $\lambda$ in the right half-plane for each value of the dimension $D$. The Choptuik exponent as a function of the dimension is then given by
\eq{
\gamma(D) = \frac{1}{\lambda(D)}\,.
}{eq:choptuikexponent}

\subsection{Observables}\label{sec:3.6}

Before we display some formulas for relevant observables we make a general comment on the Choptuik scaling of these quantities. Whenever a quantity $Q_c$ evaluated for a CSC configuration is not periodic but has instead some scaling behavior of the form $Q_c=e^{-\beta\tau}\cdot(\textrm{periodic})$ with some real (typically integer) coefficient $\beta$, this quantity close to the critical point scales at late times as 
\eq{
Q\sim|p-p_\ast|^{\beta\gamma}
}{eq:choptuikgeneral}
which follows essentially from a dimensional analysis (see, e.g., Section 2.3 in \cite{Gundlach:2007gc}).

As an example, we explain the factor $(D-3)$ in the Choptuik exponent of the mass \eqref{eq:2} using these considerations. Operationally, the black hole mass is determined from the Misner--Sharp mass $m$ defined by $1-2m/\Phi^{D-3}=(\partial \Phi)^2$, yielding
\eq{
m=\frac{1}{2}e^{-(D-3)\tau}x^{D-3}\big(1-e^{-\omega}\big)
}{eq:misnersharp}
such that an apparent horizon forms at finite $\tau$ and $x$ when $\omega\to+\infty$. Of course, this cannot happen for our CSCs, but it does happen for supercritical solutions above the threshold of black hole formation. Comparing with our general remarks above, we read off the coefficient $\beta=D-3$, which recovers the Choptuik scaling \eqref{eq:2} when inserted into the general expression \eqref{eq:choptuikgeneral}. The same logic applies to all the observables below, so we do not display their Choptuik scalings. 

We list here analytic expressions for the main observables, starting with the $D$-di\-men\-sio\-nal Ricci scalar
\eq{
R^{(D)} = -(D-2)\,e^{2\tau}\,e^{-\omega}\,\frac{\psi_+\psi_-}{x^2} \,.
}{eq:RD}
By contrast, the 2d Ricci scalar is given by
\eq{
    R = -(D-2)\,e^{2\tau}\,e^{-\omega}\,\frac{\psi_+\psi_- + (D-3)\big(1-e^\omega\big)}{x^2}  \,.
}{eq:R2}
Due to our regularity conditions \eqref{eq:regularityorigin}, $R^{(D)}$ and $R$ are regular in the center $x=0$.

Spherically symmetric spacetimes in any spacetime dimension allow up to four algebraically independent scalar curvature invariants \cite{Narlikar:1949}. We choose as a basis of invariants the dilaton $X$, the $D$-dimensional Ricci scalar $R^{(D)}$, and the matter quantities $\psi_\pm$. 

The 2d energy-momentum tensor
\eq{
T_{\mu\nu} = \frac{D-2}{4}\, e^{-(D-2)\tau} x^{D-4} \begin{pmatrix}
    (f-x)^2\psi_+^2+(f+x)^2\psi_-^2 & (f-x)\psi_+^2-(f+x)\psi_-^2\\
    (f-x)\psi_+^2-(f+x)\psi_-^2 & \psi_+^2 + \psi_-^2
\end{pmatrix}
}{eq:emt2d}
has to be supplemented by the scalar quantity
\eq{
T_X = -\frac{(D-2)\,e^{2\tau}}{2x^2}\,e^{-\omega}\psi_+\psi_- = \frac12\,R^{(D)}\,.
}{eq:Tscalar}
The 2d stress tensor \eqref{eq:emt2d} has vanishing trace and generically non-zero determinant,
\eq{
\sqrt{\det T_{\mu\nu}} = \frac{D-2}{2}\,e^{-(D-2)\tau} x^{D-2}\,\frac{\psi_+\psi_-}{x^2} \,.
}{eq:detT}

The NEC inequalities
\eq{
T_{\mu\nu}v^\mu v^\nu = (v^\mu\partial_\mu\psi)^2 \geq 0 \qquad\qquad T_{\mu\nu}u^\mu u^\nu = (u^\mu\partial_\mu\psi)^2 \geq 0 
}{eq:NEC1}
always hold, as expected on general grounds. The condition that at least one of the NEC inequalities saturates reduces to
\eq{
\textrm{NEC\;saturation:}\qquad \frac{\psi_+\psi_-}{x^2} = 0\,.
}{eq:NEC2}

As discussed in \cite{Ecker:2024haw}, there are four lines in each fundamental domain where $R^{(D)}$ vanishes (see inset of Fig.~\ref{fig:1}). They intersect pairwise at the origin. Comparing \eqref{eq:RD} with \eqref{eq:NEC2} we deduce that these lines coincide with lines of NEC saturation. Therefore, we can use the intersection angle of these NEC lines as a gauge invariant geometric observable. We call it the NEC angle, denote it by $\alpha$, and, using the Taylor expansion near the origin, we are able to calculate it exactly, as described in \cite{Ecker:2024haw}. The result is
\eq{
\alpha = \textrm{gd}(\xi) = 2\,\textrm{arccot}(D-1)
}{eq:NECangle}
where $\xi$ is the relative rapidity between the two spacelike vectors tangent to the NEC lines that intersect in the center and gd is the Gudermannian.


\section{Numerical procedure}\label{sec:4}

This Section summarizes our numerical procedure to construct CSCs in arbitrary dimensions and extract their echoing periods $\Delta(D)$ and critical exponents $\gamma(D)$. Our algorithm closely follows the one in~\cite{Martin-Garcia:2003xgm,Gundlach:2007gc}, with the notable difference that we promote the spacetime dimension to a free parameter. Our C{}\verb!++! implementation, including source code, build instructions, and examples, is openly available at \url{https://github.com/tobjec/parallel-critical-collapse}, while the plotting scripts and corresponding data files used to reproduce the figures in our manuscript can be found at \url{https://github.com/EckerChristian/CriticalSpacetimeCrystal}.
In the following, we outline the essential components of the algorithm and refer readers interested in the details of the error budget and convergence checks of the implementation to Appendix~\ref{app:Error}.

\subsection[Critical spacetime crystals and echoing period \texorpdfstring{$\Delta$}{Delta}]{Critical spacetime crystals and echoing period \texorpdfstring{$\mathbf{\Delta}$}{Delta}}\label{sec:4.1} 

To obtain the CSC and its echoing period for a given dimension we solve the system of hyperbolic non-linear PDEs \eqref{eq:eom1}-\eqref{eq:constr} in one fundamental domain of the past patch. As in $D = 4$, it turns out that for any $D$ there exists a unique solution for a specific value of the echoing period $\Delta$. To construct these solutions, we formulate the system of equations as a boundary value problem for the functions $\{f,\omega,\psi_-,\psi_+\}$.

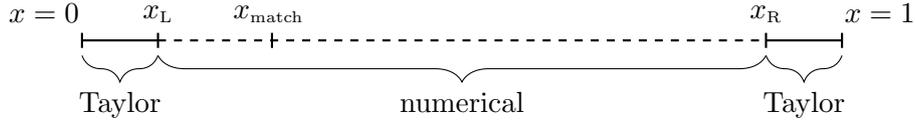
\begin{figure}[h!]
 \centering
 \begin{tikzpicture}
  \draw[thick] (0,0) -- (1,0);
  \draw[thick,dashed] (1,0) -- (9,0);
  \draw[thick] (9,0) -- (10,0);
  \foreach \x in {0, 1,2.5, 9, 10} {
   \draw[thick] (\x,0.1) -- (\x,-0.1); 
  }
  \node[above] at (-0.5,0.1) {$x=0$};
  \node[above] at (1,0.1) {$\xl$};
  \node[above] at (2.5,0.1) {$x_{\textrm{\tiny match }}$};
  \node[above] at (9,0.1) {$\xr$};
  \node[above] at (10.5,0.1) {$x=1$};
  \draw[decorate, decoration={brace, amplitude=10pt,mirror}] (1,-0.2) -- (9,-0.2) node[midway, below=10pt] {numerical};
  \draw[decorate, decoration={brace, amplitude=10pt,mirror}] (0,-0.2) -- (1,-0.2) node[midway, below=10pt] {Taylor};
  \draw[decorate, decoration={brace, amplitude=10pt,mirror}] (9,-0.2) -- (10,-0.2) node[midway, below=10pt] {Taylor};
 \end{tikzpicture}
 \caption[Decomposition of $x$-domain for numerics]{Decomposition of $x$-domain into near-boundary regions (solid lines; solution approximated by Taylor series) and interior region (dashed lines; equations solved numerically from cutoffs $\xl$ and $\xr$ toward matching surface $\xm$).}
 \label{fig:x_segments}
\end{figure}

The boundary conditions are fixed by imposing the regularity conditions \eqref{eq:regularityorigin} and by specifying three free functions together with their periodicity $\Delta$, namely $f_c(\tau)$ and $\Psi_c(\tau)$ defined in \eqref{eq:inidata_center} at the center as well as $\psi_{-p}(\tau)$ defined in \eqref{eq:up_def} at the SSH. Since the system \eqref{eq:eom1}–\eqref{eq:constr} has regular singular points\footnote{%
This jargon is borrowed from linear ODE theory. Around such points, one can still expand the solution even though the ODE looks singular. In our case, since we explicitly assume analyticity, the expansions reduce to Taylor expansions.
}
there, we impose the boundary conditions for the numerical treatment by evaluating the corresponding series expansions (see Appendix \ref{app:Taylor}) at cutoff surfaces placed a small distance away from the boundaries. The locations of the left and right cutoff surfaces are chosen as
\begin{align}
   \xl =\begin{cases}
   \;\;\;~10^{-2}\;\;,\;\; 4.8\leq D\leq 5.5 \\
   \;\;\;~10^{-3}\;\;,\;\; 3.2\leq D<4.8\\
    \;\;\;~10^{-4}\;\;,\;\; 3.1\leq D< 3.2\\
    5\cdot 10^{-5} \;\;,\;\;3.05\leq D<3.1
    \end{cases}
    \qquad \xr = \begin{cases}
        1-10^{-2}\;\;,\;\; 4.8\leq D \leq 5.5\\
        1-10^{-3}\;\;,\;\; 3.1\leq D<4.8 
    \end{cases}\,.
\end{align}
The Taylor series from the left includes terms up to $\mathcal{O}(\xl^5)$ while those from the right up to $\mathcal{O}((1-\xr)^2)$ for $D<4.8$ and up to $\mathcal{O}((1-\xr)^3)$ for $4.8\leq D\leq 5.5$. The final solution is thus expected to converge with respect to the parameters $\xl$ and $\xr$ at these rates as verified in Appendix \ref{app:Error}. We chose the expansion order at the center higher than at the SSH since the functions tend to have steeper gradients there slowing down the convergence of the series. For the same reason, we have to choose the cutoff $\xl $ smaller at smaller dimensions. The expansion order around the SSH has to be increased above $D=4.8$ such that the singular solution (discussed in Appendix \ref{app:SingularBranch}) can still be resolved to leading order (LO) in $\xr$. The different regions in $x$ are depicted schematically in Fig.~\ref{fig:x_segments}, where $\xm$ denotes the location of a matching surface, separating the computational domain into two patches $x\in [\xl,\xm]$ and $x\in [\xm,\xr]$ on which the equations are solved independently. We place the matching surface closer to the center at $\xm=0.1$ to improve convergence.

For the numeric treatment, the computational domain is discretized. In $x$-direction we use a non-uniform grid which is constructed by taking a fixed number of points  $N_x$ and distributing them logarithmically around the boundaries in $x$. This is done by transforming a uniform grid in an auxiliary variable $\bar x$ to $x$ such that grid points accumulate near the boundaries,
\eq{
    x=\frac{e^{\bar x}}{1+e^{\bar x}} ~.
}{eq:whatever}
At the boundary $\xr$ this is necessary for stability of the $x$-integration and at $\xl $ the functions develop steep gradients which otherwise cannot be resolved. For the latter reason we also need larger values of $N_x$ the smaller the dimension gets to keep the integration routine stable. At $D=3.05$ we choose $N_x=1.55\cdot 10^5$ which gradually decreases until $N_x=8000$ at $D\geq 4$.

In the $\tau$-direction we employ a spectral decomposition which implements the periodic boundary conditions $\tau \sim \tau +\Delta $ for all fields exactly. More explicitly, we perform at each instance of $x$ a Fourier decomposition 
\eq{
    Z(\tau_n,x)=\hat{Z}_0(x)+\hat{Z}_{\frac{N_\tau}{2}}(x)\cos\Big(\frac{\pi N_\tau}{\Delta}\tau_n\Big)
    +\sum _{k=1}^{\frac{N_\tau}{2}-1}\big(\hat{Z}_k(x)e^{-\frac{2\pi ik}{\Delta}\tau_n}+\hat{Z}_{N_\tau -k}(x)e^{\frac{2\pi ik}{\Delta }\tau_n}\big)
}{eq:fourier}
where $Z$ stands for either of the functions $\{\omega ,f,\psi _+,\psi _-\}$ and $\tau _n=n\Delta /N_\tau $ for $n=0,...,N_\tau -1$.
The number of modes $N_\tau$ required to accurately resolve the solution depends on the chosen dimension. We find $N_\tau = 512$ to be sufficient for $D \gtrsim 3.2$, whereas at least $N_\tau = 1024$ is needed for lower dimensions, because some of the functions develop narrow features and steep gradients in $\tau$ as $D\to3$.

Since all functions $Z$ are real, their Fourier coefficients satisfy
\begin{align}
    \hat{Z}_k^\ast (x) =\hat{Z}_{N_\tau -k}(x)~.
\end{align}
Moreover, whenever $Z$ is an even function in $\tau $ we have $\hat{Z}_{2l+1}(x)=0$ for $l\in \mathbb{Z}$ while $\hat{Z}_{2l}(x)=0$ for $Z$ odd. This permits storing the Fourier coefficients of the functions $\{f,\psi _+,\psi _-\}$ in a single complex vector of length $N_\tau $
\begin{align}
    \hat{Y}_k=\hat{\psi }_{-,k}+i(\hat{\psi }_{+,k}+\hat{f}_k) \qquad \qquad k=0,...,N_\tau -1
\end{align}
which due to the mentioned symmetries carries $3N_\tau /2$ independent real degrees of freedom. Since $f$ is even and $\psi_\pm $ are odd, one can always extract them by taking combinations of the even/odd, respectively real/imaginary parts of the coefficients. We use the vector $\hat{Y}_k$ as the variables of the shooting. All Fourier transforms are computed using the Fast Fourier Transform implemented in the FFTW library~\cite{FFTW05}.

The shooting from the two cutoff surfaces towards $\xm$ is performed with a two-stage implicit Runge--Kutta algorithm of order four which is needed to keep the evolution as stable as possible without making it computationally too expensive. Additionally, the evolution is constrained: at each integration step in $x$, Eq.~\eqref{eq:constr} is solved for $\hat{\omega }_k(x)$, and the result is substituted into \eqref{eq:eom2}-\eqref{eq:eom4} to compute $\hat{Y}_k$ at the next grid point. The advantage of this procedure is that at a fixed $x$, \eqref{eq:constr} can be solved exactly for $\omega(\tau,x)$. Indeed, this equation can always be brought into the form
\eq{
    \partial_\tau y+ay+b=0
}{eq:typical_ODE}
with $a(\tau)$ and $b(\tau)$ two periodic functions. Whenever $\Bar{a}:=\int _0^\Delta a \neq 0$, which turns out to be the case for most of our purposes (see the end of Subsection \ref{sec:chop_num} for an exception), the equation has a unique solution satisfying periodic boundary conditions given by
\begin{align}\label{eq:sol_ode}
    y(\tau)=\frac{1}{\mu}\Big(\frac{1}{1-e^{\Bar{a}}}\int\limits_0^\Delta \extd z \,\mu (z)b(z)-\int\limits_0^\tau \extd z\, \mu (z)b(z)\Big) && \mu (\tau )=\exp{\int\limits_0^\tau \extd z\, a(z)} ~. 
\end{align}

Since all the equations involve products between periodic fields, one has to make sure that possible high-frequency components of these products are not aliased into the low-frequency range. To mitigate the resulting error, we perform anti-aliasing by padding the Fourier modes with zeros (doubling their number) before transforming to real space, carrying out the multiplication with $2N_\tau$ components there, and then discarding the spurious high-frequency modes after transforming back to Fourier space.

Once the initial data have been evolved to $x_{\textrm{\tiny match}}$ from left and right, we compute the mismatch vector between the functions $\{f,\psi_+,\psi_-\}$ in the fundamental interval $\tau \in [0,\Delta/2)$, i.e.,
\begin{align}
    M_i&=\big(f_{0},...,f_{N_\tau/2-1},\psi _{+,0},...,\psi _{+,N_\tau/2-1},\psi _{-,0},...,\psi_{-,N_\tau /2-1}\big)\big |_{x=x_{\textrm{\tiny match }}}^{{\rm L}
    }\nonumber\\
   &\quad -\big(f_{0},...,f_{N_\tau/2-1},\psi _{+,0},...,\psi _{+,N_\tau/2-1},\psi_{-,0},...,\psi _{-,N_\tau/2-1}\big)\big |_{x=x_{\textrm{\tiny match}}}^{{\rm R}
   }
\end{align}
which is thus a vector of length $3N_\tau/2$. Its components are nonlinear functions of the initial data $f_c$, $\Psi _c$, $\psi _{-p}$ as well as $\Delta$ which are also $3N_\tau/2$ variables denoted by $z_j$. Importantly, this balance is achieved since the real part of the Fourier mode $\hat{f}_{c,2}$ is fixed by the condition~\eqref{eq:gaugef2} and is replaced by $\Delta$ in the mismatch vector.

Starting from an initial guess $z_j=z_j^{(0)}$, the converged solution is obtained by minimizing the $\ell^2$-norm of $M_i$ with a Newton algorithm down to machine precision. A single Newton step is thereby performed by inverting the linear relation 
\begin{align}
J_{ij}\delta z_j^{(n)}=-M_i(z_k^{(n)}) && J_{ij}=\frac{\partial M_i}{\partial z_j}\approx \frac{M_i(\ldots ,z_j^{(n)}+\epsilon,\ldots)-M_i(\ldots , z_j^{(n)}, \ldots)}{\epsilon}
\end{align}
where we use a constant finite-difference step size $\epsilon=10^{-10}$ to approximate derivatives in the Jacobian $J_{ij}$. The Jacobian is then inverted, using the Linear Algebra PACKage (LAPACK)~\cite{anderson_lapack_1999}, and multiplied with the mismatch vector to obtain $\delta z_i^{(n)}$ which is then used to update the boundary data according to
\begin{align}
z_i^{(n+1)}=z_i^{(n)}+\eta\,\delta z_i^{(n)} ~.
\end{align}
Here, a damping factor $\eta=0.1$ is introduced pragmatically to ensure monotonic decrease of the mismatch norm $M$. 

Finally, let us outline our strategy for obtaining solutions at a given dimension by starting from a solution at a nearby value of $D$. For the Newton method to converge it is essential to have an initial guess that is already close to the solution which turns out to be one of the main obstacles in this procedure. Starting from the known result \cite{Martin-Garcia:2003xgm} at $D=4$ we have to compute the solutions towards larger and smaller $D$ sequentially, using always the closest converged result as an initial guess for the next value of $D$. Once several solutions have been obtained, we begin using quadratic extrapolation to construct the next initial guess at a typical step size of $\delta D \approx 0.01$ from the previous dimension.

\subsection[Choptuik exponent \texorpdfstring{$\gamma$}{gamma}]{Choptuik exponent \texorpdfstring{$\boldsymbol{\gamma}$}{gamma}}
\label{sec:chop_num}
As explained in Section~\ref{sec:3.5}, the computation of the critical exponent associated to the unstable mode of the CSC can be phrased in terms of an eigenvalue problem for the linearized equations \eqref{eq:lin3}-\eqref{eq:lin5}. Since the linearized fields are chosen to have the same symmetries and satisfy the same boundary conditions as the background, the algorithm of the previous Section can be reused to a large extent. Also, as we already know that the $D=4$ solution admits a single real $\lambda >0$ we shall take this as a working assumption for general $D$ and assume $\lambda$ to be real from the start, making the computation considerably less expensive. Furthermore, we find that computing $\lambda$ is less sensitive to the choice of initial guess than the background calculation. Consequently, once the background solutions have been obtained, the computation can be carried out independently for each dimension.

The procedure is outlined as follows. Starting from some initial data $(\delta f_c,\delta \Psi _c,\delta  \psi_{-p})$ as well as a guess for $\lambda $ we evolve the fields from both cutoff surfaces towards the matching surface using the integration scheme detailed in the previous Subsection. Here we have to allow $\delta \text{Re}(\hat{f}_{c,2})\neq 0$ since there is no freedom left to fix this mode. At the matching surface, we compute the mismatch of the fields $\delta _iZ_{L}(x_{\textrm{\tiny match }})-\delta _iZ_{R}(x_{\textrm{\tiny match }})=A(\lambda )\cdot (\delta f_c,\delta \Psi _c,\delta \psi_{-p})$ which has to be linear in the initial data with the matrix $A(\lambda)$ being a non-linear function of $\lambda $. In order for the system to admit a smooth solution the mismatch has to vanish which is equivalent to 
\eq{
    \det A(\lambda)=0 ~.
}{eq:detA}
Therefore, the choice of initial data is not entering the result for $\lambda $ and should just be made conveniently such that the numeric integration has good stability properties, which we find to be the case after rescaling the converged background data by a factor $10^{-5}$.

To solve \eqref{eq:detA}, we begin with an initial guess for $\lambda$ and bracket its root using steps of $\Delta \lambda = 10^{-2}$. Once the root is bracketed, we apply Brent’s method~\cite{numerical_recipes_f77} to converge to it with a precision of $10^{-7}$ or better. Since the matrix $A$ has dimensions $3N_\tau/2 \times 3N_\tau/2$ and its entries span several orders of magnitude, we normalize it appropriately so that its determinant remains representable within double-precision arithmetic $A\to sA$ with $s=\exp{(-\frac{2}{3N_\tau}\sum_j\ln |\nu _j|)}$, where $\nu _j$ are the eigenvalues of $A$. The factor $s$ is computed with every guess of $\lambda $ until a root is bracketed. After that it stays the same but still ensures that $\det (sA)$ is approximately of $\mathcal{O}(1)$ while allowing Brent's method to converge. From the converged value for $\lambda(D)$ the Choptuik exponent $\gamma(D)$ is then computed via~\eqref{eq:choptuikexponent}.

There is one caveat when it comes to the evolution of the linearized initial data $(\delta f_c,\delta \Psi _c,\delta  \psi_{-p})$. Since for each $x$ the linearized constraint equation \eqref{eq:lin5} also takes the form \eqref{eq:typical_ODE} with $a=\lambda -(D-3)e^\omega$, in general, we are only able to solve the evolution for a value of $\lambda$ that satisfies
\eq{
    \lambda \neq \frac{D-3}{\Delta}\int_0^\Delta \dd \tau \; e^\omega \qquad \qquad \forall x \in [0,1]~.
}{eq:lin_cond_viol}
Otherwise, the zero mode of $a$ vanishes at some $x$ and a real periodic solution of the constraint does not exist any more. For most dimensions, the value of $\lambda $ corresponding to the unstable mode satisfies this condition but in the range $4.81\leq D \leq 5.38$ it turns out to be violated. In that case, the perturbations cannot be computed any more with the current method. We attribute this behavior to the chosen boundary conditions of the linearized problem. In other words, it is a mere artifact of how the transition matrix $A(\lambda)$ is defined and may indeed be changed by, e.g., linearizing in the function $\delta (e^{2\omega }\psi _-)$ instead of $\delta \psi _-$. In that case, it can be shown that the condition \eqref{eq:lin_cond_viol} is changed to $\lambda \neq \frac{1}{\Delta }\int _0^\Delta \dd \tau \; ((D-3)e^\omega -2\psi _-^2(1+f/x)) $ which is only violated for $5.41\leq D$. Since the critical value of $\lambda $ should not be affected by this shift one can just choose the boundary conditions such that the evolution works consistently.


\section{Numerical results}\label{sec:5}

In this Section, we present the numerical results obtained using the method outlined in Section \ref{sec:4}. We begin in Subsection~\ref{sec:5.1} with a discussion of the geometric properties of CSCs in various dimensions, followed by a detailed analysis of the echoing period $\Delta(D)$ in Subsection~\ref{sec:5.2}, the Choptuik exponent $\gamma(D)$ in Subsection~\ref{sec:5.42}, and end with results for the NEC lines in Subsection~\ref{sec:5.3}. Our data are available at \cite{CSCdata:2026}.

\subsection{Shapes of critical spacetime crystals in various dimensions}\label{sec:5.1}

\begin{figure}[h!tb]
\centering
  \includegraphics[width=1\linewidth]{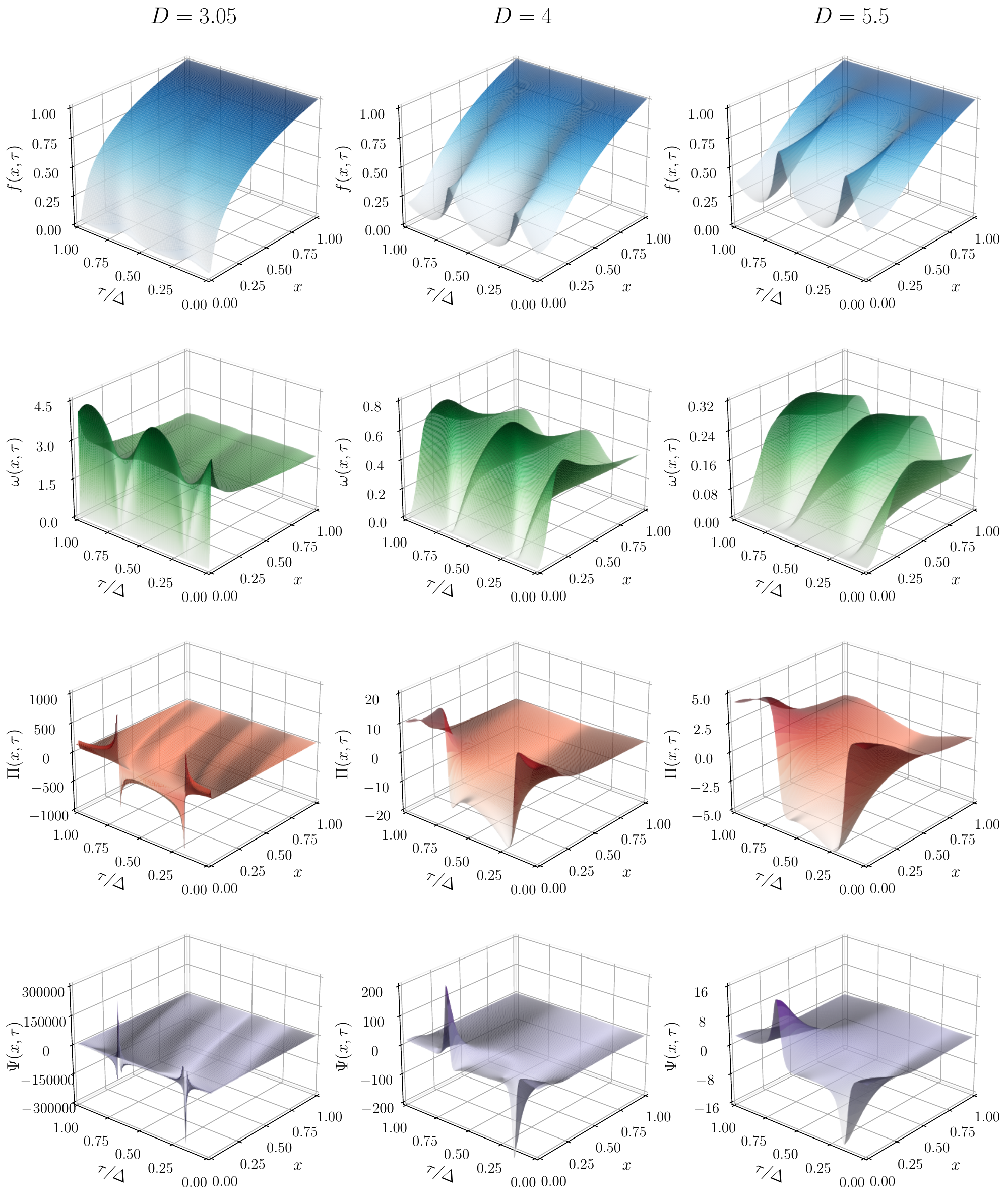}
  \caption[\textbf{3d plots of physical fields in $\boldsymbol{D=3.05}$, $\boldsymbol{D=4}$, and $\boldsymbol{D=5.5}$}]{Comparison of various fields (top to bottom) constituting CSCs in $D = 3.05$, $4$, $5.5$ (left to right) on fundamental domain $\tau/\Delta\in[0,1]$, $x\in[0,1]$.}
  \label{fig:fields}
\end{figure}

Figure~\ref{fig:fields} presents a comparison of fields that constitute numerical solutions of CSCs in three different representative dimensions, namely the smallest ($D=3.05$) and largest ($D=5.5$) dimensions obtained with our numerical procedure, alongside the well-known result in $D=4$~\cite{Martin-Garcia:2003xgm}. 
For this comparison, we normalize the $\tau$-coordinate by the corresponding value of the echoing period $\Delta$, whose dependence on $D$ will be discussed in detail in the next Subsection. 

We first discuss how the global features depend on the spacetime dimension. A common trend is that large gradients in the $x$-coordinate increase in amplitude and become increasingly concentrated toward the center ($x=0$) as $D\to3^{+}$. This trend is clearly visible from the narrow peak structure emerging most prominently in the fields $\Pi$ and in particular $\Psi$ in the vicinity of the center at $D=3.05$. This trend is less pronounced but still present for $f$, whose boundary conditions at the SSH ($x=1$) are fixed to unity, while the function decreases toward zero at the center and develops an increasingly homogeneous profile in the $\tau$-direction as the dimension approaches $D\to3^{+}$. The small width and large amplitude of the peaks in the function $\Psi$ in low dimensions is the key limiting factor preventing the convergence of our numerical routines. Qualitatively, it is easy to understand why this is the case: in low dimensions we need an ever-increasing number of Fourier modes to resolve the UV drama inherent to $\Psi$ and, additionally, have to deal with the large amplitude of $\Psi$ in the center, while its amplitude remains $\mathcal{O}(1)$ at the SSH. Moreover, the peaks in $\Psi$ fall off rapidly as we move away from the center, which implies that we need a large number of grid points in $x$-direction. Taken together, we found it numerically too costly to go below $D=3.05$ using the algorithm we described in the previous Section.

\begin{figure}[h!tb]
\centering
  \includegraphics[width=1\linewidth]{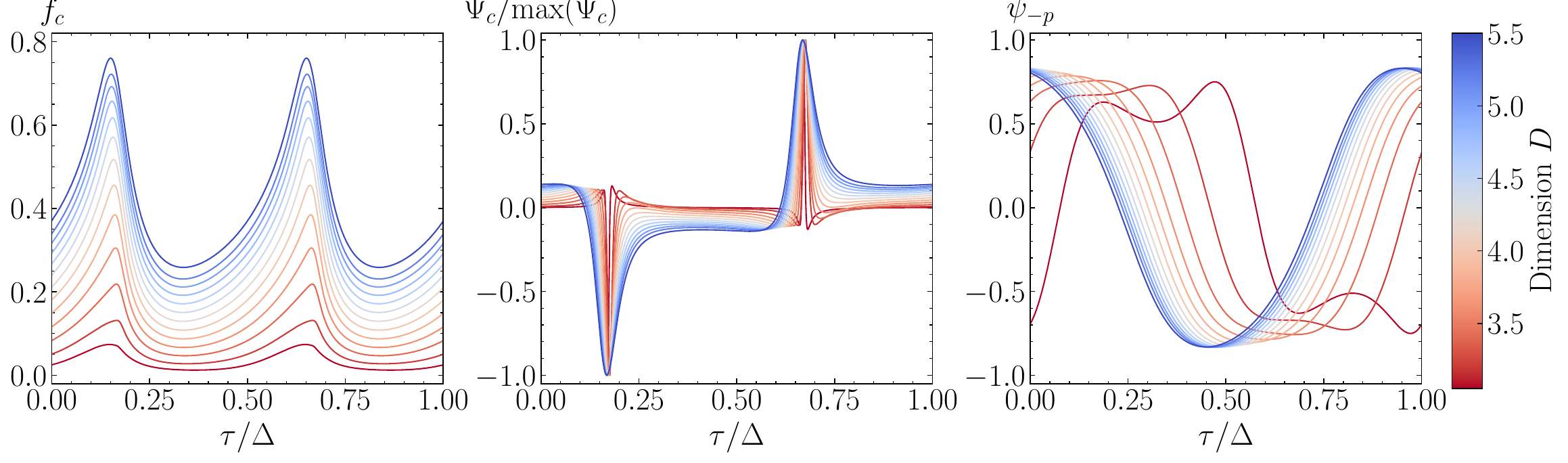}
  \caption[Converged boundary data]{Converged boundary values $f_c$ (left), $\Psi_c$ (middle), and $\psi_{-p}$ for approximately equidistant $D\in[3.05,5.5]$, with $\Psi_c$ normalized by ${\rm max}\,(\Psi_c)$.}
  \label{fig:BC}
\end{figure}

Next, we report in Fig.~\ref{fig:BC} the converged boundary values ($f_c$, $\Psi_c$) of $f$ and $\Psi$ at the center, as well as $\psi_-$ at the SSH ($\psi_{-p}$).
The SSH function in the center, $f_c$ in the left Fig.~\ref{fig:BC}, retains its characteristic double-peak structure that is also observed in exact solutions in the large-$D$ approximation \cite{Ecker:2026akf}. Consistently with their NNLO analysis, see their Eq.~(22), the maxima of $f_c$ take larger values in higher dimensions. While our numerical values of $D$ are not extremely large, $D\in[3.05,5.5]$, our results for $f_c$ in these dimensions are compatible with the large-$D$ prediction $\lim_{D\to\infty}\textrm{max}f_c(\tau)=1-\mathcal{O}(1/D^2)$ \cite{Ecker:2026akf}. The matter function in the center, $\Psi_c$ in the middle Fig.~\ref{fig:BC}, has the aforementioned peaks that get very narrow in low dimensions. They also have an increasing amplitude as we lower the dimension, which is why we normalized the result by the maximum value max$\,(\Psi_c)$. In addition, we see that there is some overshooting to the left and the right of the downward peaks (and a corresponding overshooting to the upward peaks). These features emerge when we get closer to $D=3.05$ and make it hard to come up with a reliable guess for the shape of $\Psi_c$ in the limit $D\to 3^+$. It is suggestive that such a limit may produce a $\delta$-like distribution but to test this we would need data closer to $D=3$. The matter function at the SSH $\psi_{-p}$ in the right Fig.~\ref{fig:BC}, has a rather mundane sinusoidal form at larger dimensions. As we decrease the dimension we see two effects: a relative phase shift, i.e., the extrema drift, and the appearance of additional extrema. Again, it is unclear what this suggests about the limit $D\to3^+$. Our guess is that the limiting function might be a periodic sign-function but again, we would need data much closer to $D=3$ to test this.

\begin{figure}[h!tb]
\centering
  \includegraphics[width=0.48\linewidth]{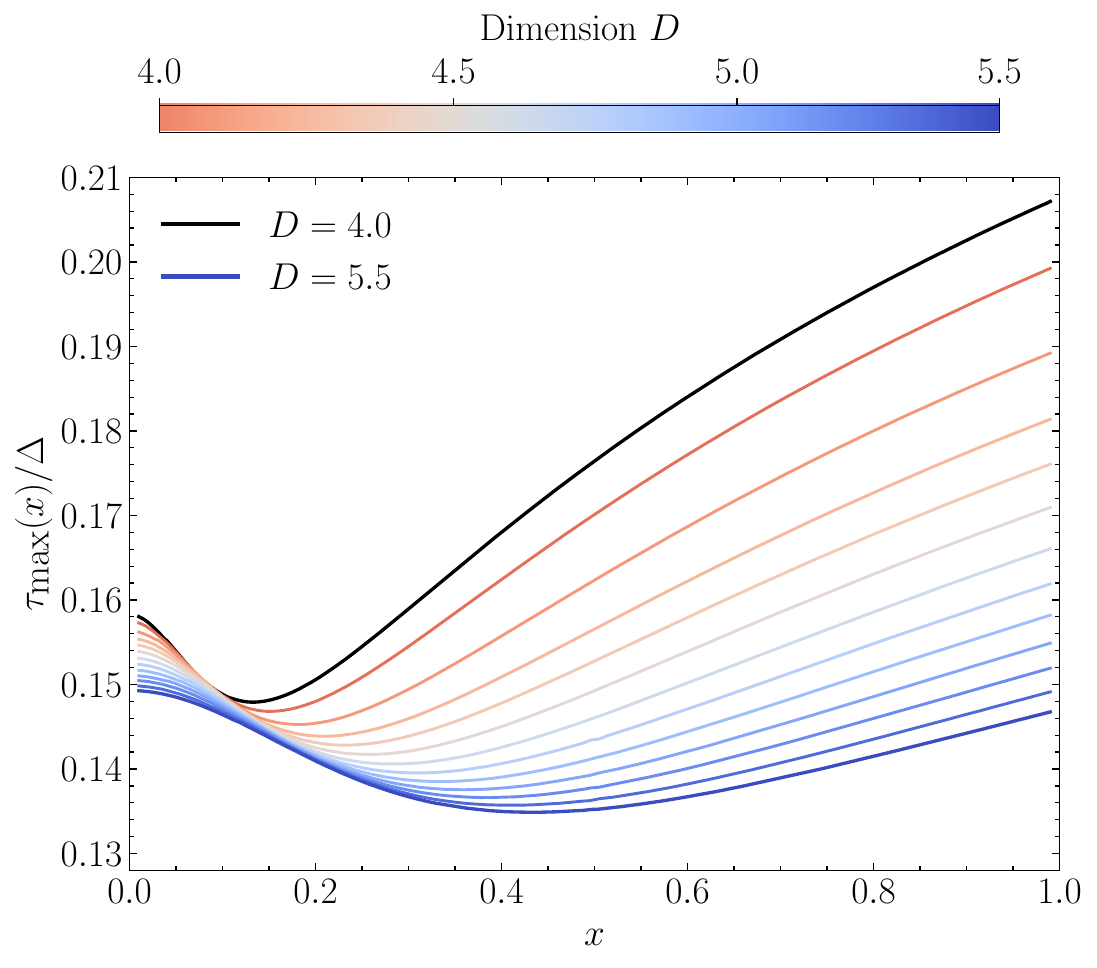}\quad
  \includegraphics[width=0.48\linewidth]{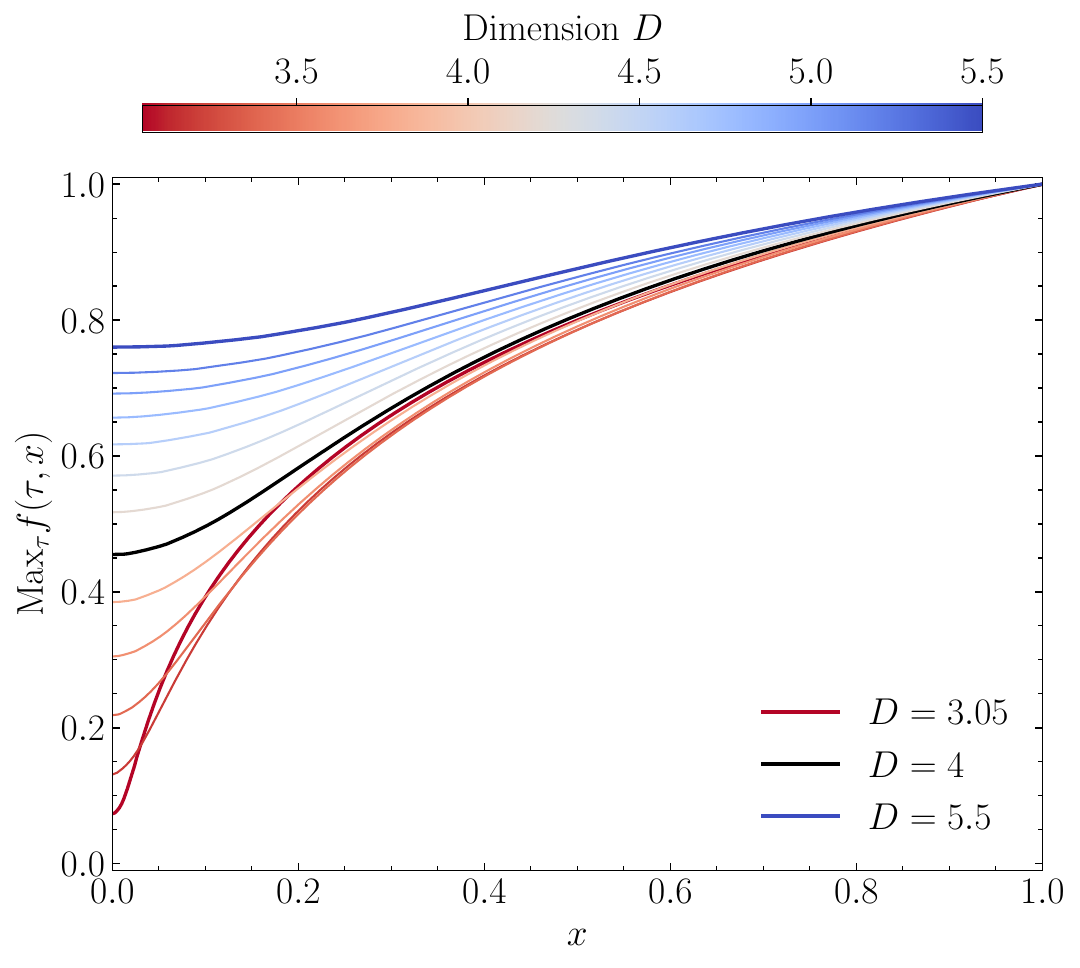}
  \caption[Maxima of SSH function]{\textbf{Left:} Contours of maxima of $f$ in $(\tau,x)$-plane with vertical axis normalized to interval $[0,1]$.
  \textbf{Right:} Maxima of $f$ as function of $x$ for various $D$.} 
  \label{fig:fmaxlines}
\end{figure}

Figure~\ref{fig:fmaxlines} shows how the $\tau$-dependence (left) and the amplitude (right) of the global maxima of $f$ vary as a function of $D$.
The left Fig.~\ref{fig:fmaxlines} shows the shape of the maximum curves in the $(\tau,x)$-plane in a part of the fundamental domain between the center and the SSH and for times $\tau$ around the NEC vertex time, which approximately aligns with the time scale where the SSH function has its maximum. This property is predicted generically from the large-$D$ expansion \cite{Ecker:2026akf}. For the specific example presented in that paper, the Supplemental Material investigates curves analogous to the left Fig.~\ref{fig:fmaxlines}, see their Fig.~5 in the \href{https://arxiv.org/abs/2601.14358}{\texttt{arXiv}} version. While the detailed shapes naturally look quite different (the model discussed therein requires dimensions $D\geq 52$), we confirm numerically the trend observed in their plots: the $x$-value where the maximum curves take the smallest value in $\tau$ move further towards the SSH in larger dimensions. In the left Fig.~\ref{fig:fmaxlines} this point moves from $x<0.2$ for $D=4$ to $x>0.4$ for $D=5.5$, whereas in the Supplemental Material of \cite{Ecker:2026akf} it moves from $x<0.8$ for $D=52$ all the way to the SSH, $x=1$, for $D=300$ or higher. The right Fig.~\ref{fig:fmaxlines} shows the height of the maxima of $f$ (with respect to $\tau$) for all values of $x$ from the center to the horizon. At the SSH, this number has to tend to $1$ due to our SSH condition $f(\tau,1)=1$, and it must increase monotonically away from the center due to the convexity conditions \eqref{eq:inequalities}. The value of the maxima in the center is, therefore, always between $0$ and $1$ in any finite dimension. As mentioned above, the trend observed in the right Fig.~\ref{fig:fmaxlines} is compatible with the large-$D$ prediction, i.e., for higher dimensions the value of the maximum of $f$ gets increasingly closer to $1$.

\begin{figure}[h!tb]
\centering
  \includegraphics[width=0.5\linewidth]{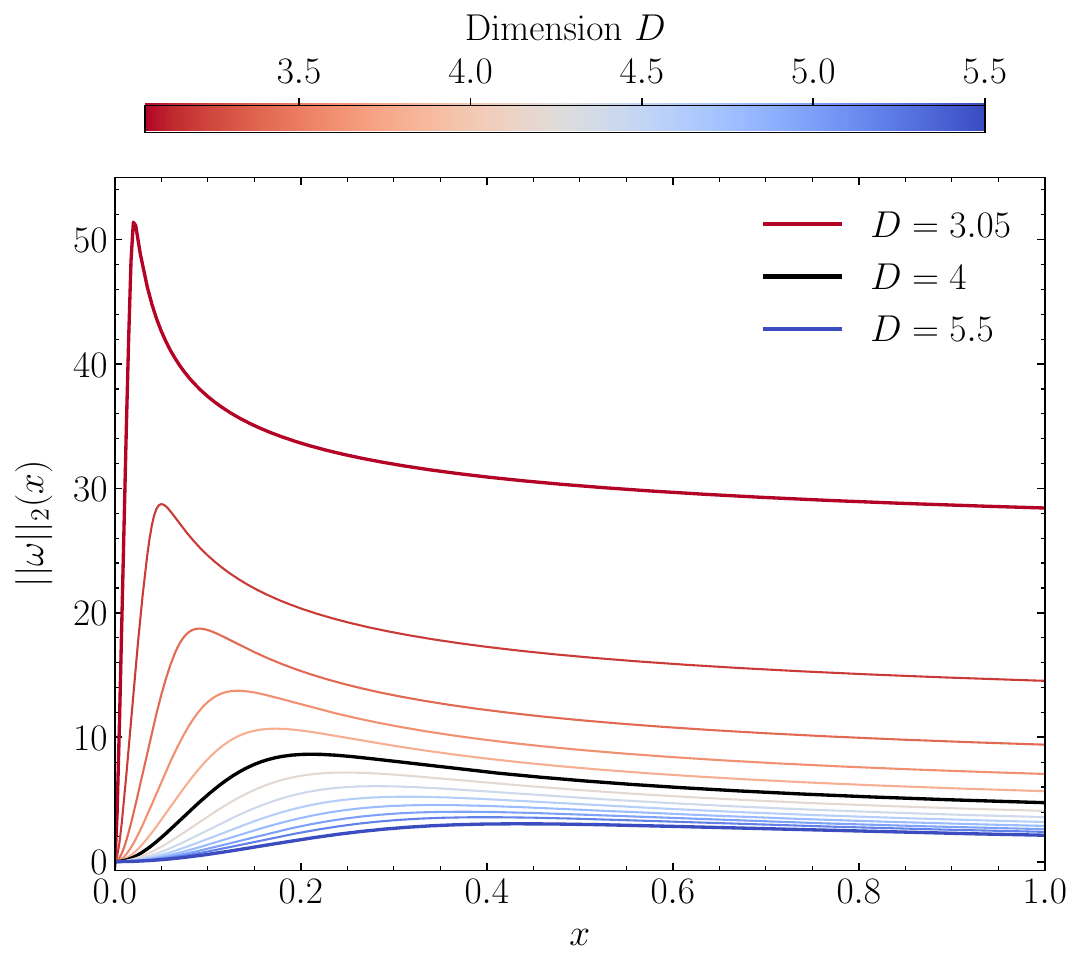}
  \caption[$\ell^2$-norm of Weyl factor]{$\ell^2$-norm of Weyl factor $\omega(\tau,x)$ taken over $\tau$ samples.    
  }
  \label{fig:l2norm}
\end{figure}

Finally, in Fig.~\ref{fig:l2norm} we show the $\ell^2$-norm in $\tau$ of $\omega$. At large $D$, it is plausible that this function tends to zero, in agreement with the ansatz used in the large-$D$ limit \cite{Rozali:2018yrv,Ecker:2026akf}.
An additional feature at lower dimensions is the appearance of a sharp peak around the maximum value of the $\ell^2$-norm at a decreasing value of $x$, rather close to the center. This may indicate the emergence of a scaling regime in low dimensions. We shall test this hypothesis in Section \ref{sec:6.3}.

\subsection[Echoing period \texorpdfstring{$\Delta$}{Delta}]{Echoing period \texorpdfstring{$\boldsymbol{\Delta}$}{Delta}}\label{sec:5.2}

The echoing period $\Delta$ is the key number for CSCs, and determining its dependence on the spacetime dimension $D$ is our main goal.

\begin{figure}[h!]
\centering
  \includegraphics[width=0.9\linewidth]{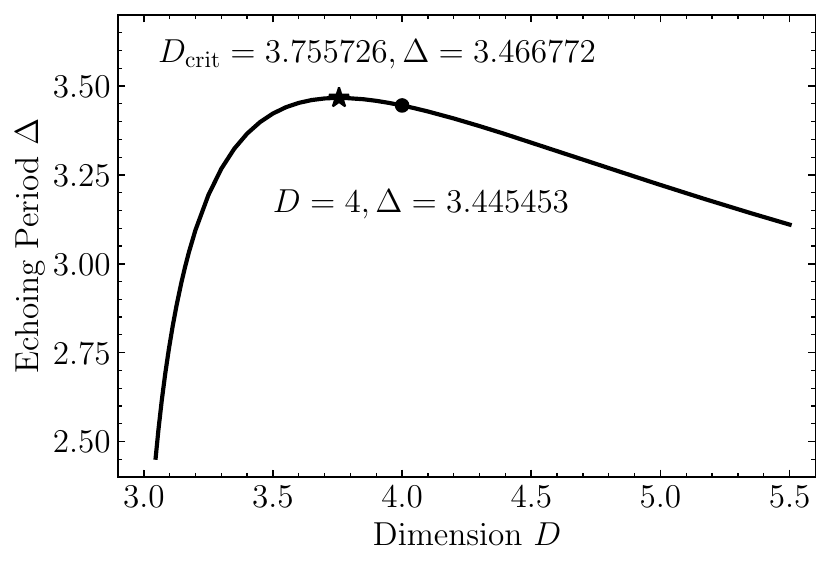}
  \includegraphics[width=0.9\linewidth]{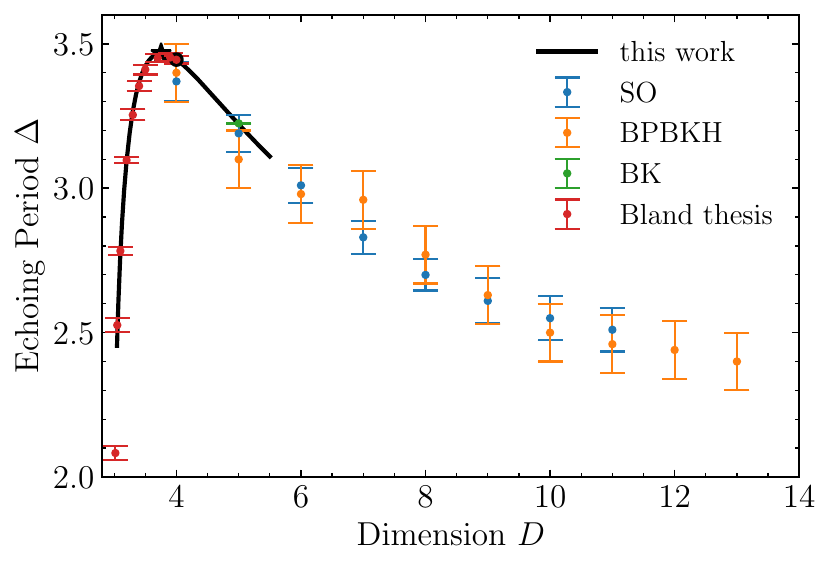}
  \caption[\textbf{Echoing period $\boldsymbol{\Delta}$ as function of $\boldsymbol{D}$}]{\textbf{Above:} Echoing period as continuous function of $D$, with values at $D=4$ and at local maximum indicated by circle and star, respectively.
  \textbf{Below:} Comparison to discrete data in literature, with acronyms SO \cite{Sorkin:2005vz}, BPBKH \cite{Bland:2005kk}, BK \cite{Bland:2007sg} and Bland's PhD thesis \cite{blandthesis}. See also Appendix \ref{app:numbers} for our data on $\Delta$.
  }
  \label{fig:Delta}
\end{figure}
Figure~\ref{fig:Delta} displays our results for $\Delta$ together with results from earlier publications at discrete (typically integer) values of $D$, including their error bars. The numerical error bars of our data are smaller than the thickness of the curves. The upper plot shows the region that we evaluated numerically, $D\in[3.05,5.5]$. The bullet point denotes the starting point of our numerical evaluation, $D=4$, where we obtain $\Delta=3.445453$. For comparison, Martin-Garc\'ia and Gundlach obtained for $D=4$ the echoing period $\Delta=3.445452$, where we displayed the same number of digits. We deviate only in the last displayed digit and attribute this to the larger number of Fourier modes that we used: while the results for $\Delta$ always converge to machine precision for any number of Fourier modes, these results will not agree precisely with the continuum limit of infinitely many Fourier modes. Given that our respective results agree to the first six digits and deviate only slightly in the seventh digit, we expect that both of our results are accurate to approximately one part in a million, i.e., we expect the correct continuum result to be given by  
\eq{
D=4:\qquad\qquad \Delta=3.445453 \pm 10^{-6}\,,
}{eq:Delta4}
which is also consistent with the result by Reiterer and Trubowitz \cite{Reiterer:2012hnr}. Details on the convergence of our numerics can be found in Appendix \ref{app:Error}, in particular \ref{app:fourier}. 

The second highlighted value in the upper Fig.~\ref{fig:Delta} is denoted by a star and gives both the dimension at the maximal value for $\Delta$ and the value of this maximum. We call this the ``critical dimension'' $D_{\textrm{\tiny crit}}$ because $\extd\Delta/\extd D=0$ at $D=D_{\textrm{\tiny crit}}$. Assuming the same accuracy as for $D=4$, we can therefore read off the results
\eq{
D_{\textrm{\tiny crit}}=3.755726\pm 10^{-6}:\qquad\qquad \Delta=3.466772\pm 10^{-6}\,.
}{eq:Deltacrit}
We have not found any indications for other extrema in the function $\Delta(D)$ and find it plausible that there are none besides the maximum discussed above. If correct, then $\Delta$ must decrease strictly monotonically both for $D>D_{\textrm{\tiny crit}}$ and $D<D_{\textrm{\tiny crit}}$. Such a behavior is compatible with the conjectures $\Delta\to 0$ both for $D\to\infty$ and for $D\to 3^+$. 

The lower plot in Fig.~\ref{fig:Delta} shows all the existing data on $\Delta$ in various dimensions. Our own data is again displayed as a continuous black curve, with errors smaller than the thickness of the curve. It passes through all existing data points within their estimated $1\sigma$-error bars (those error bars were taken from the corresponding publication indicated in the caption). In particular, the result by Reiterer and Trubowitz \cite{Reiterer:2012hnr} at $D = 4$ is contained, but since their error is $\sim 10^{-80}$ the error bar is not visible. All the other results were obtained by dynamically evolving critical initial data and measuring the periodicity of the scalar field directly. The main source of these error bars in that case comes from fitting a perodic function to the scalar field. Our larger-$D$ data suggest that the curve $\Delta(D)$ has an inflection point around $D\approx4.7$. See also Appendix \ref{app:numbers}, which contains our data on $\Delta$ and $\gamma$.

\subsection[Choptuik exponent \texorpdfstring{$\gamma$}{gamma}]{Choptuik exponent \texorpdfstring{$\boldsymbol{\gamma}$}{gamma}}\label{sec:5.42}

For physical observables slightly above and below criticality, $\gamma$ is the most visible aspect of criticality since it gets imprinted on essentially all observables, most famously the mass of black holes above the threshold \eqref{eq:2}. Despite this phenomenological significance, conceptually, the Choptuik exponent $\gamma$ is a secondary quantity of CSCs in the sense that it requires linearized perturbations above CSCs. 

\begin{figure}[h!]
\centering
  \includegraphics[width=0.85\linewidth]{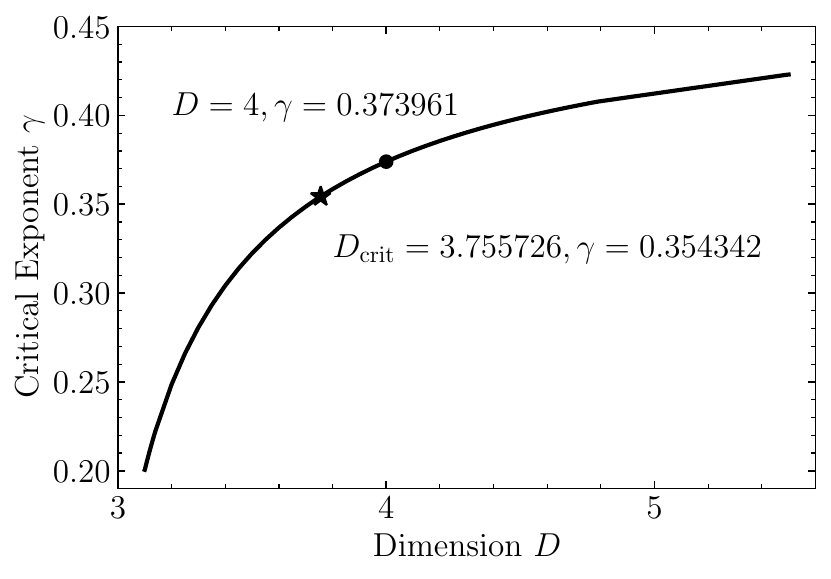}
  \includegraphics[width=0.85\linewidth]{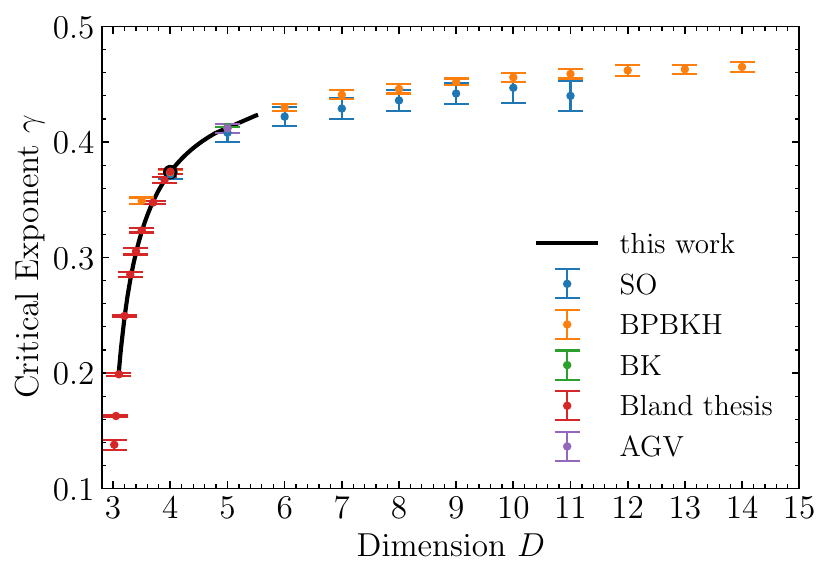}
  \caption[\textbf{Choptuik critical exponent $\boldsymbol{\gamma}$ as function of $\boldsymbol{D}$}]{\textbf{Above:} Critical exponent as continuous function of $D$, with $D=4$ indicated by a circle.
  \textbf{Below:} Comparison to discrete data in literature, with acronyms SO \cite{Sorkin:2005vz}, BPBKH \cite{Bland:2005kk}, BK \cite{Bland:2007sg}, Bland's PhD thesis \cite{blandthesis}, and AGV~\cite{Alvarez-Gaume:2006klm}. See also Appendix \ref{app:numbers} for our data on $\gamma$.}
  \label{fig:gamma}
\end{figure}
Figure \ref{fig:gamma} displays our results for $\gamma$ together with results from earlier publications at discrete (typically integer) values of $D$, including their error bars. The numerical error bars of our data are smaller than the thickness of the curves. The upper Fig.~\ref{fig:gamma} shows again our data in the range $D\in[3.05,5.5]$ and the lower Fig.~\ref{fig:gamma} all existing data. Again, we agree with all data within the respective error bars. Our result for Choptuik's original setup in four spacetime dimensions is
\eq{
D=4:\qquad\qquad\gamma=0.373961\pm 10^{-6}
}{eq:gamma4}
where we estimated the error as being similar to the one in \eqref{eq:Delta4}. The result \eqref{eq:gamma4} is compatible with $\gamma=0.374\pm0.001$ reported in \cite{Gundlach:1996eg} using the same methods as the present work but fewer Fourier modes combined with a less accurate determination of the background solution. The large-$D$ behavior displayed in the lower Fig.~\ref{fig:gamma} is compatible with the conjecture $\lim_{D\to\infty}\gamma=\frac12$, see for instance \cite{Emparan:2020inr} and Refs.~therein.

A periodic finestructure in the critical mass scaling \eqref{eq:2} was predicted in \cite{Gundlach:1996eg} and independently predicted and confirmed numerically in \cite{Hod:1996az}, 
\eq{
M_{\textrm{\tiny BH}} = M_0\,\big(p-p_\ast\big)^{(D-3)\gamma}\,e^{\mu[\ln(p-p_\ast)]}\qquad\qquad \mu:\textrm{\;periodic\;with\;period\;} \frac{\Delta}{2\gamma} \,.
}{eq:finestructure}
The period of the finestructure function $\mu$ is given by $\Delta/(2\gamma)\approx 4.61$ in $D=4$.

\begin{figure}[h!]
\centering
  \includegraphics[width=0.5\linewidth]{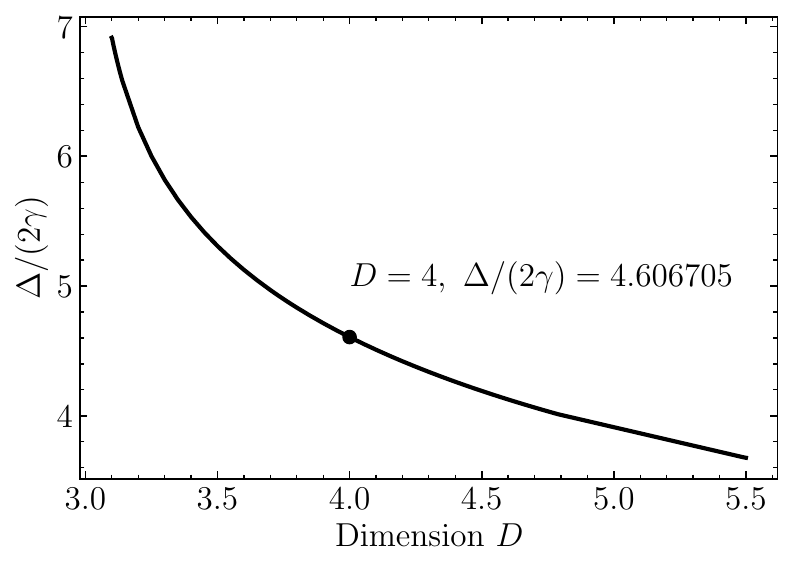}
  \caption[Finestructure period $\Delta/(2\gamma)$ as function of $D$]{Finestructure period $\Delta/(2\gamma)$ as function of $D$.}
  \label{fig:finestructure}
\end{figure}

We plot in Fig.~\ref{fig:finestructure} the value of the period of the finestructure function $\mu$ for dimensions in the interval $D\in[3.05,5.5]$. Our results for $\Delta/(2\gamma)$ indicate that the period of the finestructure is a monotonically decreasing function of $D$, decaying to zero at large $D$ and increasing rapidly as $D\to 3^+$. The large-$D$ behavior is easily understood noting the conjectured relations $\lim_{D\to\infty}\gamma=\tfrac12$ and $\lim_{D\to\infty}\Delta=0$. Since for $D\to3^+$ both $\gamma$ and $\Delta$ approach zero, the result for $\lim_{D\to3^+}\Delta/(2\gamma)=?$ is less obvious; Fig.~\ref{fig:finestructure} indicates that $\gamma$ approaches zero faster than $\Delta$ as $D\to3^+$, so we expect this limit to be infinite.

\subsection{NEC saturation lines}\label{sec:5.3}

The NEC angle was proposed as another critical parameter in \cite{Ecker:2024haw}. The associated NEC saturation lines, given by solutions of \eqref{eq:NEC2}, coincide with lines of vanishing curvature and provide a useful diagnostic tool of CSCs.

\begin{figure}[h!]
\centering
  \includegraphics[width=\linewidth]{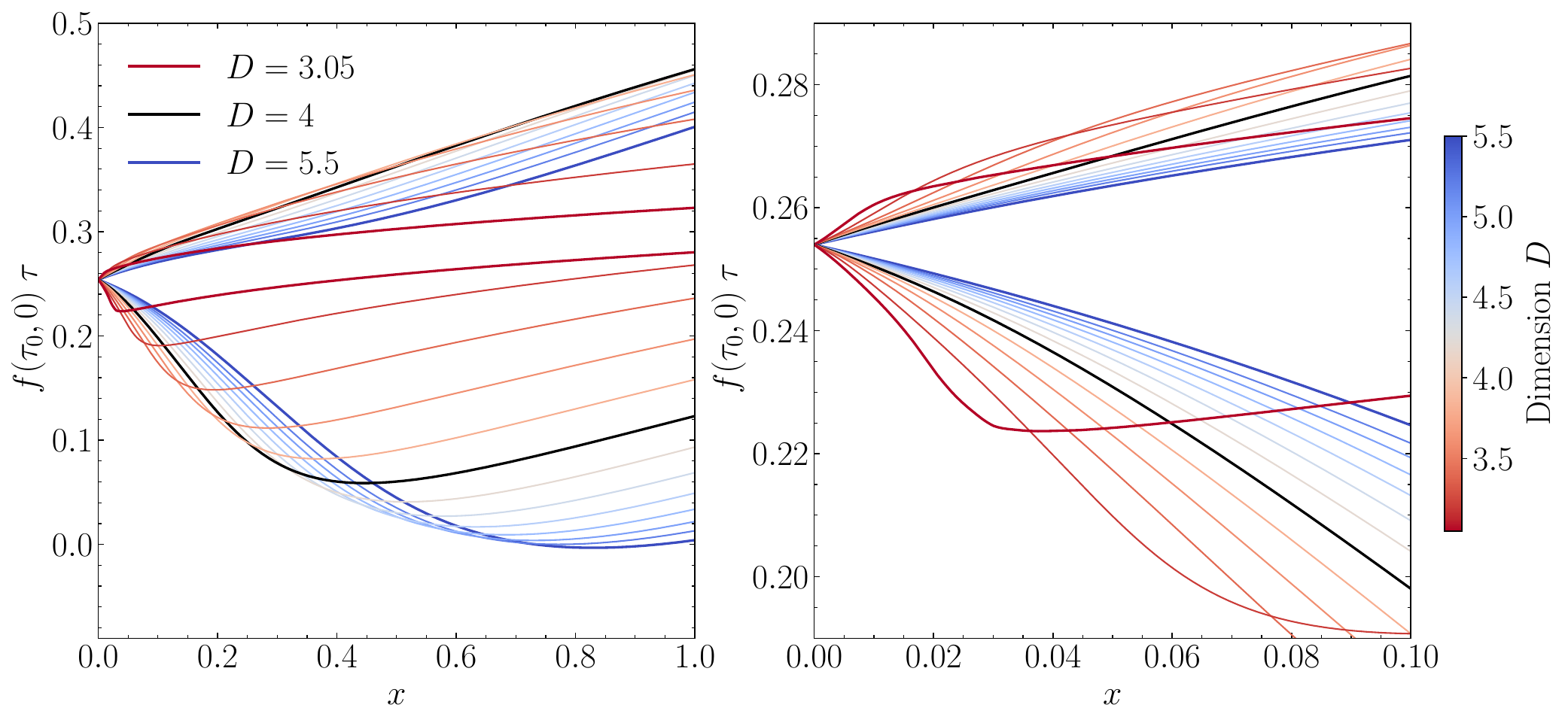}
  \caption[\textbf{NEC saturation lines for various dimensions}]{\textbf{Left:} NEC saturation lines. \textbf{Right:} Zoom into center. \textbf{Both:} Vertical axes rescaled to proper time at $x=0$ and shifted so NEC vertices coincide.}
  \label{fig:NEClines}
\end{figure}
Figure~\ref{fig:NEClines} displays the NEC lines for a discrete set of dimensions between $D=3.05$ and $D=5.5$. Besides these two endpoints, we highlight $D=4$ in the left and right Figs.~as black lines. The left Fig.~\ref{fig:NEClines} shows the NEC lines in the lower half of a fundamental domain of our CSC (there is an equivalent copy of the NEC lines in the upper half). Since we have normalized the time $\tau$ with the echoing period $\Delta$, all solutions share the uniform periodicity $\tau\to\tau+1$ in these graphs. This is why the lower half of the fundamental domain is $\tau\in[0,0.5)$ and $x\in[0,1]$. The right Fig.~\ref{fig:NEClines} zooms into a region closer to the center to highlight the change of the NEC angle --- always consistent with the exact result \eqref{eq:NECangle} --- and the pronounced change of shape between lower and higher dimensions. Indeed, in low dimensions, the NEC lines quickly spread from the center due to the large NEC angle and then flatten out towards the SSH, while in high dimensions, the NEC lines are well-approximated by quadratic functions sufficiently close to the center.

\begin{figure}[h!]
\centering
  \includegraphics[height=0.28\linewidth]{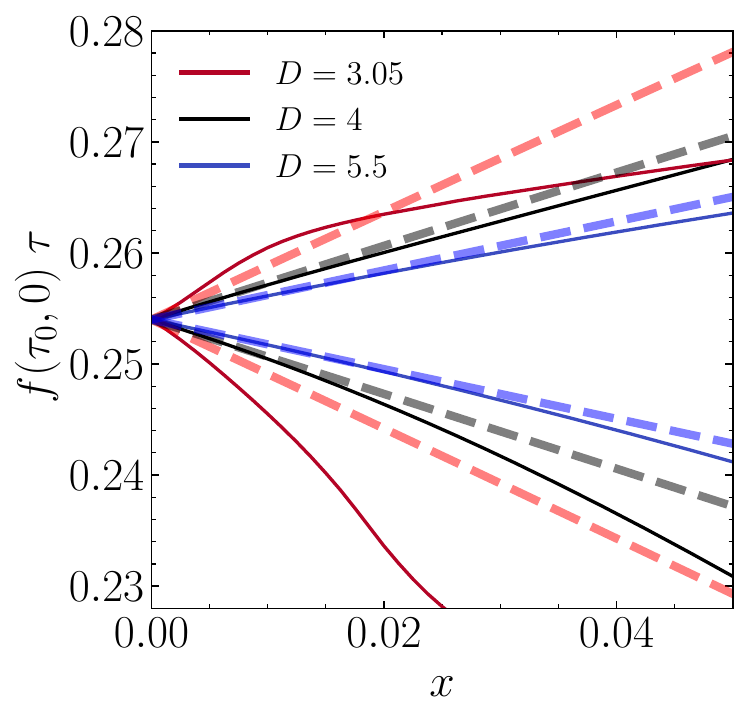}
  \includegraphics[height=0.28\linewidth]{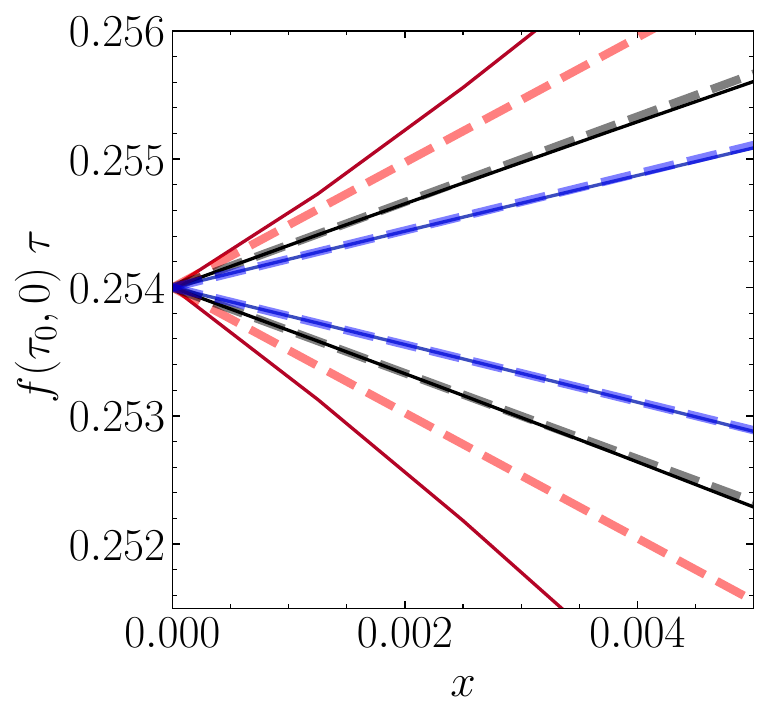}
  \includegraphics[height=0.28\linewidth]{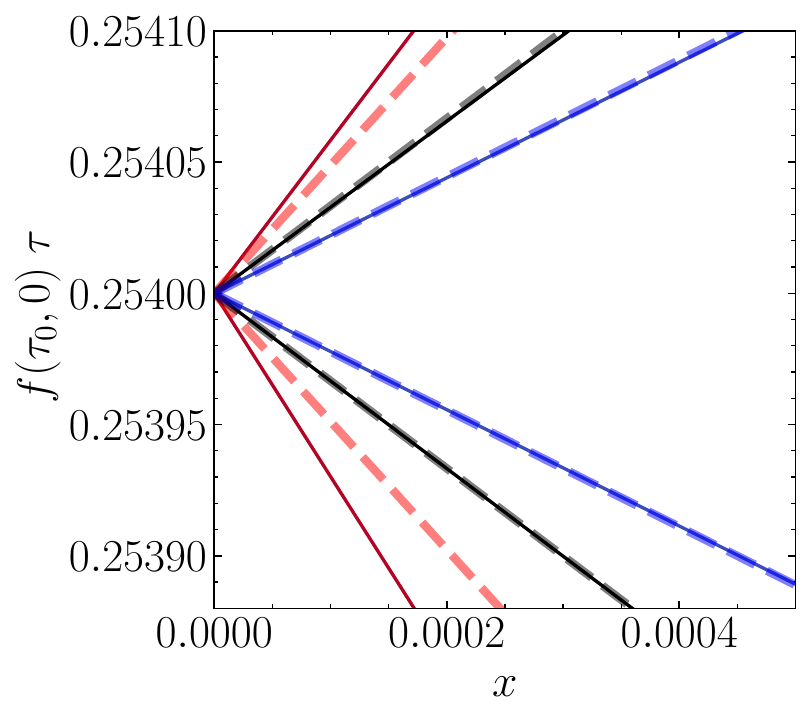}
  \caption[NEC line comparison to NNLO large-$D$ approximation]{NEC line comparison to NNLO large-$D$ approximation (dashed lines).}
  \label{fig:NEClargeD}
\end{figure}

To quantify how well (or poorly) the large-$D$ approximation of \cite{Ecker:2026akf} does, we compare their exact NNLO results to our data in Fig.~\ref{fig:NEClargeD}.
In the three plots we successively zoom closer towards the center, presenting in each case results for the dimensions $D=3.05$, $D=4$, and $D=5.5$ (solid curves in red, black, and blue, respectively). We also show the NNLO results for the NEC lines as dashed lines in the same colors. The NNLO results are taken from the Supplemental Material of  \cite{Ecker:2026akf}, see their Eq.~(S21), where we drop their quartic term in $x$ since we only care about the region close to $x=0$. In this approximation, their NNLO NEC lines simplify to a universal quadratic relation that generates the dashed lines in Fig.~\ref{fig:NEClargeD}. Note, in particular, that the free integration function $\intfct(\tau)$ that characterizes the large-$D$ DSS solutions of \cite{Ecker:2026akf} does not appear in this relation: 
\eq{
\textrm{NNLO\;NEC\;lines}:\qquad\qquad \tau=\tau_0\pm\frac{x}{D-1}-\frac{x^2}{2(D-1)^2} + \mathcal{O}(x^4)
}{eq:NNLONEC}
Thus, our NNLO NEC lines near the center are universal in the sense that they cannot be changed by adjusting $\intfct(\tau)$. Two conclusions can be drawn from Fig.~\ref{fig:NEClargeD}: 1.~Unsurprisingly, the large-$D$ approximation works better in higher dimensions: For $D=5.5$, the deviations between solid and dashed blue curves are small in the whole interval, whereas for $D=3.05$ the deviations between solid and dashed red curves are pronounced. 2.~Perhaps surprisingly, the large-$D$ approximation works better the more we zoom into the central region; even for a dimension as low as $D=4$, the deviations between the dashed and solid black curves become small in the middle and right Fig.~\ref{fig:NEClargeD}. 

\begin{figure}[h!]
\centering
  \includegraphics[width=0.5\linewidth]{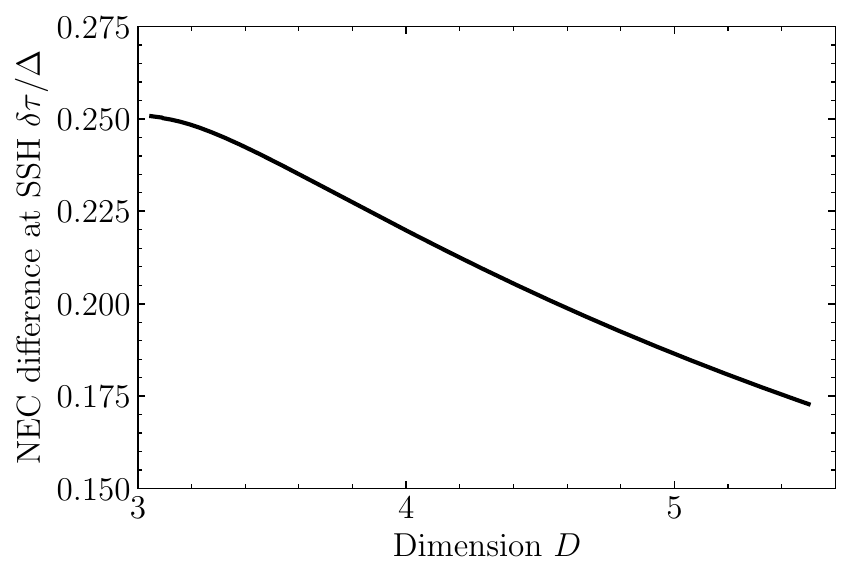}
  \caption[NEC line distance at SSH as function of $D$]{Time-difference between NEC lines at SSH. Vertical axis in units $\tau/\Delta$.}
  \label{fig:diffSSHoverDel}
\end{figure}

It is also illuminating to focus on the other side of the fundamental domain, the SSH. In Fig.~\ref{fig:diffSSHoverDel}, we plot the normalized time-distance between two consecutive NEC lines as a function of $D$ so that a value of $1$ corresponds to a full period and a value of $0.25$ to a quarter period.
Remarkably, as we approach $D\to 3^+$ the normalized time-distance between two consecutive NEC lines approaches a number close to $0.25$. Geometrically, this means that close to the SSH for dimensions slightly above $D=3$ the NEC lines make a regular pattern and the regions of positive and negative curvature alternate with approximately the same sizes. The SSH would then have two regions of positive curvature [by adjusting the phase, these intervals are $\tau\in(0,\tfrac14)$ and $\tau\in(\tfrac12,\tfrac34)$] and two regions of negative curvature [for $\tau\in(\tfrac14,\tfrac12)$ and for $\tau\in(\tfrac34,1)$] of equal size. We do not have an explanation for the apparent regularity of this pattern in low dimensions. 

By contrast, as the dimension increases, the time-distance between the two NEC lines emanating from the same NEC vertex becomes smaller and plausibly tends to zero at infinite $D$. The large-$D$ prediction to NNLO \eqref{eq:NNLONEC},
\eq{
\frac{\delta\tau}{\Delta}=\frac{2}{(D-1)\,\Delta}+\mathcal{O}(1/D^3)
}{eq:largeDDeltatau}
does not work very accurately for our values of the dimension: e.g., for $D=5.5$ the NNLO formula \eqref{eq:largeDDeltatau} predicts $\delta\tau/\Delta\approx0.14$, which is below the value $\delta\tau/\Delta\approx0.175$ read off from the right part of Fig.~\ref{fig:diffSSHoverDel}. We attribute this discrepancy to the slower convergence of the large-$D$ expansion near the SSH noted already in \cite{Ecker:2026akf}. To accurately determine the asymptotic behavior of \eqref{eq:largeDDeltatau} we would need to know the asymptotic behavior of $\Delta$. We address this issue in the next Section when elaborating on the large-$D$ expansion.


\section{Analytic results}\label{sec:6}

While the numerical simulations become difficult as $D$ approaches $3$ from above or tends to infinity, precisely in these two limits, it is possible to employ analytic methods. 

It is useful to label different regions in a fundamental domain according to the relative values of the spatial coordinate $x$ and the normalized echoing period $\Delta/(4\pi)$: 
\begin{enumerate}
    \item Near-center: $x\ll\frac{\Delta}{4\pi}$
    \item Center-bulk transition: $x\sim{\cal O}(\frac{\Delta}{4\pi})$
    \item Bulk: $x\gg\frac{\Delta}{4\pi}$, $1-x\gg\frac{\Delta}{4\pi}$
    \item Bulk-SSH transition: $1-x\sim{\cal O}(\frac{\Delta}{4\pi})$
    \item Near-SSH: $1-x\ll\frac{\Delta}{4\pi}$
\end{enumerate}
Note that the regions 2.-4.~all merge to a single one at finite $D$, because $\Delta/(4\pi)\sim\mathcal O(1)$. However, $\Delta/(4\pi)$ is expected to become very small at large $D$ and small $(D-3)$ so that the labeling above is meaningful in these limits.

In the near-center and near-SSH regimes, we can use Taylor expansions in $x$ and $(1-x)$, respectively, as displayed in Appendix \ref{app:Taylor}. However, in the bulk and its two transition regimes this no longer works and we need other methods, which we develop in this Section.

A tricky aspect of any such expansion is that at finite $D$, we are free to rescale our fields by arbitrary powers of $D-3$. However, when this quantity becomes small or large, the qualitative structure of the equations may change. Thus, it is important to have a sound working hypothesis of how the various functions scale in such limits. We are guided here by the structure of the equations \eqref{eq:eom}, by our numerical results, and by the discovery of exact DSS solutions in the large-$D$ limit in \cite{Ecker:2026akf}.

Especially in large dimensions, it is convenient to re-express the matter variables in terms of their sums and differences \eqref{eq:pipsi} and to redefine the Weyl factor $\omega$ as
\eq{
\Omega=\frac{D-3}{x^2}\,\big(e^\omega-1\big)\,.
}{eq:largeD101}
All these quantities are finite at $x=0$ and have Taylor series in even powers of $x$, as does $f$. Therefore, we also redefine $y=x^2$ and express the EOM \eqref{eq:eom} in these variables 
\begin{subequations}
    \label{eq:eomPiPsi}
\begin{align}
    \partial_y\Omega &= \frac{1}{2y}\,\big((D-3)\,\Pi^2-(D-1)\,\Omega\big)+\frac{\Omega}{2}\,\big(\Pi^2-\Omega+y\,\Psi^2\big)+\frac{D-3}{2}\,\Psi^2 \\
   \partial_y f &=\frac{\Omega\,f}{2} \\
   \big(\partial_\tau + 2y\,\partial_y\big)\,\Pi &= 2f\,y\,\partial_y \Psi + \big(D-1+y\,\Omega\big)\,f\,\Psi-\Pi\\
   \big(\partial_\tau + 2y\,\partial_y\big)\,\Psi &= 2f\,\partial_y \Pi + f\,\Omega\,\Pi - 2\Psi \\
   \big(\partial_\tau + 2y\,\partial_y\big)\,\Omega &= 2\big(D-3+y\,\Omega\big)\,f\,\Pi\,\Psi-2\Omega\,.
\end{align}
\end{subequations} 
Regularity in the center is guaranteed by
\eq{
y=0:\qquad \Omega = \frac{D-3}{D-1}\,\Pi^2\,.
}{eq:regularity}
Additional relations in the center are
\eq{
y=0:\qquad \Psi=\frac{\partial_\tau\Pi+\Pi}{(D-1)\,f}
}{eq:center1}
and
\eq{
y=0:\qquad \partial_\tau^2\Pi+3\partial_\tau\Pi+2\Pi=\big(\partial_\tau\Pi+\Pi\big)\,\partial_\tau\ln f + (D-3)f^2\Pi^3+2(D-1)f^2\partial_y\Pi\,.
}{eq:center2}

\subsection[Large-\texorpdfstring{$D$}{D} expansion]{Large-\texorpdfstring{$\boldsymbol{D}$}{D} expansion}\label{sec:6.1}

The large $D$ limit of general relativity allows using $1/D$ as a small expansion parameter, see \cite{Emparan:2020inr} and Refs.~therein. As suggested already in \cite{Emparan:2013xia}, the large-$D$ limit could help in producing analytic results for critical collapse. It was shown in \cite{Ecker:2026akf} that this is indeed the case by solving exactly the EOM \eqref{eq:eom} in the large $D$ limit. In this Subsection, we elaborate on this solution and focus on an aspect that was not addressed in detail in that work, namely the scaling of the echoing period with $D$.

\subsubsection[Large-\texorpdfstring{$D$}{D} scaling of \texorpdfstring{$\Omega$}{Omega} and \texorpdfstring{$\Delta$}{Delta}]{Large-\texorpdfstring{$\boldsymbol{D}$}{D} scaling of \texorpdfstring{$\boldsymbol{\Omega}$}{Omega} and \texorpdfstring{$\boldsymbol{\Delta}$}{Delta}}\label{sec:6.1.1}

In the limit of large dimension, $D\gg 1$, the Weyl factor $\omega$ tends to zero, as evident from the numerical plots in the previous Section. We also see this analytically from the EOM \eqref{eq:eom2}: unless $\omega$ scales like $1/D$, the quantity $x\,\partial_x\ln f$ is infinite, but we need $f$ to be bounded in the interval $(0,1]$ to get a spacetime crystal with a regular SSH. This is why we introduced $\Omega$ in \eqref{eq:largeD101}, which has the added bonus of remaining nonzero at $x=0$.

This large-$D$ scaling of $\omega$ was first exploited by Rozali and Way in \cite{Rozali:2018yrv} (see their Section 2.2), and we follow their lead using the expansions
\begin{align}
\Omega &= \Omega_{\textrm{\tiny LO}}(\tau,x)  + \cal{O}(\epsilon) &
f &= f_{\textrm{\tiny LO}}(\tau,x) + \cal{O}(\epsilon) &
\psi_\pm &= \psi_{\pm\textrm{\tiny LO}}(\tau,x)+\cal{O}(\epsilon)
\label{eq:largeD1}
\end{align}
where $\epsilon=D^{-1/N}$ is a small parameter and $N$ is some fixed integer. We introduce $\epsilon$ to parameterize the echoing period $\Delta$, which we expect to tend to zero at large $D$. While we do not know if it decays monomially with some rational power of $1/D$, for the moment we assume this to be the case, i.e.,
\eq{
\Delta = \frac{\epsilon}{A}\qquad\qquad\epsilon=D^{-1/N}\qquad\qquad N\in\mathbb{Z}^+
}{eq:largeD100}
with some positive constant $A\sim\mathcal{O}(1)$. We restrict to positive integer $N$ since only then can $\partial_\tau$-terms in the EOM compete to the same order in $\epsilon$ as non-derivative terms.

We rescale time
\eq{
\tau\to\Delta\tau \qquad\Rightarrow \qquad \partial_\tau\to\frac{A}{\epsilon}\,\partial_\tau
}{eq:largeD102}
so that we have the uniform periodicity of $1$ regardless of the values of $\epsilon$ or $A$.

We have investigated the EOM \eqref{eq:eom} for the values $N=1,2,3,4,5,6,\infty$, which we explicate in the following Subsubsections. By contrast to \cite{Ecker:2026akf}, we consider here only solutions up to terms of order $o(1/D)$, i.e., neglect $1/D$ terms. Thus, while we will take into account subleading corrections in $\epsilon$ for finite $N$, we will never consider terms of order $\mathcal{O}(\epsilon^N)$ or higher.

\subsubsection[\texorpdfstring{$N=1$}{N=1}]{\texorpdfstring{$\boldsymbol{N=1}$}{N=1}}\label{sec:6.1.2}

All of the EOM but the last one can be solved in closed form to LO, yielding
\begin{subequations}
\begin{align}
    \Omega_{\textrm{\tiny LO}} &= \Pi^2_{\textrm{\tiny LO}} + x^2\,\Psi^2_{\textrm{\tiny LO}} \\
    \Psi_{\textrm{\tiny LO}}&=\frac{A\,\partial_\tau\Pi_{\textrm{\tiny LO}}}{f_{\textrm{\tiny LO}}} \\
    f_{\textrm{\tiny LO}} &= f_0(x)\,\partial_\tau \Pi_{\textrm{\tiny LO}} \label{eq:largeD107}\\
    \partial_\tau\partial_x\Pi_{\textrm{\tiny LO}} &= \frac{\partial_\tau\Pi_{\textrm{\tiny LO}}}{f_0^2(x)}\,\big(Ax^3+xf_0^2(x)\Pi_{\textrm{\tiny LO}}^2-f_0(x)f_0^\prime(x)\big)
\end{align}
\end{subequations}
where $f_0(x)$ is an arbitrary integration function and the constraint equation \eqref{eq:constr} is fulfilled automatically. The last equation is still a non-linear PDE, but we do not need to solve it.

This is so because from \eqref{eq:largeD107} we deduce that none of the $N=1$ solutions has an SSH: It is impossible to achieve the boundary condition $f(\tau,x=1)=1$ for any choice of $f_0(x)$ and any periodic function $\Pi_{\textrm{\tiny LO}}$. One would need to switch on the winding mode of the scalar field \eqref{eq:windings} to get such solutions, and we do not address such a generalization in the present work. 

Therefore, we can rule out the case $N=1$ and deduce that $\Delta$ cannot decay faster than $1/\sqrt{D}$ at large $D$.

\subsubsection[\texorpdfstring{$N=2$}{N=2}]{\texorpdfstring{$\boldsymbol{N=2}$}{N=2}}\label{sec:6.1.3}

The EOM remain coupled nonlinear PDEs, but now allow a family of solutions with SSH. To each order in $\epsilon$, we get one free integration function of $\tau$. To LO, we get
\begin{subequations}
\label{eq:ngleichzwei}
\begin{align}
     \Omega_{\textrm{\tiny LO}} &= \Pi_{\textrm{\tiny LO}}^2 \\
    \Psi_{\textrm{\tiny LO}}&=\epsilon\,\frac{A\,\partial_\tau \Pi_{\textrm{\tiny LO}}}{f_{\textrm{\tiny LO}}} \\
    \partial_x f_{\textrm{\tiny LO}} &= x\,\Pi_{\textrm{\tiny LO}}^2 f_{\textrm{\tiny LO}} \label{eq:largeD108} \\
    \partial_x\Pi_{\textrm{\tiny LO}}&= \frac{A^2x}{f_{\textrm{\tiny LO}}}\,\partial_\tau\bigg(\frac{\partial_\tau\Pi_{\textrm{\tiny LO}}}{f_{\textrm{\tiny LO}}}\bigg) -x\,\Pi_{\textrm{\tiny LO}}^3 \label{eq:largeD109}
\end{align}
\end{subequations}
and the constraint equation is fulfilled automatically again. The system of PDEs \eqref{eq:ngleichzwei} differs from the large-$D$ system \cite{Ecker:2026akf}: apart from some relatively minor change in $\Psi_{\textrm{\tiny LO}}$, the last PDE \eqref{eq:largeD109} involves both temporal and spatial derivatives, a bit like a non-linear heat equation with reversed roles of space and time. By contrast, the corresponding PDE in \cite{Ecker:2026akf} does not feature the term with $\tau$-derivatives, see their Eq.~(9).

The field $\Psi$ is now suppressed by one order in $\epsilon$, and the last two equations form a coupled system of nonlinear PDEs for the two functions $\Pi_{\textrm{\tiny LO}}$ and $f_{\textrm{\tiny LO}}$. Applying the same methods as in our numerical Sections, we solve the EOM in Taylor expansions near the center, $x=0$, and the SSH, $x=1$, with the boundary condition $f_{\textrm{\tiny LO}}(\tau,x=1)=1$. In this way, we have two free functions of $\tau$ from the center-expansion,  and one free function of $\tau$ from the SSH-expansion. However, at some matching surface $x=x_{\textrm{match}}$, we only get two matching conditions, and thus there remains one free integration function that labels our solutions. This is the same property as discovered in \cite{Ecker:2026akf}, who also find exactly one free function of $\tau$ in their large-$D$ solution.

An alternative route to reach the same conclusion is to eliminate algebraically $\Pi_{\textrm{\tiny LO}}$ using \eqref{eq:largeD108} and inserting it into \eqref{eq:largeD109}. This gives a single non-linear PDE for $f_{\textrm{\tiny LO}}$, second order in $x$-derivatives, that has constant coefficients if we use $y=x^2$ as the radial evolution parameter. This PDE can be solved by Taylor expanding from the SSH with the boundary conditions $f_{\textrm{\tiny LO}}(\tau,y=1)=1$ and $\partial_yf_{\textrm{\tiny LO}}(\tau,y=1)=f_1(\tau)^2$, with some free function $f_1(\tau)$.

\begin{figure}
    \centering
        \includegraphics[width=0.8\linewidth]{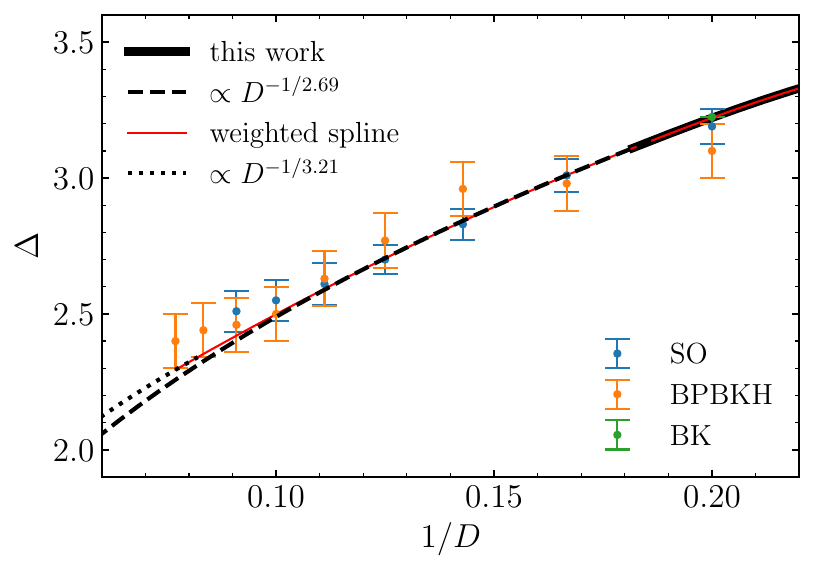}
    \caption[Echoing period $\Delta$ as function of $1/D$]{Echoing period $\Delta$ as function of $1/D$ for $D>4$. Solid black curve: Our numerical results. Symbols with error bars: Data at $D>4$ from SO~\cite{Sorkin:2005vz}, BPBKH~\cite{Bland:2005kk}, and BK~\cite{Bland:2007sg}. Red curve: Weighted quadratic spline interpolation of combined dataset. Dashed/dotted black lines: Power-law fits to small-$1/D$ end of our numerical data/spline, respectively.
    }
    \label{fig:logDeltalargeD}
\end{figure}

While the $N=2$ case is mathematically interesting and allows for an SSH, it is more complicated than any of the higher $N$ cases. To see how $\Delta$ behaves at large-$D$ we collected available data in Fig.~\ref{fig:logDeltalargeD}.
The numerical fits shown in the plot allow us to infer a lower bound on $N$: $N>2.69$ when fitting only the largest-$D$ data from our simulations, and $N>3.21$ when combining our results with available literature data for $D>4$. Given that our fit of the large-$D$ data suggests $N\geq 3$, we therefore move on.

\subsubsection[\texorpdfstring{$N=3$}{N=3}]{\texorpdfstring{$\boldsymbol{N=3}$}{N=3}}\label{sec:6.1.4}

The equations decouple, and we can solve algebraically order-by-order in $\epsilon$ for all the expansion coefficients of $\Omega$ and $\psi_-$, and get linear inhomogeneous PDEs (with a single $x$-derivative) for all the expansion coefficients of $\psi_+$ and $f$ that can be solved in closed form. Like for $N=2$, to each order we get one free integration function of $\tau$ while the other integration function is fixed by the SSH condition $f(\tau,x=1)=1$. To LO, these equations reduce to 
\begin{align}
\Omega_{\textrm{\tiny LO}} &= \Pi_{\textrm{\tiny LO}}^2 &
\partial_x f_{\textrm{\tiny LO}} &= x\,\Pi_{\textrm{\tiny LO}}^2\,f_{\textrm{\tiny LO}} &
\partial_x \Pi_{\textrm{\tiny LO}} &= -x\,\Pi_{\textrm{\tiny LO}}^3
\label{eq:largeD2}
\end{align} 
as well as
\eq{
\Psi_{\textrm{\tiny LO}} = \frac{A\,\partial_\tau\Pi_{\textrm{\tiny LO}}}{D\epsilon\,f_{\textrm{\tiny LO}}}\,.
}{eq:largeD3}  
The key difference to the $N=2$ case is the differential equation for $\Pi_{\textrm{\tiny LO}}$, which now decouples from the others as opposed to \eqref{eq:largeD109}. The only difference to the LO large-$D$ equations of \cite{Ecker:2026akf} is \eqref{eq:largeD3}, i.e., except for $\Psi_{\textrm{\tiny LO}}$ the $N=3$ LO solution above coincides with the LO solution of \cite{Ecker:2026akf}.

The EOM \eqref{eq:largeD2}, \eqref{eq:largeD3} are solved by
\begin{subequations}
\label{eq:largeD4}
\begin{align}
\Omega_{\textrm{\tiny LO}}(\tau,x) &= \frac{\intfct^2(\tau)}{1+\intfct^2(\tau)x^2}  &
f_{\textrm{\tiny LO}}(\tau,x) &= \sqrt{\frac{1+\intfct^2(\tau)x^2}{1+\intfct^2(\tau)}} \\
\Pi_{\textrm{\tiny LO}} &= \frac{\intfct(\tau)}{\sqrt{1+\intfct^2(\tau)x^2}} &
\Psi_{\textrm{\tiny LO}}(\tau,x) &=\frac{A\,\sqrt{1+\intfct^2(\tau)}}{D\epsilon}\,\frac{\partial_\tau\intfct(\tau)}{(1+\intfct^2(\tau)x^2)^2}  \label{eq:angelinajolie}
\end{align}
\end{subequations}
with the LO integration function $\intfct(\tau)$. 

Perhaps the most notable aspect of the large $D$ solution \eqref{eq:largeD4} is that the function $\intfct(\tau)$ is not constrained to be periodic. Still, it can be chosen as periodic with uniformized period $1$. This allows using the solution above (up to any desired order in $\epsilon$) as an approximate solution at finite $D$ by tuning $\Delta$ to the value appropriate for a given finite dimension. All these features were observed already in \cite{Ecker:2026akf}.

The $D$-dimensional Ricci scalar (expressed in terms of our original time $\tau$)
\eq{
R_{\textrm{\tiny LO}} = -e^{2\tau}\,\frac{D\,\intfct^2(\tau/\Delta)}{1+\intfct^2(\tau/\Delta)x^2}
}{eq:app2}
scales linearly with the dimension and is negative almost everywhere to LO, except at zeros of the function $\intfct$. As anticipated for critical solutions, the Ricci scalar diverges in the limit $\tau\to\infty$ if $\intfct(\tau)$ is a (non-zero) periodic function but has no singularities at finite $\tau$ in the whole interval between the center $x=0$ and the SSH $x=1$. 

According to the general formula \eqref{eq:NECangle}, the NEC angle goes to zero at large $D$ as
\eq{
\alpha \approx \frac{2}{D-1} + \mathcal{O}(1/D^3)\,.
}{eq:largeD42}
We check now whether this is compatible with the exact solution \eqref{eq:largeD4}, assuming that the function $\intfct(\tau)$ has at least one single zero and linearize around it, $\intfct(\tau)\approx\intfct_0\,(\tau-\tau_0)$. In the notation of this Section, the two NEC lines emanating  from the vertex $(\tau=\tau_0,x=0)$ are given by linearized solutions of $\Pi_{\textrm{\tiny LO}} = \pm x\,\Psi_{\textrm{\tiny LO}}$ where we linearize both sides in the original (unrescaled) time $\tau-\tau_0$ and in $x$, i.e., $\Pi_{\textrm{\tiny LO}}\approx\intfct_0\,\frac1\Delta\,(\tau-\tau_0)$ and $\Psi_{\textrm{\tiny LO}}\approx\intfct_0\,\frac{1}{\Delta D}$. The two spacelike vectors pointing from the NEC vertex along the NEC lines in this approximation are given by $n_\pm^\mu\partial_\mu = \pm\partial_\tau + D\,\partial_x$. Their relative rapidity $\xi$ is determined from their normalized inner product,
\eq{
\cosh\xi = \frac{n_+^\mu n_-^\nu g_{\mu\nu}}{\sqrt{n_+^\mu n_+^\nu g_{\mu\nu} n_-^\lambda n_-^\sigma g_{\lambda\sigma}}} \approx 1 + \frac{2}{D^2}
}{eq:largeD40}
yielding $\xi\approx 2/D$. The NEC angle, being the Gudermannian of the rapidity, is then given by 
\eq{
\alpha \approx \frac{2}{D} + \mathcal{O}(1/D^2)
}{eq:largeD39}
which to LO is compatible with the expansion \eqref{eq:largeD42} of the exact result. Moreover, as shown in the Supplemental Material of \cite{Ecker:2026akf}, the $\mathcal{O}(1/D^2)$ vanishes, as it should.

\subsubsection[\texorpdfstring{$N\geq 4$}{N>=4}]{\texorpdfstring{$\boldsymbol{N\geq 4}$}{N>=4}}\label{sec:6.1.5}

The LO solution is precisely the same as for $N=3$. Moreover, all subleading terms at any given order in $\epsilon$ are identical to each other, regardless of the choice of $N$, up to (including) the order $\mathcal{O}(\epsilon^{N-3})$. Starting with the order $\mathcal{O}(\epsilon^{N-2})$, in addition to a universal term present for any $N\geq 3$ there is an $N$-dependent contribution depending on $A$ defined in \eqref{eq:angelinajolie}; the higher the order, the higher the powers of $A$ that can appear. Apart from this, the scheme works exactly as for $N=3$ and we get a free integration function in $\tau$ at each order in $\epsilon$.

As an example beyond LO, we present the universal NLO results for all functions for any $N\geq 4$,
\begin{subequations}
    \label{eq:largeD110}
\begin{align}
\Omega &= \Omega_{\textrm{\tiny LO}} + \frac{2\epsilon\,x^2\intfct\intfct_1}{(1+\intfct^2x^2)^2} + \mathcal{O}(\epsilon^2)\\
f &= f_{\textrm{\tiny LO}} -\frac{\epsilon\,(1-x^2)\,\intfct\intfct_1}{(1+\intfct^2)^{3/2}\sqrt{1+\intfct^2x^2}} + \mathcal{O}(\epsilon^2) \\
\Pi &= \Pi_{\textrm{\tiny LO}} + \frac{\epsilon\,\intfct_1}{(1+\intfct^2x^2)^{3/2}} + \mathcal{O}(\epsilon^2) \\
\Psi &= \Psi_{\textrm{\tiny LO}} + \frac{(1+\intfct^2)(1+\intfct^2x^2)(\intfct+A\intfct^\prime_1)+A\intfct\intfct^\prime\intfct_1(1-4x^2-3\intfct^2x^2)}{D\,\sqrt{1+\intfct^2}(1+\intfct^2x^2)^3}\, + \mathcal{O}(\epsilon/D)
\end{align}
\end{subequations}
where $\intfct=\intfct(\tau)$ is the LO integration function and $\intfct_1=\intfct_1(\tau)$ the NLO integration function. The contributions with index LO are given in \eqref{eq:largeD4}. The Ricci scalar expands as
\eq{
R=R_{\textrm{\tiny LO}} - e^{2\tau}\,\frac{2\epsilon\,D\,\intfct(\tau/\Delta)\intfct_1(\tau/\Delta)}{(1+\intfct(\tau/\Delta)^2x^2)^2}  + \mathcal{O}(\epsilon^2\,D)
}{eq:largeD111}
where again we are using the original time $\tau$ and not the rescaled one.

A crucial aspect in the construction of the subleading contributions is to avoid poles $1/\intfct(\tau)$ if $\intfct$ is chosen to have at least one zero, a necessary feature to model any finite $D$ spacetime crystal. We found this always to be possible by suitably defining the integration functions. In fact, it is always possible (and convenient) to the define the integration functions $\intfct_n(\tau)$ such that
\eq{
\Pi(\tau,0) = \intfct(\tau) + \sum_{n=1}^\infty\intfct_n(\tau)\,\epsilon^n
}{eq:largeD106}
which turns out to avoid all such poles.

\subsubsection[\texorpdfstring{$N\to\infty$}{N->Infinity}]{\texorpdfstring{$\boldsymbol{N\to\infty}$}{N->Infinity}}\label{sec:6.1.6}

In this case, $\Delta$ does either not decay or decays slower than any monomial. Thus, we no longer expand in $\epsilon$ but instead expand in $1/D$. Moreover, we do not rescale $\tau$ anymore. The only change of the LO solution compared to \eqref{eq:largeD4} is 
\eq{
\Psi_{\textrm{\tiny LO}}(\tau,x) =\frac{\sqrt{1+\intfct^2(\tau)}}{D}\,\frac{\intfct(\tau)+\partial_\tau\intfct(\tau)}{(1+\intfct^2(\tau)x^2)^2} \,.
}{eq:largeD104}
This is so because $\partial_\tau$ derivatives no longer are enhanced by a monomial factor $1/\epsilon$. 

This is the case studied in \cite{Ecker:2026akf}, i.e., the large-$D$ solution presented therein, including their subleading corrections, requires the condition
\eq{
\lim_{D\to\infty}\Delta\,D^{1/N}\to\infty \qquad\forall N\in\mathbb{Z}^+\,. 
}{eq:DeltalargeD}
Since we additionally conjecture that $\Delta$ vanishes for $D\to\infty$ this means that $\Delta$ decays to $0$ more slowly than any positive power of $D$. A decay behavior allowed by these conditions would be, for example, $\Delta\propto 1/\ln{D}$ with some $\mathcal{O}(1)$ proportionality constant.

While we do not have sufficient numerical data to be more precise about the large-$D$ behavior of the echoing period $\Delta$, the data collected in Fig.~\ref{fig:logDeltalargeD} is compatible with the $N\to\infty$ case: if we make polynomial fits to the existing data we find the exponent $1/N$ keeps decreasing if we include data with higher values of $D$ and increasing if we exclude them. The only statement we can make confidently is that $N\geq 3$, see the discussion around Fig.~\ref{fig:logDeltalargeD}.

In the $N\to\infty$ case we can thus apply the large-$D$ expansion presented in \cite{Ecker:2026akf}, including subleading corrections. We compare their results with our numerical data in the previous Section and conclude the following. In regions 1., 2., and 3., i.e., in the near-center, the center-bulk-transition, and the bulk, the large-$D$ expansion works very well and shares many qualitative and quantitative features (such as NEC lines or the behavior of the maxima of the SSH function) with solutions at moderate values of $D$ ($4$ and above). Moreover, in this region some of the features of the large-$D$ solution do not depend on the choice of the free function $\intfct$, see, e.g., Fig.~\ref{fig:NEClargeD}. By contrast, in regions 4. and 5., i.e., in the bulk-SSH transition and the near-SSH, the large-$D$ expansion requires detailed knowledge of the free function $\intfct$. For instance, the $\mathcal{O}(x^4)$ term in the NNLO NEC lines \eqref{eq:NNLONEC} depends on $\intfct(\tau)$. This does not mean the large-$D$ expansion of \cite{Ecker:2026akf} does not work near the SSH. However, it means that to make it work one needs to come up with good choices for the function $\intfct(\tau)$, and the optimal choice may depend on the specific dimension one tries to approximate by a large-$D$ expansion. We comment more on this issue in Section \ref{sec:8}.

\subsubsection[Large-\texorpdfstring{$D$}{D} vs.~Taylor expansions]{Large-\texorpdfstring{$\boldsymbol{D}$}{D} vs.~Taylor expansions}
The large-$D$ expansion of \cite{Ecker:2026akf} uses $\epsilon=1/(D-1)$ as an expansion parameter instead of $1/D$. This was motivated by the observation \cite{Ecker:2024haw} that the NEC angle receives no $\mathcal{O}(\epsilon^2)$ corrections but only $\mathcal{O}(\epsilon^3)$ corrections for precisely this choice of $\epsilon$, i.e., with respect to the NEC angle the LO large-$D$ expansion is accurate even to NLO. We provide here an alternative reason for choosing  $\epsilon=1/(D-1)$ as optimal for a large-$D$ expansion by comparing the latter with the Taylor expansions of Appendix \ref{app:Taylor}. 

We start by noting that the dimensional scalings introduced in \cite{Ecker:2026akf} match perfectly the Taylor coefficients in the center, thus explaining why the large-$D$ expansion converges so well in this region: the LO quantities $f$, $\Omega$ and $\Pi$ were assumed to be $\mathcal{O}(1)$, while $\Psi$ was assumed to be $\mathcal{O}(\epsilon)$, which agrees with the $1/(D-1)$-factors in the Taylor coefficients \eqref{eq:TaylorSwift}.

Near the SSH, the expansion coefficients do not feature such factors, which explains why convergence may be slower there. However, we stress that the near-SSH expansion is still compatible with the large-$D$ expansion and there is no order-of-limits issue. To see this explicitly, note that the potentially dangerous terms linear in $D$ in the near-SSH expansion coefficients \eqref{eq:TaylorSSH} cancel precisely to LO due to the properties $\bar\omega_0=\mathcal{O}(\epsilon)$ and $\bar\psi_{+0}=\bar\psi_{-0}+\mathcal{O}(\epsilon)$.

Taking $\epsilon=1/(D-1)$ as large-$D$ expansion parameter implies that \eqref{eq:largeD104} receives a correction, replacing the factor $1/D$ by $1/(D-1)$. Using the LO property $f_0(\tau)=f_c(\tau)=f(\tau,0)=1/\sqrt{1+\intfct^2(\tau)}$ this expression evaluates in the center to
\eq{
\Psi_{\textrm{\tiny LO}}(\tau,0) =\frac{\partial_\tau\intfct(\tau)+\intfct(\tau)}{f_0(\tau)\,(D-1)}
}{eq:newsubsub1}
which should be compared with the Taylor expression \eqref{eq:psicenter}. We see that both expressions coincide if we identify $\intfct(\tau)=\psi_{+1}(\tau)$, which indeed holds to all orders in the large-$D$ expansion \cite{Ecker:2026akf}.

The key difference between the LO large-$D$ and near-center Taylor expansion is that in the latter case $f_0(\tau)$ and $\intfct(\tau)=\psi_{+1}(\tau)$ are independent boundary data, see Eq.~\eqref{eq:inidata_center}, whereas in the former case they are not. In other words, the boundary functions $f_c$ and $\Psi_c$ (see the left and middle Fig.~\ref{fig:BC}) are related in a specific way at large-$D$. This property permits a quantitative check of how well (or poorly) the large-$D$ expansion does: We take the converged results for $\psi_{+1}(\tau)$, which in the large-$D$ expansion is the function $\intfct(\tau)$ to all(!) orders \cite{Ecker:2026akf}. Then we plot the LO result $f_c(\tau)=1/\sqrt{1+\intfct^2(\tau)}$ and compare it with the converged result $f_0(\tau)$. We know already one key difference: The LO result for $f_c$ necessarily has maxima at $f_c=1$, which is not true at finite values of $D$. Additionally, the large-$D$ approximation also provides a prediction for the boundary data at the SSH, see Eq.~\eqref{eq:up_def}, which can be checked in a similar manner, using the LO large-$D$ result $\psi_{-p}=\intfct(\tau)/\sqrt{1+\intfct^2(\tau)}$.

\begin{figure}
\centering
\includegraphics[width=0.49\linewidth]{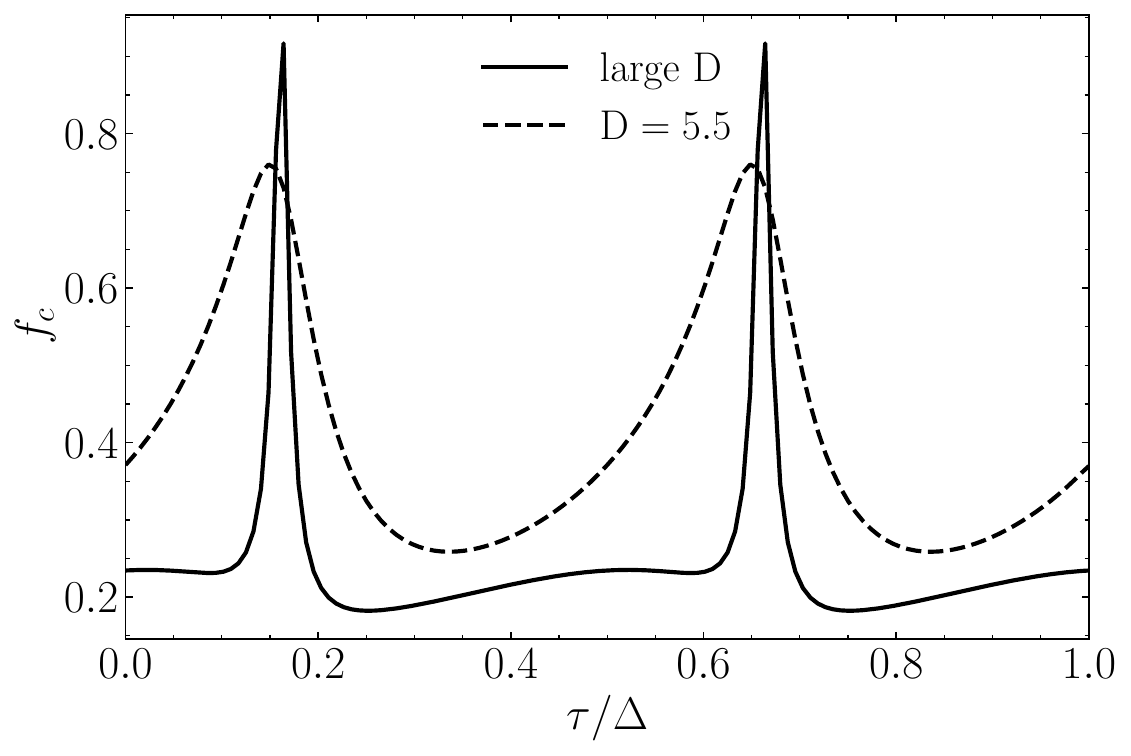}
\includegraphics[width=0.49\linewidth]{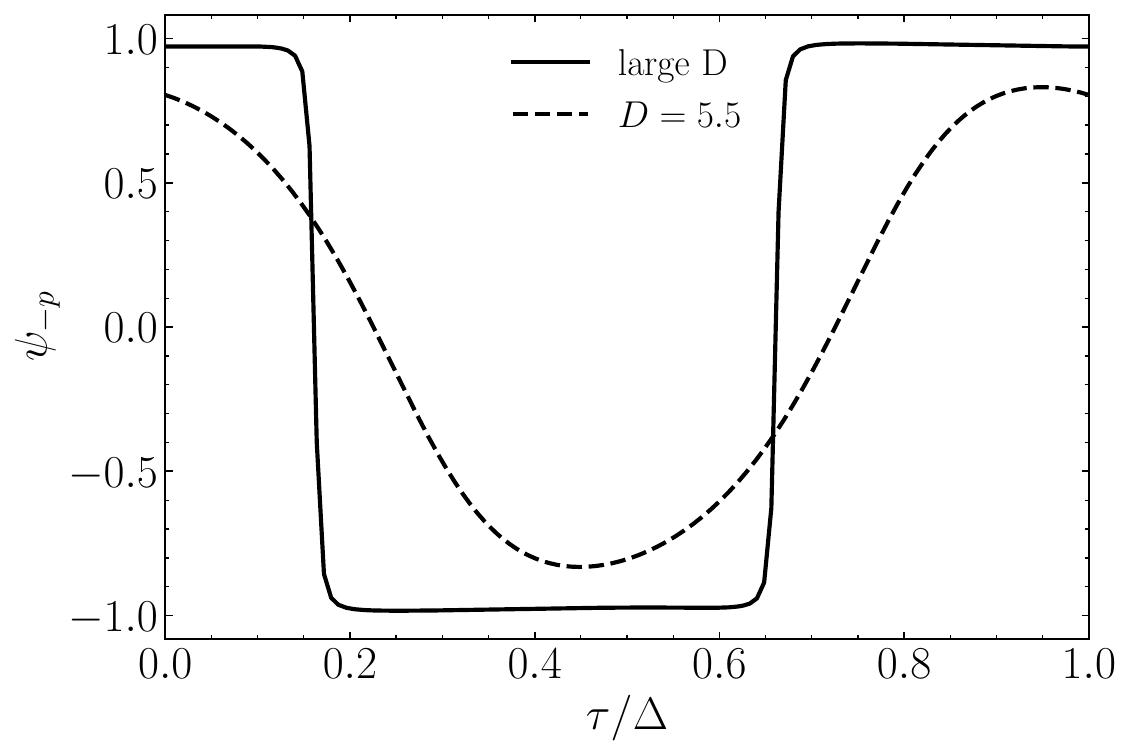}
\caption[Large-$D$ comparison of boundary data]{Comparison of converged numerical boundary data (dashed) with LO large-$D$ approximation (solid) for $D=5.5$. \emph{Left}: Boundary value $f_c$ at center ($x=0$). \emph{Right}: Boundary value $\psi_{-p}$ at SSH ($x=1$).}
\label{fig:blackjack}
\end{figure}

We have performed this rudimentary check for $D=5.5$ and plot the results in Fig.~\ref{fig:blackjack}.
The curves do not match, which we attribute to the dimension $D=5.5$ not being large enough for the LO approximation to work well. The only aspects that match qualitatively are the presence of two maxima located at approximately the NEC vertex time in the left plot, the approximate location and signs of the exrema in the right plot, and the approximate scale of the amplitudes in both plots.

An important caveat about the large-$D$ expansion is that the systematic improvement from adding subleading corrections comes with consistency conditions that are not necessarily obeyed by the numerical data at moderate values for $D$. Indeed, the explicit example worked out in \cite{Ecker:2026akf} does not obey these consistency conditions for $D<52$, which is a much larger dimension than studied in numerical simulations so far \cite{Sorkin:2005vz,Bland:2005kk}.

\subsection[Small-\texorpdfstring{$(D-3)$}{(D-3)} expansion]{Small-\texorpdfstring{$\boldsymbol{(D-3)}$}{(D-3)} expansion}\label{sec:6.2}

The small $(D-3)$-expansion of general relativity is not fully developed yet. In this Subsection, we provide the first steps, confined to the context of the present work. We start by showing that to LO, the equations of motion linearize and simplify to a Fuchsian system that we analyze in detail. The key take-away from this analysis is that the LO solution is always singular, either in the center or at the SSH. This motivates us to push further and search for a scaling regime where we simultaneously scale the dimension $D$, the echoing period $\Delta$, and the radial coordinate $x$ to small values.

\subsubsection{Fuchsian system}

For $D=3+\epsilon$ with positive $\epsilon\ll1$ we have a similar situation as at large $D$, except that it is now the function $f$ rather than $\omega$ that becomes approximately constant.
\begin{subequations}
\label{eq:smallD1}
\begin{align}
\omega(\tau,x) &= \omega_{\textrm{\tiny LO}}(\tau,x) + \omega_{\textrm{\tiny NLO}}(\tau,x)\,\epsilon  + {\cal O}(\epsilon^2) \\
f(\tau,x) &= 1 + f_{\textrm{\tiny NLO}}(\tau,x)\,\epsilon + {\cal O}(\epsilon^2) \\
\psi_-(\tau,x) &= \psi_{-\textrm{\tiny LO}}(\tau,x)+\psi_{-\textrm{\tiny NLO}}(\tau,x)\,\epsilon+{\cal O}(\epsilon^2) \\
\psi_+(\tau,x) &= \psi_{+\textrm{\tiny LO}}(\tau,x)+\psi_{+\textrm{\tiny NLO}}(\tau,x)\,\epsilon+{\cal O}(\epsilon^2)
\end{align}
\end{subequations}

The EOM \eqref{eq:eom} to LO in $\epsilon$ simplify to 
\begin{subequations}
\begin{align}
    x\,\big(\partial_\tau\pm(1\pm x)\partial_x\big)\,\omega_{\textrm{\tiny LO}} &= \pm\psi_{\pm\textrm{\tiny LO}}^2  \qquad\qquad
    x\partial_xf_{\textrm{\tiny NLO}} = e^{\omega_\textrm{\tiny LO}}-1 
\label{eq2:smallD2} \\
    2x\,\big(\partial_\tau\pm(1\pm x)\partial_x\big)\,\psi_{\mp\textrm{\tiny LO}} &= \pm\psi_{+\textrm{\tiny LO}}\pm\psi_{-\textrm{\tiny LO}} \label{eq:smallD200} 
\end{align}
\end{subequations}
Remarkably, the matter equations \eqref{eq:smallD200} decouple from the geometric equations \eqref{eq2:smallD2} and can be solved independently. Due to their linearity, we can solve separately for each Fourier mode. Again, it is convenient to use $\Pi_{\textrm{\tiny LO}}$ and $\Psi_{\textrm{\tiny LO}}$ defined as
\eq{
\Pi_{\textrm{\tiny LO}} = \frac{\psi_{+\textrm{\tiny LO}}+\psi_{-\textrm{\tiny LO}}}{2x}\qquad\qquad\Psi_{\textrm{\tiny LO}} =\frac{\psi_{+\textrm{\tiny LO}}-\psi_{-\textrm{\tiny LO}}}{2x^2}
}{eq:smallD3}
and Fourier expand them,
\eq{
\Pi_{\textrm{\tiny LO}}(\tau,x)=\sum_{k\in\mathbb{Z}} e^{i\frac{2\pi}{\Delta}k\tau}\pi_k(x)\qquad\qquad
\Psi_{\textrm{\tiny LO}}(\tau,x)=\sum_{k\in\mathbb{Z}} e^{i\frac{2\pi}{\Delta}k\tau}\hat\psi_k(x)
}{eq:smallD4}
assuming some fixed but undetermined echoing period $\Delta$.

Diagonalizing the first-order PDEs \eqref{eq:smallD200} and using the Fourier modes \eqref{eq:smallD4} yields a second-order Fuchsian ODE for each $\pi_k$,
\eq{
\pi_k'' + \frac{1-(4+\frac{4\pi i}{\Delta}k)\,x^2}{x(1-x^2)}\,\pi_k' + \frac{\frac{4\pi^2}{\Delta^2}k^2-2-\frac{6\pi i}{\Delta}k}{1-x^2}\,\pi_k = 0
}{eq:smallD5}
which has regular singular points at $x=0,1,-1$, two of which lie at the boundary of the domain we are concerned with, the center $x=0$ and the SSH $x=1$. 

Even though we know the zero-mode solution of \eqref{eq:smallD5}, 
\eq{
\pi_0 = \frac{c_0}{\sqrt{1-x^2}}+\frac{b_0\,\textrm{arctan}\sqrt{1-x^2}}{\sqrt{1-x^2}}
}{eq:smallD33}
is not switched on, we gain some insights that turn out to be true for generic non-zero modes, too: In the center, one of the solutions is regular ($c_0\neq0=b_0$) while the other one has a log-singularity ($c_0=0\neq b_0$). Conversely, the solution that is regular in the center has a one-over-square-root divergence at the SSH, $\pi_0\approx c_0/\sqrt{2(1-x)} + b_0\,\mathcal{O}(1)$. This matches with the numerical analysis of the singular branch near the SSH in Appendix \ref{app:SingularBranch}.

Using standard Fuchsian methods, the general solution for $\pi_k$ is given by
\eq{
\pi_k = c_k\,_2F_1\Big(1+\frac{i\pi k}{\Delta},\,\frac12+\frac{i\pi k}{\Delta},\,1;\,x^2\Big)+b_k\,_2F_1\Big(1+\frac{i\pi k}{\Delta},\,\frac12+\frac{i\pi k}{\Delta},\,\frac{3}{2}+\frac{i2\pi k}{\Delta};\,1-x^2\Big)\,.
}{eq:smallD6}
For general values of the integration constants $c_k,b_k$ the solution is singular in the center and at the SSH. Near the center, $\pi_k$ expands as
\eq{
\pi_k = -b_k\,(\ln x)\,\frac{2^{1+\frac{2\pi ik}{\Delta}}\Gamma(\frac32+\frac{2\pi ik}{\Delta})}{\sqrt{\pi}\,\Gamma(1+\frac{2\pi ik}{\Delta})}+ {\cal O}(1)
}{eq:smallD7}
and thus, the solution is regular in the center only if $b_k$ vanishes for every $k$. Therefore, we set $b_k=0$, in which case the Taylor expansion 
\eq{
\pi_k = c_k\,\Big(1 - \frac14\,\big(\tfrac{2\pi k}{\Delta}-i\big)\big(\tfrac{2\pi k}{\Delta}-2i\big)\,x^2 + {\cal} O(x^4)\Big)
}{eq:smallD17}
is regular in the center and compatible with all regularity conditions \eqref{eq:regularityorigin}.

The Fourier coefficient functions $\hat\psi_k$ can be obtained from the solution for $\pi_k$,
\eq{
 \hat\psi_k=\frac{\Delta}{2\pi i k}\,\frac{1-x^2}{x}\,\pi_k^\prime-\Big(1+\frac{\Delta}{2\pi ik}\Big)\,\pi_k
}{eq:smallD10}
and are also regular in the center since $\pi_k^\prime\propto x$. 

Using the solution regular at the center allows calculating the NEC angle as a cross-check. The near-center expansion \eqref{eq:smallD17} together with \eqref{eq:smallD10} imply the relation $\Psi_{\textrm{\tiny LO}}=\frac12\partial_\tau\Pi_{\textrm{\tiny LO}}$ at $x=0$. This yields the two spacelike NEC vectors $n_\pm^\mu\partial_\mu=\pm\partial_\tau+2\partial_x$ whose relative rapidity is then given by $\cosh\xi=5/3$ or $\xi=\ln 3$, yielding the NEC angle $\alpha=\textrm{gd}(\xi)=2\,\textrm{arccot}\,2$, in agreement with the general expression \eqref{eq:NECangle} for $D=3$. 

Having solutions for $\pi_k,\hat\psi_k$ yields solutions for $\psi_\pm$ by virtue of \eqref{eq:smallD3} and \eqref{eq:smallD4}. Their near-center expansion (we define for brevity $\hat k := \frac{2\pi}{\Delta}\,k$)
\eq{
\psi_{\pm}(\tau,x) = \sum_{k\in\mathbb Z} c_k\,e^{i\hat k\tau}\,\Big(x \pm  \frac{x^2}{2}\,\big(1+\hat k\big) + \frac{x^3}{2}\,\big(1+\frac{i\hat k}{2}\big)\big(1+i\hat k\big) + \mathcal{O}(x^4)\Big) 
}{eq:smallD38}
is regular by construction [since we set $b_k=0$ in \eqref{eq:smallD6}]. However, while $\psi_-(\tau,x)$ remains finite at the SSH, $\psi_+(\tau,x)$ has the same type of divergence as the zero mode,
\eq{
\psi_+(\tau,x\to1^-) = \sum_{k\in\mathbb{Z}}\frac{c_k\,e^{i\hat k\tau}}{(1 - x)^{\frac12+i\hat k}}\,\frac{\sqrt{2\pi}}{\cosh(\hat k\pi)\Gamma(\frac12-i\hat k)\Gamma(1+i\hat k)} + {\cal O}(1) 
}{eq:smallD37}
i.e., its absolute value diverges like $1/\sqrt{1-x}$ near the SSH. In Appendix \ref{app:SingularBranch}, we show that this is consistent with the expected behavior of the singular branch near the SSH extracted from numerics.

This means, much like in some quantum mechanical models, imposing regularity at both boundaries over-constrains the system. Unlike these quantum mechanical systems, there is no quantization condition on the free parameter $\Delta$ that would eliminate the singularity. On physical grounds, it may have been anticipated that at $\epsilon=0$ there can be no critical solution at the threshold of black hole formation since there are no black holes in 3d flat space Einstein gravity, see, e.g., \cite{Ida:2000jh}.

However, we still can get a glimpse of how the singularity could be resolved by looking at the $\Delta$-dependence of the singular term in \eqref{eq:smallD37}: in the limit $\Delta\to 0$, the whole expression vanishes due to the exponential damping from the $\cosh$-term in the denominator. While sending $\Delta\to 0^+$ does not commute with taking the near SSH limit $x\to 1^-$ --- nor the near-center limit $x\to 0^+$, see \eqref{eq:smallD38} --- these considerations suggest a different route, namely, to focus on scaling regimes of the PDEs where both $\Delta$ and $D-3$ are small. We investigate such regimes in Subsubsection \ref{sec:6.3}.

\begin{figure}
    \centering
    \includegraphics[width=0.6\linewidth]{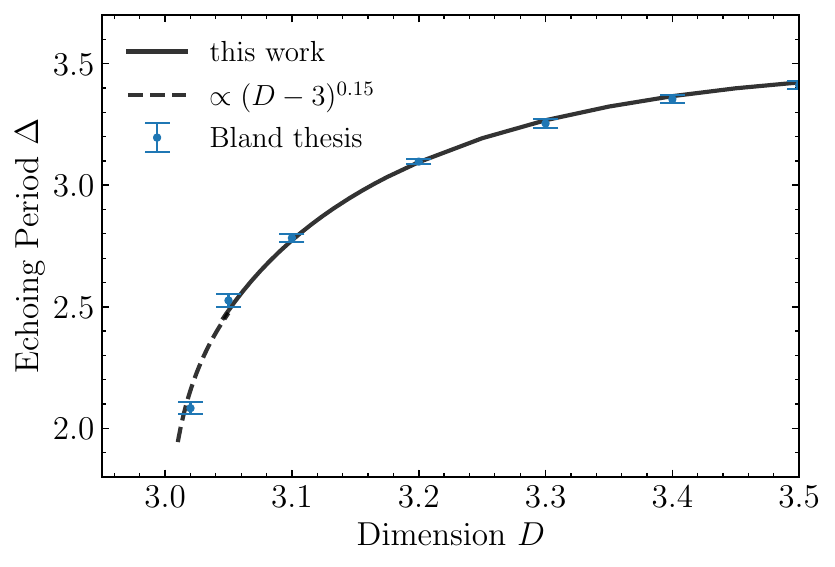}
    \caption[Echoing period $\Delta$ as function of $D-3$]{Echoing period $\Delta$ near $D=3$. Solid line: Our numerical results. Dashed line: Best-fit extrapolation $a\,(D-3)^{\alpha}$ with $a \approx 3.88$ and $\alpha \approx 0.15$. 
    Blue: Data points from Ref.~\cite{blandthesis} with error bars.
    }
    \label{fig:logDeltasmallD}
\end{figure}

Before analyzing scaling regimes, we display Fig.~\ref{fig:logDeltasmallD}, the low-dimensional counterpart of Fig.~\ref{fig:logDeltalargeD}. 
We can take away two lessons from this Fig.: First, if our conjecture is correct and $\Delta$ vanishes as $D\to3^+$ then it should do so as
\eq{
\lim_{D\to3^+}\Delta \propto (D-3)^\alpha \qquad \alpha\approx 0.15\,.
}{eq:Deltazero}
The exponent $\alpha$ does not have to be rational, and the precise form of $\Delta$ may contain non-polynomial terms in $D-3$. Our estimate $\alpha\approx 0.15$ is obtained from a best fit of our numerical data at $D\approx 3.05$ to the power law~\eqref{eq:Deltazero}. As seen in the Fig., the data point $\Delta=0.1379\pm0.0042$ at $D=3.02$ from Ref.~\cite{blandthesis} suggests an even faster fall-off. Taking this into account, our result may be interpreted as providing a lower bound, $\alpha\gtrsim 0.15$. However, for a conclusive determination of $\alpha$, data much closer to $D=3$ are required to reliably assess the scaling of $\Delta$ with $D-3$; our current lowest value, $D=3.05$, is clearly not sufficiently close. As discussed in previous sections, obtaining such data will require a new algorithm, since the present one becomes prohibitively expensive in this regime (see also Section~\ref{sec:8}).

\subsubsection{Scaling regime}\label{sec:6.3}

\newcommand{\scalecoord}{y}

We have seen above that we can get exact solutions for $D\to\infty$ and $D\to3^+$, but with deficiencies not shared by solutions at finite $D\in(3,\infty)$: the large-$D$ solutions are not unique and have an undetermined period, while the $D\to 3^+$ solutions are singular in the center or at the SSH unless we take an additional scaling limit where the echoing period tends to zero. This motivates us to search for scaling regimes where we send multiple quantities to zero at certain finetuned rates.

We focus here on $D\to 3^+$ and search for a scaling regime of the equations of motion \eqref{eq:eom}. Inspired by the behavior of the Weyl factor in Fig.~\ref{fig:l2norm} for small $\epsilon=D-3$, we consider the EOM \eqref{eq:eom} in a scaling regime where $x\sim\mathcal{O}(\epsilon^\delta)$ with some positive $\delta$. To account for this, we introduce the new spatial coordinate $\scalecoord=x/\epsilon^\delta$ and assume $\scalecoord\sim\mathcal{O}(1)$. We again uniformize the time by rescaling it as $\tilde\tau=\tau/\Delta$ so that $\tilde\tau\sim\tilde\tau+1$. Assuming there is a non-trivial solution in the scaling regime, we deduce 
\eq{
e^{\omega(\tau,\,x)} = \frac{a(\tau,\,x)}{\epsilon}\qquad\qquad a(\tau,\,x)\sim\mathcal{O}(1)
}{eq:scale1}
so that \eqref{eq:eom2} is consistent in the scaling regime [note that both sides of this equation are linear and homogeneous in $f$, so whatever its scaling is, $\omega$ must obey \eqref{eq:scale1}]. 

To LO in $\epsilon$, the EOM in the scaling regime simplify to
\begin{align}
    \scalecoord\,\partial_\scalecoord\ln a &= - a+\frac12\,\big(\psi_+^2+\psi_-^2\big) \label{eq:scaleeom1}\\
    \scalecoord\,\partial_\scalecoord\ln f &= a \label{eq:scaleeom2}\\
    \bigg(\frac{\epsilon^\delta}{f\,\Delta}\,\partial_{\tilde\tau}\mp\partial_\scalecoord\bigg)\psi_\pm &= \mp \frac{\psi_++\psi_- - 2a\psi_\pm}{2\scalecoord}\,.\label{eq:scaleeom3}
\end{align}
For consistency of the first equation, we require $\psi_\pm\sim\mathcal{O}(1)$. The second equation is solved by $f\sim\mathcal{O}(\epsilon^\beta)$ with an arbitrary exponent $\beta$.\footnote{%
In principle, $f$ could be non-monomial, so we cannot prove monomiality of $\Delta$ either. Nevertheless, for simplicity we assume monomial $f$.
} 
The last two equations are too trivial if either the time- or the space-derivative dominates over the other one. So, to get a sufficiently nontrivial solution in the scaling regime, we need to require a certain scaling of $\Delta$, namely
\eq{
\Delta\sim\mathcal{O}(\epsilon^{\delta-\beta})\,.
}{eq:scale2}
Thus, if $\beta<\delta$, the echoing period $\Delta$ tends to zero monomially in $\epsilon$ with the exponent given in \eqref{eq:scale2}.

We have checked our data using the $\ell^\infty$-norm and find the scaling $e^\omega\sim1/\epsilon$ as in \eqref{eq:scale1} is compatible with the data if we choose the scaling exponent $\delta\approx0.6$. We just used the ten lowest-dimensional data points (equidistant $D=3.05-3.056$) for this fit, as a compromise between wanting the dimension to be as close to 3 as possible and needing as many data points as possible. Using this value of $\delta$, we find the other scaling exponent $\beta\approx0.3$. From \eqref{eq:scale2} we deduce the scaling for the echoing period as $\Delta\sim\epsilon^\alpha$ with $\alpha\approx0.3$. When comparing with our data (see Fig.~\ref{fig:logDeltasmallD}), we find about half this value, $\alpha\gtrsim0.15$. As mentioned above, our lowest dimension, $D=3.05$, is apparently not small enough to get a more precise determination of $\alpha$ and to verify that we have reached this scaling regime.


\section{Critical collapse in 2d dilaton gravity}\label{sec:7}

There is a larger class of models described by actions of the form \eqref{eq:2d_SRG}, where the potentials $U(X)$ and $V(X)$ are given by expressions different from \eqref{eq:UV_srg}. While these models do not correspond to effective descriptions of higher-dimensional spherically symmetric Einstein gravity, they are still interesting in their own right. 

One prominent example is the Jackiw--Teitelboim (JT) model given by $U(X)=0$ and $V(X)=\rm const.$ \cite{Teitelboim:1983ux,Jackiw:1985je}. Both, spherically reduced models and the JT-model are part of a larger class, the so-called $ab$-family (see \cite{Grumiller:2002nm} for a review) given by 
\eq{
    U(X)=-\frac{a}{X} \qquad \qquad V(X)=-\frac{B}{2}X^{a+b}
}{eq:ab_potentials}
where $B\in \mathbb{R}$ is a scale parameter that can be fixed conveniently for each model. One can easily check that the spherically reduced models lie on a line $a=1+b$ with $a=(D-3)/(D-2)$. Even for other values of $a$ on this line one can show that all solutions are asymptotically flat \cite{Grumiller:2002nm}. The line $a=1-b$ contains the JT-model at $a=0$ but also more general models with asymptotically constant curvature and an (A)dS$_2$ ground state.   

A priori it is not clear that models different from Einstein gravity could exhibit critical collapse when coupled to a massless scalar field. However, one can at least ask whether some of these models support DSS solutions, which would be a first hint into this direction. We show now that this is generically the case and derive the echoing period for all models in the $ab$-family.

Consider a change of variables in the action \eqref{eq:2d_SRG} corresponding to a dilaton-dependent Weyl rescaling. Writing the new metric as $\hat{g}_{\alpha\beta}=X^{-\sigma}g_{\alpha\beta}$, $\sigma\in\mathbb{R}$ and choosing the potentials \eqref{eq:ab_potentials} the action reads
\begin{align}
    S_{\textrm{\tiny 2d}}[X,\hat{g}_{\alpha \beta},\psi ]&=\frac{1}{2}\,\int\extd^2x\sqrt{-\hat{g}}\,\Big(X\hat{R}-\sigma \hat{\nabla }^2X +\frac{\sigma +a}{X}(\partial X)^2+BX^{\sigma +a+b}\Big)\nonumber\\
    &\,-\frac{1}{2} \int\extd^2 x\sqrt{-\hat{g}}\,X(\partial \psi)^2
    \label{eq:abaction}
\end{align}
where the matter part was not modified because of its conformal coupling in 2d. Dropping the total derivative term one can readily see that this action is again part of the $ab$-family with a shifted value for the parameter $a$,
\eq{
    a\to a+\sigma ~.
}{eq:ashift}
Moreover, since all we did was a change of variables the solution spaces of the two different models are mapped onto each other. 

Let us now look at the example of spherically reduced Einstein gravity in $D=4$ and see where the critical solution is mapped to. Since the diffeomorphism $\phi$ associated to DSS satisfies
\eq{
    \phi_\ast g_{ab}=e^{-2\Delta}g_{ab} \qquad \qquad \phi_\ast X=e^{-2\Delta}X
}{eq:2dDSS}
it follows that the hatted metric satisfies
\eq{
 \phi_\ast \hat{g}_{ab}=e^{-2\Delta (1-\sigma)}\hat{g}_{ab} =:e^{-2\hat{\Delta}}\hat{g}_{ab}
}{eq:wtf}
and so we find again a DSS spacetime, albeit with different echoing period
\eq{
\hat{\Delta}=\Delta (1-\sigma)
}{eq:Deltasigma}
as can be seen from the metric
\eq{
    \extd \hat{s}^2=e^{-2\tau (1-\sigma)}\big(x^{2\sigma}\Tilde{g}_{\alpha \beta}(\tau,x) \extd x^\alpha \extd x^\beta\big)2=e^{-2\hat\tau}\big(x^{2\sigma}\hat{g}_{\alpha \beta}(\hat\tau,x) \extd x^\alpha \extd x^\beta\big)
}{eq:2dmetric}
where $\Tilde{g}_{\alpha\beta}$ is still given by \eqref{eq:lalapetz}. Since it has an unchanged echoing period $\tau\sim\tau+\Delta$ we see that in this new Weyl frame we should replace the coordinate $\tau$ by $\hat\tau=\tau(1-\sigma)$ which then readjusts the echoing period to $\hat\Delta$ when the fields are expressed as functions of $\hat\tau$.

From the point of view of the new model, the hatted quantities, however, may have very different physical behavior and satisfy different regularity conditions. As an example for this, the 2d Ricci scalar transforms like
\eq{
    \hat{R}=X^\sigma \big(R+\sigma \nabla ^2 \ln X\big) ~.
}{eq:hatR}
Using the boundary conditions for the critical solution in $D=4$ we can expand it around $x=0$ and find
\eq{
    \hat{R}(\tau ,x)\approx x^{-2(1-\sigma )}\big(-2\sigma e^{2\hat\tau}+\mathcal{O}(x^2)\big)
}{eq:rahrahrah}
which has a pole for $\sigma <1$. Since a Weyl-rescaling of the form we are considering cannot change the presence/absence of an apparent horizon, this is still a naked singularity which now extends over the whole $x=0$ line. 

By contrast, choosing $\sigma=1$ we find
\eq{
    \hat{R}=x^2e^{-2\tau}R-2e^{-\omega}\,\big(1-x\partial_x\ln f\big)
}{eq:Rhat}
which stays finite in the whole past domain, i.e. does not even form a naked singularity at $\tau \to \infty $ because $R\sim e^{2\tau }$. In this model, the metric of the critical solution is continuously self-similar (CSS) instead of DSS, as evident from \eqref{eq:Deltasigma}. For $\sigma>1$ there is still a naked singularity present at $\tau\to\infty$ but the $x=0$ line stays regular for finite $\tau$.

The discussion above can be generalized to arbitrary starting dimensions $D>3$. Performing similar Weyl rescalings one finds naked singularities at $x=0$ whenever $\sigma <\frac{2}{D-2}$, no naked singularity at all for $\sigma =\frac{2}{D-2}$, and a naked singularity only at $\tau \to \infty $ for $\sigma >\frac{2}{D-2}$. Continuing $D$ even further to values $D<3$, one finds similar bounds from the other sides. The Weyl transformed echoing period
\eq{
 \hat\Delta = \Delta \,\Big(1-\frac{(D-2)\,\sigma}{2}\Big)
}{eq:hatDelta}
yields a general formula for the echoing period for all $ab$-models, $\Delta_{a,\,b}$ in terms of the spherically reduced ones,
\eq{
 \Delta_{a,\,b} = \Delta_{D=2-1/b}\,\frac{a+b-1}{2b}
}{eq:abDelta}
where the echoing period on the right-hand-side, $\Delta_{D=2-1/b}$, can be read off from Fig.~\ref{fig:Delta} for $b=-1/(D-2)$ between approximately $-0.28$ and $-0.95$.

For the Minkwoski ground state models, $a=1+b$, we recover, of course, $\Delta_{1+b,\,b}=\Delta_{D=2-1/b}$. More interestingly, for (A)dS$_2$ ground state models, $a=1-b$, we get $\Delta_{1-b,\,b}=0$. This means that all (A)dS$_2$ ground state models have CSS rather than DSS solutions. A particular example is the JT model, Weyl related to the $D=1$ Minkowski ground state model. For general $a,b$ note that many DSS solutions are singular at the origin and hence not suitable candidates for CSCs, so despite the generality of our result \eqref{eq:abDelta} it is of limited use in applications.

\begin{figure}
\centering
\includegraphics[width=0.6\linewidth]{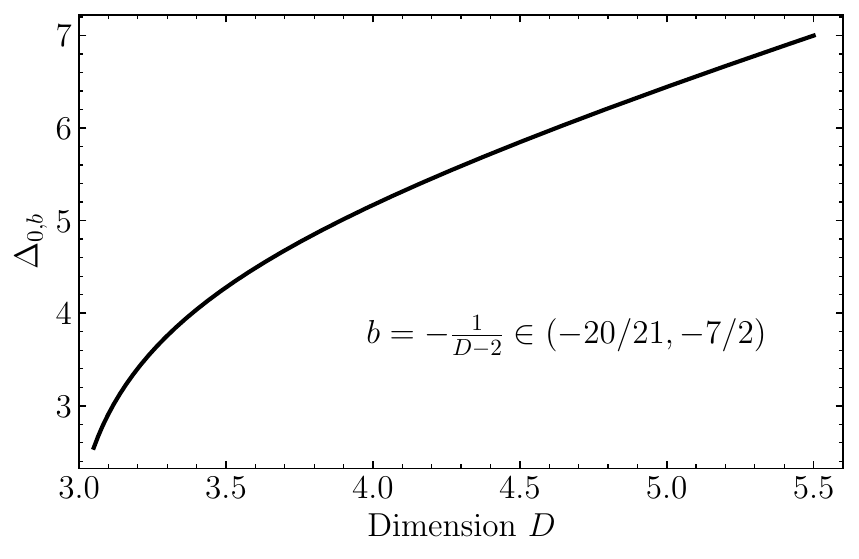}
\caption[Echoing period for 2d dilaton gravity without kinetic term]{Echoing period for 2d dilaton gravity without kinetic term.}
\label{fig:a0}
\end{figure}

An often-used class of models omits the kinetic term, $a=0$, see, e.g., \cite{Louis-Martinez:1993bge}. Restricting to this class within the $ab$-family yields $\Delta_{0,\,b}=\Delta_{D=2-1/b}\,\frac{D-1}{2}$. We plot the echoing period in Fig.~\ref{fig:a0}.
As opposed to Fig.~\ref{fig:Delta}, the echoing period is strictly monotonically increasing with the dimension. The behavior near $D=3$ is identical to spherically reduced gravity because the prefactor $(D-1)/2$ in the relation between the two echoing periods tends to $1$. At large $D$, the behavior is quite different since the linear prefactor dominates over the relatively weak asymptotic falloff in Fig.~\ref{fig:Delta}. 

\begin{figure}[htb]
\centering
\begin{tikzpicture}[scale=2.5]

    \fill[gray!30] (-1.2, -2.2) -- (3.2, -2.2) -- (1, 0) -- (-1.2,0) -- cycle;

    \fill[gray!30] (1, 0) -- (3.2, 0) -- (3.2, 2.2) -- (-1.2,2.2) -- cycle;

    \fill[orange, opacity=0.5] (-1.2,-0.28571) rectangle (3.2,-0.95238);

    \draw[->,thick] (-1.4, 0) -- (3.4, 0) node[right] {$a$}; 
    \draw[->,thick] (0, -2.4) -- (0, 2.4) node[above] {$b$}; 

    \foreach \x in {2} {
        \draw (\x,0.1) -- (\x,-0.1) node[below] {$\x$};
    }
    
    \foreach \y in {-1,1} {
        \draw (0.1,\y) -- (-0.1,\y) node[left] {$\y$};
    }

    \draw[thick, black] (1.5, 0.5) -- (3.2, 2.2) node[anchor=south west] {$a = 1 + b$}; 
    \draw[thick, black] (-1.2, -2.2) -- (0,-1) node[anchor=south west] {}; 
    \draw[thick, dotted] (1, 0) -- (1.5, 0.5) node[anchor=south west] {}; 

    \draw[thick, blue]  (3.2, -2.2) -- (-1.2, 2.2)  node[anchor=south west,rotate=-45] {}; 

    \node[] at (1,-1.8) {\small{NS along $x=0$}};
   \node[blue, rotate=-45] at (-0.7,1.9) {\small{No NS}};
    
    \coordinate (A) at (0,-1);
    \coordinate (B) at (0.5,-0.5);
    \coordinate (C) at (1,0);
    \coordinate (D) at (1.5,0.5);
    
    \draw[thick, red] (A) -- (C);
    \draw[dashed, black] (0,1) -- (2,1);
    \draw[black] (1,0.3) -- (1,0.1);

    \draw[->, thick] (-0.5, -1.3) -- (-1, -1.8) node[above=3pt,right=-10pt,rotate=45] {$D\to 2^+$};

    \draw[->, thick] (2.5, 1.3) -- (3, 1.8) node[left=-10pt,rotate=45] {$D\to 2^-$};

   \draw[thick,red] (0, -1) circle (1pt);
    \node[anchor=west] at (0, -1) {$D=3$};  
    \draw[fill=black] (0.5, -0.5) circle (0.8pt);
    \node[anchor=west] at (0.5, -0.5) {$D=4$};  
    \draw[thick,red] (1, 0) circle (1pt);
    \node[anchor=north] at (1, 0.6) {$D\to \infty $};  
    \draw[fill=black] (0, 1) circle (0.8pt);
    \node[anchor=east] at (0.5, 1.2) {JT}; 
    \draw[fill=black] (1.5, 0.5) circle (0.8pt);
    \node[anchor=west] at (1.5, 0.5) {$D=0$};
    \draw[fill=black] (2, 1) circle (0.8pt);
    \node[anchor=west] at (2, 1) {$D=1$};
\end{tikzpicture}
\caption[Two-parameter family of 2d dilaton gravity models]{Parameter space of $ab$-family. Red segment: spherically reduced Einstein gravity ($3<D<\infty$). Blue line: (A)dS$_2$ ground state models with CSS. Weyl rescalings preserve $b$ but shift $a$. Gray regions (except diagonal lines): DSS solutions with naked singularity along $x=0$. Orange rectangle: $ab$-models accessible with our data with echoing period $\Delta$ given by \eqref{eq:abDelta}.}
\label{fig:ab_plot}
\end{figure}
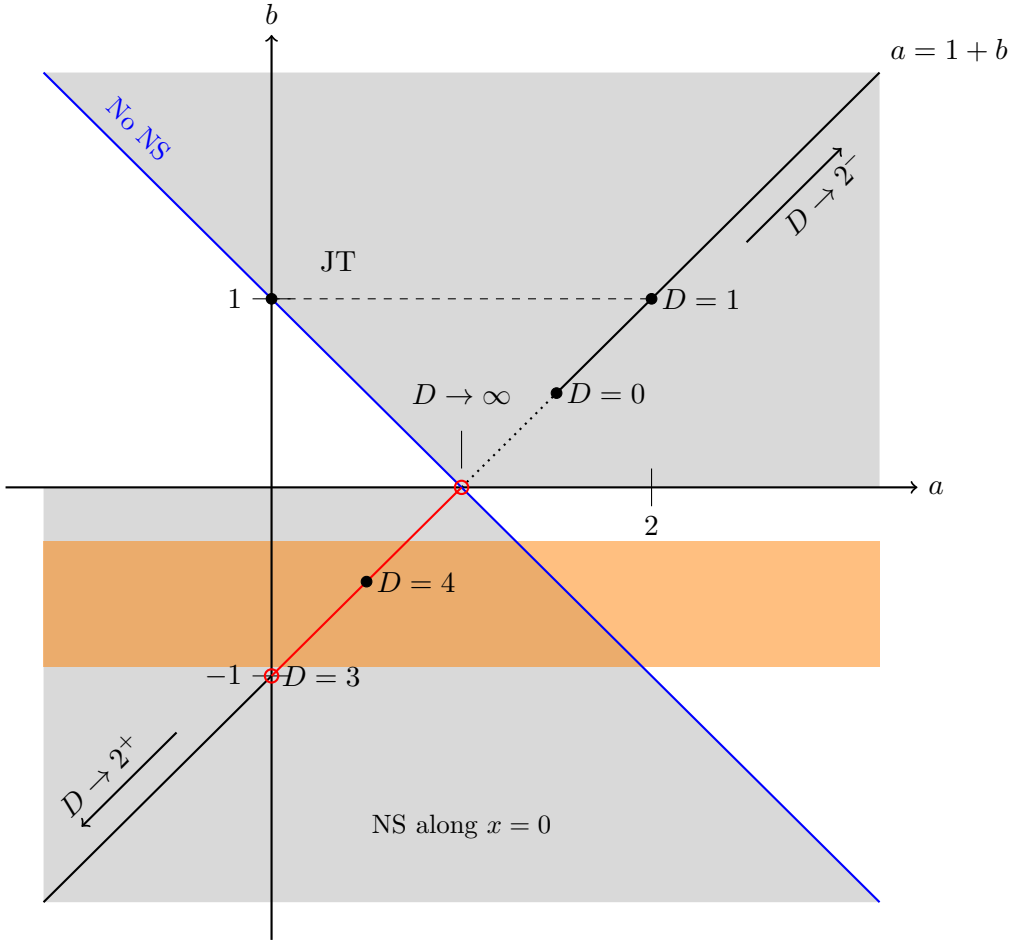

The full parameter space is summarized in Fig.~\ref{fig:ab_plot}.
The models on the red segment were the focus of the first six Sections of our paper. Weyl rescalings move horizontally in this model space, i.e., $a$ changes but $b$ remains the same. Such Weyl rescalings typically lead to singularities at either $x=0$ (gray shaded regions) or $x=\infty$ (white shaded regions). The orange rectangle covers all models whose echoing period we can determine using \eqref{eq:abDelta} and our data in Fig.~\ref{fig:Delta}. The diagonal extending the red segment describes models with a Minkowski ground state. The blue diagonal describes model with (A)dS$_2$ ground state, e.g., JT. An interesting aspect for our large-$D$ considerations is that the blue line intersects the diagonal with Minkowski ground state models precisely at the location of the Witten black hole, which in turn arises as effective description of general relativity at infinite $D$ \cite{Emparan:2013xia}. Since the (A)dS$_2$ ground state models on the blue line have vanishing $\Delta$, this confirms our conjecture $\lim_{D\to\infty}\Delta=0$. 

Table \ref{tab:1} summarizes all models within the $ab$-family that have a Minkowski ground state, i.e., $a=1+b=(D-3)/(D-2)$. These models lie on the black/red diagonal in Fig.~\ref{fig:ab_plot}.
\begin{table}[h!tb]
\centering
\begin{tabular}{|l|l|l|l|}\hline
range of $D$ & model/black hole name \& Refs. & asymp.~flat? & $X=0$ at $x=0$? \\ \hline
$D=\infty$ & Witten \cite{Mandal:1991tz,Elitzur:1991cb,Witten:1991yr} & yes & yes \\
$3<D<\infty $ & Schwarzschild--Tangherlini \cite{Schwarzschild:1916uq,Tangherlini:1963bw} & yes & yes \\
$D=3+0$ & 3+$\epsilon$ \cite{Grumiller:2003hq}, $B\neq 0$ in \eqref{eq:ab_potentials} & no & yes \\
$D=3$ & no BH \cite{Ida:2000jh}, $B=0$ in \eqref{eq:ab_potentials} & --- & --- \\
$2<D<3$ & asymp.~mass dominated \cite{Bagchi:2014ava} & no & yes\\
$D=2+0$ & Liouville \cite{Mann:1992ar,Grumiller:2007wb} & no & yes \\
$D=2$ & theory does not exist & --- & --- \\
$1<D<2$ & exotic \cite{Katanaev:1996ni} & no & no\\
$D=1$ & conformal to JT \cite{Ecker:2022vkr} & no & no\\
$-\infty<D<1$ & exotic \cite{Katanaev:1996ni} & no & no \\\hline
\end{tabular}
\caption[Zoo of 2d dilaton gravity models with Minkowski ground state]{Zoo of 2d dilaton gravity models with Minkowski ground state.}
\label{tab:1}
\end{table}
Our focus was on $3<D<\infty$, the Schwarzschild--Tangherlini black holes. From the Table it is evident that the dimensions $D=1,2,3,\infty$ are special. This is also obvious from the analysis in previous Sections: For $D=3$ various terms in the equations of motion \eqref{eq:eom} vanish. Moreover, in $D=3$ we do not have black hole solutions. In the range $2<D<3$ various terms in the equations of motion \eqref{eq:eom} change their signs so that, for instance, the convexity conditions \eqref{eq:inequalities} no longer hold. In this range, black hole solutions exist but they are not asymptotically flat in the strict sense: While they still have a Minkowski ground state, the metric does not asymptote to the Minkowski metric at large values of the dilaton, reminiscent of Rindler space. Indeed, for the lower endpoint of that interval, $D=2$, finite mass solutions yield 2d Rindler space instead of black holes. From our analysis in Section \ref{sec:3} it is also evident that $D=2$ is special, since again several terms in the equations of motion \eqref{eq:eom} vanish and the rescaling of the matter variables \eqref{eq:psipm} becomes singular. Moreover, the NEC angle \eqref{eq:NECangle} is a right angle at $D=2$, indicating that the NEC lines become null near the center. For $-\infty<D<2$ the relation \eqref{eq:dilaton} between dilaton and radial coordinate implies that the geometric center, $x=0$, no longer corresponds to vanishing dilaton but rather to infinite dilaton. This reversal between weak and strong coupling regions is well-documented in the 2d dilaton gravity literature, see, e.g., \cite{Bagchi:2014ava,Ecker:2022vkr}. A special point in that range is $D=1$, the Minkowski ground state model conformal to JT. Our expressions for the Taylor expansion near the center become singular in this case, see the various expressions in Appendix \ref{app:Taylor} with poles $1/(D-1)$. Another way of explaining why $D=1$ is special is to consider the NEC angle $\alpha$ presented in \eqref{eq:NECangle}, which purely geometrically should obey $\alpha\leq\pi$; at $D=1$ this inequality saturates. Finally, the singular limit $D\to\infty$ yields the Witten black hole \cite{Mandal:1991tz,Elitzur:1991cb,Witten:1991yr}, which exhibits asymptotically flat black hole solutions. 


\section{Conclusions}\label{sec:8}

In this last Section, we conclude. In Subsection \ref{sec:8.1}, we summarize key results and plots. In Subsection \ref{sec:8.2}, we formulate conjectures based on our numerical and analytic results. In Subsection \ref{sec:8.3}, we provide an outlook to future developments.

\subsection{Summary}\label{sec:8.1}

In this paper, we constructed CSCs in dimensions $D$ between $3.05$ and $5.5$. Starting from the massless Einstein--Klein--Gordon action \eqref{eq:EMKG_ddim} in some integer spacetime dimensions $D>3$, we assumed spherical symmetry to obtain the 2d dilaton gravity action \eqref{eq:2d_SRG} where $D$ can be analytically continued to arbitrary real numbers. We derived the equations of motions \eqref{eq:eom} using suitable gauge-fixing and some additional assumptions. In particular, we imposed the DSS property \eqref{eq:1}, adapted the gauge so that we cover the full past patch in Figs.~\ref{fig:illustration_crystal} and \ref{fig:1}, and demanded the absence of winding modes in \eqref{eq:winding}. We exploited the periodicity of our functions with the echoing period $\Delta$ by Fourier decomposing all functions. A key aspect was the generalized symmetry \eqref{eq:gen_sym} that permitted to fix one Fourier component, in some sense making room for the echoing period $\Delta$ in our optimization algorithm.

We solved the equations of motion \eqref{eq:eom} after imposing parity and regularity conditions. Our algorithm involved imposing boundary conditions on the SSH function $f$ and the matter function $\Psi$ in the center and shooting from there to some matching surface, and boundary conditions for the matter function $\psi_-$ at the SSH and shooting from there to the same matching surface. At the matching surface, we took the free Fourier components of these three boundary functions and the echoing period $\Delta$ as parameters to be optimized by a Newton algorithm, described in more detail in Section \ref{sec:4}.

Our results for spacetime dimensions $D\in[3.05,5]$ are contained in Section \ref{sec:5}. The key information is contained in four Figures: Fig.~\ref{fig:fields} provides 3d plots of all geometric and matter functions for the dimensions $D=3.05$, $D=4$, and $D=5.5$. Fig.~\ref{fig:Delta} plots the prime observable of CSCs, the echoing period $\Delta$, as a function of $D$. Fig.~\ref{fig:gamma} plots the key secondary observable of CSCs, the Choptuik exponent $\gamma$, as a function of $D$. Finally, Fig.~\ref{fig:NEClines} plots the NEC saturation lines for various dimensions, which coincide with lines of vanishing Ricci scalar.

A spin-off for $D=4$ are our precise determination of the echoing period \eqref{eq:Delta4} and the Choptuik exponent \eqref{eq:gamma4} to about one part in a million. We also highlight the critical dimension \eqref{eq:Deltacrit} where the echoing period has a maximum (and possibly even a supremum). Since also $D=5$ is of interest given the numerous applications of AdS$_5$/CFT$_4$ (see \cite{Alvarez-Gaume:2006klm} for a pertinent example) we display now explicitly our results for the echoing period $\Delta$ and the Choptuik critical exponent $\gamma$ for this dimension:\footnote{%
The result for $\Delta$ is from our CSC data with error estimates discussed in Appendix \ref{app:Error}. The result for $\gamma$ is obtained from interpolating between results in our linearized data. The reason our algorithm does not produce results in dimensions $D\in[4.81,5.38]$ is explained in the last paragraph of Section \ref{sec:chop_num}.
}
\eq{
D=5:\qquad \Delta = 3.22176\pm 10^{-5}\qquad \gamma = 0.41322\pm 10^{-5}
}{eq:D5}

At the low end, we stopped our numerical evaluation at $D=3.05$ because in lower dimensions, our algorithm becomes prohibitively costly. We identified as the main source of the problem the increased amplitude and peakiness of the boundary function $\Psi_c$, see the bottom Fig.~\ref{fig:fields} and the middle Fig.~\ref{fig:BC}. It seems plausible to us that there could be better algorithms that convert this peaky structure into an asset and yield faster convergence. However, even within the framework of our algorithm there is likely room for further improvement, e.g., in optimizing the position of the matching surface or the way the spatial grid is set up. Probably, some of these improvements could even be automatized.

At the high end, we stopped our algorithm at $D=5.5$ because in higher dimensions, we no longer can reliably discriminate between the regular and the singular branches at the SSH. This issue could be resolved by going to higher orders in the Taylor expansion near the SSH, see Appendix \ref{app:Taylor}. In principle, it could be easier to go to higher dimensions than to lower dimensions with our algorithm since fewer Fourier modes are needed, see the power spectra in Fig.~\ref{fig:Nt_conv}. All our data on echoing period $\Delta$ and Choptuik exponent $\gamma$ are collected in Appendix \ref{app:numbers} and in \cite{CSCdata:2026}.

We complemented our numerical analysis by analytical results in Section \ref{sec:6}, expanding either in $1/D$ (or even better, in $1/(D-1)$ \cite{Ecker:2026akf}) or in $D-3$. These limiting cases have features qualitatively different from finite $D\in(3,\infty)$. Namely, in the large-$D$ limit we lose the uniqueness property of DSS solutions that is characteristic for CSCs, i.e., neither the echoing period nor the boundary functions are fixed uniquely. In the small-$(D-3)$ limit we lose the regularity property of DSS solutions, i.e., we have singularities either in the center or at the SSH, unless we take an additional scaling limit where we send the echoing period to zero. We discussed properties of such scaling regimes in Subsubsection \ref{sec:6.3} and concluded that our lowest dimension $D=3.05$ is not quite low enough to test the properties of these scaling regimes. This provides yet-another motivation for pushing future numerical simulations even closer to $D=3$. In Subsection \ref{sec:8.2} we combine the insights from our numerical and analytical results to present some conjectures. 

Finally, in Section \ref{sec:7} we pointed out that our results can be applied to a much larger class of theories than the spherically reduced massless Einstein--Klein--Gordon model, namely an infinite set of intrinsically 2d dilaton gravity models, see the summaries in Fig.~\ref{fig:ab_plot} and Table~\ref{tab:1}. We postpone further generalizations to Subsection \ref{sec:8.3} below.

\subsection{Conjectures}\label{sec:8.2}

We summarize now six conjectures and then discuss their respective statuses. 
\begin{enumerate}
\item Weak version: $\lim_{D\to\infty}\Delta=0^+$, Strong version: $\lim_{D\to\infty}\Delta\,D^{1/N}\to\infty$, $\forall N\in\mathbb{Z}^+$
\item $\lim_{D\to\infty}\gamma=\frac12$ 
\item $\lim_{D\to\infty}\frac{\textrm{NEC}_{\textrm{\tiny in}}}{\textrm{NEC}_{\textrm{\tiny out}}}=0$ 
\item Weak version: $\lim_{D\to 3^+}\Delta=0^+$, Strong version: $\lim_{D\to 3^+}\Delta\,(D-3)^{-\alpha}\to0$, $\alpha\gtrsim0.15$ 
\item $\lim_{D\to 3^+}\gamma\to 0^+$
\item $\lim_{D\to 3^+}\frac{\textrm{NEC}_{\textrm{\tiny in}}}{\textrm{NEC}_{\textrm{\tiny out}}}=1$ 
\end{enumerate}
In these conjectures, $\Delta$ is the echoing period, $D$ the spacetime dimension, $\gamma$ the Choptuik exponent, and NEC$_{\textrm{\tiny in}}$/NEC$_{\textrm{\tiny out}}$ is the ratio of the time interval at the SSH between NEC lines emanating from the same vertex over the time interval at the SSH between adjacent NEC lines emanating from neighboring NEC vertices. The strong versions mean that weak and strong limits are both true simultaneously, while for the weak versions only the weak limit is true.

We explain why we believe the first and fourth conjectures might be true. First of all, they are supported by the data displayed in Fig.~\ref{fig:Delta} since $\Delta$ clearly falls off both towards large $D$ and towards $D=3$, see also Figs.~\ref{fig:logDeltalargeD} and \ref{fig:logDeltasmallD} for a quantitative analysis supporting the strong versions of these conjectures. Second, in $D=3$ with negative cosmological constant there is an exact critical solution by Garfinkle \cite{Garfinkle:2000br} with the CSS property, in agreement with earlier numerical simulations by Pretorius and Choptuik \cite{Pretorius:2000yu}. Since CSS emerges from DSS in the limit $\Delta\to0$, this example provides a theoretical indication that $\Delta$ has to go to zero as $D=3$ is approached. However, there are three weak points in this argument: i)~The Garfinkle solution has non-vanishing winding number $n$ in \eqref{eq:winding} while all our CSCs require vanishing winding number, ii)~The solutions by Garfinkle and by Pretorius and Choptuik both require negative cosmological constant and we demanded vanishing cosmological constant, and iii)~As we showed in Section \ref{sec:6.2} the strict limit $D\to3$ does not exist. In the large-$D$ limit the behavior of the NEC lines uncovered in \cite{Ecker:2026akf} is at least compatible with $\Delta$ tending to zero as long as it does not do so too fast, because the NEC lines do not extend very far --- indeed, the NEC angle \eqref{eq:largeD39} tends to zero at large $D$. We quantified what we mean by ``too fast'' in Section \ref{sec:6}. Finally, the 2d dilaton gravity discussion in Section \ref{sec:7} shows that the line where the echoing period vanishes intersects spherically reduced gravity precisely at the point of the Witten black hole (see Fig.~\ref{fig:ab_plot}) where formally $D\to\infty$, supporting the weak version of the first conjecture. Having said this, our conjectures may be wrong and it would be good to test them, e.g., by pushing the numerics even closer to $D=3$ and $1/D\to 0$. This will require new numerical algorithms, see also our comments in Section~\ref{sec:8.3}.

The second conjecture is well-known in the large-$D$ literature, see \cite{Emparan:2020inr} and Refs.~therein, while the fifth one seems to be new (previous work suggested instead the limit might be either finite or minus infinity \cite{Bland:2005kk,blandthesis}). While gathering more data on large-$D$ will probably asymptotically confirm the conjecture, gathering more small-$(D-3)$ data could be very useful for testing the fifth conjecture since for $D=3.05$, $\gamma\approx0.16$ is not very close to zero yet. Moreover, such data could refine the conjecture into a stronger version that provides the decay rate of $\gamma$ as a function of $(D-3)$. As above, this will require new numerical algorithms.

The third and sixth conjectures are new and the main support from them are the data in Fig.~\ref{fig:diffSSHoverDel}. Especially the sixth conjecture is a bit unexpected and, if correct, probably has some deeper explanation. The third conjecture can be proven if the strong version of the first conjecture holds, using the large-$D$ prediction \eqref{eq:largeDDeltatau}.

\subsection{Outlook}\label{sec:8.3}

We conclude with an outlook to a variety of future research directions, besides testing the conjectures of the previous Subsection.
\begin{itemize}
    \item \textbf{Outer patch and Cauchy horizon.} We focused on constructing CSCs in the past patch of Fig.~\ref{fig:1} but it is straightforward numerically to continue these solutions into the whole outer patch all the way to the Cauchy horizon, along the lines of \cite{Martin-Garcia:2003xgm}. It could be rewarding to do this and to confront the question of how to approach (and extend beyond) the Cauchy horizon for arbitrary spacetime dimensions $D$, and to verify if there are again some simplifications for large or small values of $D-3$.
    \item \textbf{Winding modes.} In our analysis we have restricted attention to the case without winding, $n=0$, following Choptuik's original $D=4$ construction and subsequent studies in integer dimensions. However, more general DSS solutions of the form
    \eq{
    \psi(\tau,x)=n\tau+\tilde\psi(\tau,x)\qquad\qquad \tilde\psi(\tau+\Delta,x)=\tilde\psi(\tau,x)
    }{eq:windings}
    with $n\neq 0$, are mathematically allowed. While no such solutions are known in asymptotically flat spacetimes, a related example has appeared in AdS$_3$ gravity \cite{Pretorius:2000yu,Garfinkle:2000br}.\footnote{%
    This subject has some interesting history, especially regarding the various rational values of $\gamma$ found in the literature depending on the winding number, including $\gamma=\frac{4}{5}$ \cite{Husain:2000vm}, $\gamma=\frac{2}{5}$ \cite{Clement:2001ns}, $\gamma=\frac{1}{2}$ \cite{Birmingham:2001hc,Baier:2013gsa}, then generalized to $\gamma=2n/m$ with certain conditions on the positive integers $n,m$ \cite{Garfinkle:2002vn,Hirschmann:2002bw,Cavaglia:2004mt}, see also \cite{Jalmuzna:2015hoa}.
    } 
    Since we approach $D\to 3^+$ in our setup, and since AdS is natural in the context of string theory and holography, see, e.g., \cite{Chesler:2019ozd}, it would be worthwhile to revisit whether nonzero winding numbers can yield additional families of spacetime crystals also above $D=3$. 
    \item \textbf{Small $\boldsymbol{(D-3)}$.} Our lowest dimension, $D=3.05$, may seem close to $D=3$, but the echoing period at that value, $\Delta\approx2.48$ and the Choptuik exponent, $\gamma\approx0.16$, are still not close to zero. Thus, it would be good to push numerical simulations even closer to 3 to see if and how the echoing period and the Choptuik exponent approach zero. As mentioned, this will require a new type of algorithm that does not suffer as much from the peaks and high amplitudes in our boundary function $\Psi_c$, see the left purple plot in Fig.~\ref{fig:fields}. It could be that using spectral methods also in the spatial direction (e.g., with Chebyshev polynomials as basis) may improve the situation but perhaps more radical departures from our methods are required. On the analytical side, it would be useful to further investigate scaling regimes and find evidence for how the various fields have to scale as $D\to3^+$. Additionally, it may be rewarding to construct some effective gravity theory in the limit of $3+\epsilon$ dimensions, perhaps analogous to gravity in $2+\epsilon$ dimensions \cite{Mann:1992ar,Grumiller:2007wb}. Another promising idea is to transpose the large-$D$ methods of Clark and Pimentel \cite{Clark:2025tqi} to construct CSS solutions at small $(D-3)$. Finally, a complementary numerical approach would be to push critical collapse simulations to even lower dimensions than in Bland's thesis \cite{blandthesis}. Besides improving the numerical analysis at small $(D-3)$, it could be rewarding to develop theoretical aspects of gravity in $3+\epsilon$ dimensions, including a description of black holes, their thermodynamics, boundary conditions, effective actions, and addressing the (in)compatibility of the Newton limit for $\epsilon\to 0^+$. Some of these aspects will be addressed in future work.
    \item \textbf{Large $\boldsymbol{D}$.} Similarly, at large $D$ one should adapt our algorithm to push to higher dimensions to test the conjectures $\lim_{D\to\infty}\Delta=0$ and $\lim_{D\to\infty}\gamma=\frac12$. Adding more Taylor expansion coefficients near the SSH is a useful first step to better resolve the singular branch there. Ironically, the main problem is that the singular branch gets less and less singular the higher the dimension and therefore, is harder to discriminate numerically from the regular branch. Probably, our algorithm may then be extended into double digit dimensions. However, for our ultimate goal to match precisely with the large-$D$ expansion of \cite{Ecker:2026akf} it is possible that the algorithm has to be changed more drastically. Besides numerical work, the large-$D$ expansion also allows analytical treatment of CSS solutions \cite{Clark:2025tqi} and DSS solutions \cite{Ecker:2026akf,Mathematica:largeD}, see their respective outlooks for corresponding research directions.
    \item \textbf{Window $\boldsymbol{2<D<3}$ and small $\boldsymbol{(D-2)}$.} Our analysis of DSS solutions in 2d dilaton gravity leads us to believe that there could be critical collapse in dimensions between $2$ and $3$. While a bit academic, it still could be interesting to numerically construct such solutions and their echoing periods with the intention to check whether the limit $D\to3^-$ matches smoothly with the $D\to3^+$ results or not. Moreover, there could be another (semi-)analytic regime, $D\to2^+$, where perturbative methods might work, expanding in $(D-2)$.
    \item \textbf{Generic 2d dilaton gravity models and exotic $\boldsymbol{D<2}$.} To complete the scan of the whole $ab$-family one should also consider the exotic models of ``gravity in less than two dimensions'' that cover the upper half-plane in the model space of Fig.~\ref{fig:ab_plot}. One could also generalize to models \eqref{eq:2d_SRG} where the dilaton potentials are not monomials in the dilaton or to the most general 2d dilaton gravity models \cite{Grumiller:2021cwg} to see how the shape of the dilaton potentials affects the echoing period. 
    \item \textbf{Semiclassical corrections.} Simple heuristics leads to the expectation that type II critical collapse gets converted into type I critical collapse due to semiclassical corrections, since there could be a mass gap of the order of the Planck mass (see, e.g., \cite{Gundlach:2007gc} and the recent work \cite{Tomasevic:2025clf,Tomasevic:2025kqy}). We can support this heuristics by an argument based on the generalized symmetry \eqref{eq:gen_sym}: Since this is only a symmetry of the classical equations of motion but not of the action, it seems obvious that it has to be broken by quantum effects. This means that there is no ``room'' for the echoing period $\Delta$ in the matching procedure we described as we can no longer fix one of the Fourier modes and trade it for $\Delta$, suggesting that the DSS symmetry is anomalous, i.e., broken by semiclassical effects. It could be useful to make this symmetry breaking mechanism precise by studying semiclassical corrections first within the 2d dilaton gravity approach, since it is under better analytic control, see, e.g., \cite{Callan:1992rs}.
    \item \textbf{Superstring perspective.} Since the Ricci scalar of CSCs diverges at the singularity, the classical gravity approximation must eventually break down, and string or loop corrections are expected to become essential at sufficiently late times. CSCs could thus provide a controlled setting to test ideas about singularity resolution in string theory. Because the relevant dynamics takes place near the singularity, a perturbative treatment is unlikely to suffice, suggesting that a non-perturbative framework such as AdS/CFT may be required. Pursuing this line would necessitate extending our analysis to include a negative cosmological constant, which we leave for future work.
\end{itemize}
The generalizations above are mostly based on models of the type \eqref{eq:2d_SRG}. A much broader generalization would be to consider more general models with other field content and/or different symmetries and to construct within them critical spacetime crystals in continuous dimensions.


\addcontentsline{toc}{section}{Acknowledgements}
\section*{Acknowledgements}

We thank Carsten Gundlach for numerous discussions and collaboration at an early stage of this project, as well as for providing his vintage Fortran code on the construction of Choptuik's spacetime crystal in four spacetime dimensions. We thank Roberto Emparan for discussions on the large-$D$ limit of general relativity. Additionally, we thank Peter Aichelburg, Craig Clark, Maciej Maliborski, Guilherme Pimentel, and Benson Way for discussions.

CE acknowledges support by the DFG through the CRC-TR 211 ``Strong-interaction matter under extreme conditions'' -- project number 315477589 -- TRR 211.
This work was supported by the Austrian Science Fund (FWF) [Grants DOI: \href{https://www.fwf.ac.at/en/research-radar/10.55776/P33789}{10.55776/P33789}, \href{https://www.fwf.ac.at/en/research-radar/10.55776/P36619}{10.55776/P36619}, \href{https://www.fwf.ac.at/en/research-radar/10.55776/PAT1871425}{10.55776/PAT1871425}].  

The presented numerical results were partially achieved at the Vienna Scientific Cluster (VSC), project ``Critical collapse in (sm)all dimensions'' (\#72844) on VSC-5 and the ITP Supercomputing Cluster iboga at Goethe University Frankfurt. Our data are available at \cite{CSCdata:2026}.
\texttt{ChatGPT} was used to assist with the design of plotting scripts and to provide enhanced spelling and grammar checking for the text.

\bigskip
\begin{center}
{\Large\EOpo}
\end{center}


\appendix


\section{Taylor expansions near center and SSH}\label{app:Taylor}

Around the center, $x=0$, we use a Taylor series for the fields
\begin{subequations}
    \label{eq:tayser}
\begin{align}
    f(\tau,x) &= \sum_{i=0}^5f_i(\tau)x^i + \mathcal{O}(x^6) & 
    \omega(\tau,x) &= \sum_{i=1}^5\omega_i(\tau)x^i+ \mathcal{O}(x^6) \\
    \psi_+(\tau,x) &= \sum_{i=0}^5\psi_{+i}(\tau)x^i + \mathcal{O}(x^6) & 
    \psi_-(\tau,x) &= \sum_{i=0}^5\psi_{-i}(\tau)x^i + \mathcal{O}(x^6) 
\end{align}
\end{subequations}
where some of the regularity conditions \eqref{eq:regularityorigin} are already implemented explicitly while others constrain the $\tau$-dependent coefficient functions.

Inserting the Taylor series \eqref{eq:tayser} into the equations of motion \eqref{eq:eom} and dropping higher order terms we can solve the system hierarchically in terms of two functions $\{f_0(\tau),\psi_{+1}(\tau )\}$. 

Up to $\mathcal{O}(x^3)$ the solutions read
\begin{subequations}
    \label{eq:TaylorSwift}
\begin{align}
\omega (\tau ,x)&=\frac{(\psi _{+1})^2}{D-1}\,x^2+\mathcal{O}(x^4)\\
f(\tau ,x)&=f_0+\frac{(D-3)f_0(\psi _{+1})^2}{2(D-1)}\,x^2+\mathcal{O}(x^4)\\
\psi _+(\tau ,x)&=\psi _{+1}\,x+\frac{\partial _\tau \psi _{+1}+\psi _{+1}}{f_0(1-D)}\,x^2 + \psi_{+3}\,x^3 +\mathcal{O}(x^4) \\
\psi _-(\tau ,x)&=\psi_{+1}x - \frac{\partial _\tau \psi _{+1}+\psi _{+1}}{f_0(1-D)}x^2 + \psi_{+3}\,x^3 +\mathcal{O}(x^4)  
\end{align}
with 
\eq{
\psi_{+3}=\frac{f_0(\partial _\tau ^2 \psi _{+1}+3\partial _\tau \psi _{+1}+2\psi _{+1})-\partial _\tau f_0(\partial _\tau \psi _{+1}+\psi _{+1})-(D-3)f_0^3(\psi _{+1})^3}{2(D-1)f_0^3}\,.
}{eq:apptaylor1}
\end{subequations}
The configuration above obeys all the regularity constraints \eqref{eq:regularityorigin} and is fully determined by specifying $f_0(\tau)$ and $\psi_{+1}(\tau)$. This remains true perturbatively to all orders in $x^n$.

As mentioned in the main text, we parametrize the ``initial'' data at $x=0$ in terms of $\{f_c(\tau),\Psi_c(\tau)\}$. The former is just given by $f_c=f_0$ while according to \eqref{eq:inidata_center} the latter is related to $\psi_{+1}$ by
\eq{
   \Psi_c=\frac{\partial_\tau \psi_{+1}+\psi_{+1}}{f_0(D-1)} ~.
}{eq:psicenter}
Thus, given $\Psi_c$ and $f_c$, we can determine $\psi_{+1}$ by solving the above ODE in $\tau$. This can be done uniquely for periodic boundary conditions, see Eqs. \eqref{eq:typical_ODE}, \eqref{eq:sol_ode}. 

Around the SSH at $x=1$ we also use Taylor expansions,
\begin{align}\label{eq:expansion_SSH1}
    f(\tau ,x)&=1+\sum _{i=1}^2\Bar{f}_i(\tau )(x-1)^i & 
    \omega (\tau ,x)&=\sum _{i=0}^2\Bar{\omega } _i(\tau )(x-1)^i\\\label{eq:expansion_SSH2}
    \psi _+(\tau ,x)&=\sum _{i=0}^2\Bar{\psi} _{+i}(\tau )(x-1)^i & 
    \psi _-(\tau ,x)&=\sum _{i=0}^2\Bar{\psi }_{-i}(\tau )(x-1)^i
\end{align}
where we already built in the gauge condition $f(\tau ,1)=1$ and neglected terms $\mathcal{O}(1-x)^3$. Expanding again the equations of motion \eqref{eq:eom} we find two ODEs at LO, 
\begin{align}
\partial _\tau \Bar{\omega }_0-(D-3)(e^{\Bar{\omega }_0}-1)+\Bar{\psi }_{-0}^2&=0\label{eq:LO_constr}\\
    \partial _\tau \Bar{\psi }_{+0}+\frac{D-2-2(D-3)e^{\Bar{\omega }_0}}{2}\Bar{\psi }_{+0}+\frac{D-2}{2}\Bar{\psi }_{-0}&=0
\end{align}
which, being again of the form \eqref{eq:typical_ODE}, can be solved exactly once $\Bar{\psi}_{-0}$ is given. Then, starting with NLO, we solve the system algebraically for $\{\Bar{f}_i,\Bar{\omega }_i,\Bar{\psi }_{-i}\}$ and subsequently solve an ODE coming from \eqref{eq:eom4} for $\Bar{\psi}_{+i}$. This pattern can be repeated up to the desired order. 

At $\mathcal{O}(x-1)$ we get the algebraic solutions
\begin{subequations}
\label{eq:TaylorSSH}
\begin{align}
    \Bar{f}_1&=(D-3)(e^{\Bar{\omega }_0}-1)\\
    \Bar{\omega }_1&=\frac{1}{2}\Big(\Bar{\psi }_{+0}^2+\Bar{\psi }_{-0}^2-2(D-3)(e^{\Bar{\omega }_0}-1)\Big)\\
    \Bar{\psi }_{-1}&=\frac{1}{4}\Big((D-2+2(3-D)e^{\Bar{\omega }_0})\Bar{\psi }_{-0}+(D-2)\Bar{\psi }_{+0}-2\partial _\tau \Bar{\psi }_{-0}\Big)
\end{align}
as well as the ODE for $\Bar{\psi }_{+1}$,
\begin{multline}
    2\partial _\tau \Bar{\psi }_{+1}+\big(D-2\Bar{f}_1+2(3-D)e^{\Bar{\omega }_0}\big)\Bar{\psi }_{+1}\\
     +(D-2)\Bar{\psi }_{-1}-2(D-3)e^{\Bar{\omega }_0}\Bar{\psi }_{+0}\Bar{\omega }_1-2\partial _\tau \Bar{\psi }_{+0}(\Bar{f}_1-1) =0 
\end{multline}
\end{subequations}
that again can be integrated uniquely upon imposing periodicity. 

The free data for the expansion near the SSH are the Fourier modes of the function $\Bar{\psi}_{-0}$ which we store as $\psi_{-p}(\tau)$ by the identification \eqref{eq:up_def}. 


\section{Singular branch at SSH}\label{app:SingularBranch}

In our algorithm, the fields around the SSH can acquire contributions from a singular solution branch. This happens because of the finite order at which the Taylor series is truncated. Then, at the cutoff surface we have (schematically)
\begin{align}
    Z_{\textrm{\tiny Tayl}}(\tau ,\xr ,\mathrm{ord})=Z_{\textrm{\tiny reg}}(\tau ,\xr)+Z_{\textrm{\tiny sing}}(\tau ,\xr)+\mathcal{O}(\xr-1)^{\textrm{\tiny ord}+1} 
\end{align}
where the size of the singular contribution depends on the dimension $D$, the value of $\xr $ and the order at which the Taylor series is truncated. It is suppressed by construction,
\begin{align}
    Z_{\textrm{\tiny sing}}(\tau ,\xr) \sim \mathcal{O}(\xr-1)^{\textrm{\tiny ord}+1}
\end{align}
but this suppression might not be enough to ensure stability for the shooting algorithm. Depending on $D$ we might have to choose a different order of the Taylor expansion to ensure that this contribution stays small. 

The numerical data suggest that the dominant contribution of the singular solution is in the field $\psi_+$. Around $x=1$ we therefore choose a generalized Taylor ansatz
\begin{align}\label{eq:psip_ansatz}
    \psi_+(\tau,x)=\psi_{+0}(\tau)+(x-1)^\delta \psi_{+\delta}(\tau,x)+o(x-1)^\delta 
\end{align}
where $\psi_{+\delta}$ can possibly depend on $x$ logarithmically. Consistency with the gauge fixed equations of motion \eqref{eq:eom} implies
\begin{align}
    \psi _-(\tau ,x) &=\psi _{-0}(\tau )+(x-1)\psi _{-1}(\tau )+(x-1)^{1+\delta }\psi _{-\delta }(\tau ,x)+o(x-1)^{1+\delta } \\
    f(\tau ,x)&=1+(x-1)f_1(\tau )+(x-1)^{1+\delta} f_\delta (\tau ,x)+o(x-1)^{1+\delta }\\
    \omega (\tau ,x)&=\omega _0(\tau )+(x-1)\omega _1(\tau )+(x-1)^{1+\delta} \omega _\delta (\tau ,x)+o(x-1)^{1+\delta }
\end{align}
i.e., the singular terms appear at one order lower than in $\psi _+$. We leave out the overbars for the coefficients to avoid clutter. 

We determine the behavior of $\delta $ by solving the equations of motion in the vicinity of the SSH. Starting from the equation
\begin{align}\label{eq:psip_eq}
     (\partial _\tau -(f-x)\partial _x) \psi _+&=\frac{f}{2x}\big(2-D+2(D-3)e^\omega \big)\psi _++\frac{f}{2x}(2-D)\psi _- 
\end{align}
we know from the ansatz we chose that the coefficients of the differential operator on the left have to be at least once differentiable around the SSH, i.e.
\begin{align}
    f-x=:F\sim (x-1)F_1(\tau )+o(x-1) 
\end{align}
and we can extract $F_1$ from the equation of motion for $f$,
\begin{align}
    x\partial _xf&=(D-3)(e^\omega -1)f
\end{align}
which implies to LO
\begin{align}
    F_1=(D-3)(e^{\omega _0}-1) -1~.
\end{align}
The function $\omega _0$ in turn is determined by the LO constraint \eqref{eq:LO_constr} in terms of the free data $\psi _{-0}$ of the regular solution
\begin{align}\label{eq:constr_LO}
    \partial_\tau \omega_0=-\psi_{-0}^2+(D-3)(e^{\omega _0}-1) ~.
\end{align}

We solve \eqref{eq:psip_eq} by the method of characteristics. The characteristic equations read
\begin{align}
    \frac{\extd \tau}{\extd s}&=1 & \frac{\extd x}{\extd s}=-F
\end{align}
and we see that we can choose $\tau $ as a parameter along the characteristics. Then, to LO in $(x-1)$ we have from the right equation
\eq{
    \frac{\extd x}{\extd \tau}=-(x-1)F_1
}{eq:appB1}
which is solved by
\eq{
    \log (|x(\tau )-1|)=-\int\limits_{\tau_0}^\tau F_1+\log(|x_0-1|)
}{eq:appB2}
where $x_0=x(\tau _0)$ can be used to parametrize the set of characteristic curves. Explicitly, we get
\eq{
    x(\tau )=1-(1-x_0)e^{-\int\limits_{\tau_0}^\tau F_1}
}{eq:appB3}
showing that they only have a non-trivial $\tau $-dependence once we choose $x_0\neq 1$. Along these curves we can now solve \eqref{eq:psip_eq} which we first expand in the ansatz \eqref{eq:psip_ansatz}. At $\mathcal{O}(x-1)^0$ we just find the regularity condition [see also Eq.~\eqref{eq:SSH}]
\eq{
    \frac{\extd \psi _{+0}}{\extd \tau }=\frac{1}{2}\big[2-D+2(D-3)e^{\omega _0}\big]\psi _{+0}+\frac{2-D}{2}\psi _{-0}
}{eq:appB4}
but at the next order, $\mathcal{O}(x-1)^\delta$, we get
\eq{
    \frac{\extd \;(\log \psi _{+\delta })}{\extd \tau }=\delta F_1+\frac{1}{2}\big[2-D+2(D-3)e^{\omega _0}\big] ~.
}{eq:appB5}

Since we demand $\psi_{+\delta}(\tau,x)$ to be a periodic function in $\tau $ we can first look at the zero mode of this equation which gives
\eq{
    0=\frac{1}{\Delta}\int\limits_0^\Delta \Big(\delta F_1+\frac{1}{2}\big[2-D+2(D-3)e^{\omega _0}\big]\Big) ~.
}{eq:appB6}
After some lines of simplifying and using the zero mode of the LO constraint \eqref{eq:constr_LO} to express $\int _0^\Delta e^{\omega _0}$ in terms of $\int _0^\Delta \psi _{-0}^2$ we find
\eq{
    \delta =\frac{2\overline{\psi_{-0}^2}+D-4}{2-2\overline{\psi_{-0}^2}} \,.
}{eq:appB7}

With this value of $\delta $ the solution for $\psi _{+\delta }$ can be written as
\eq{
    \log \psi _{+\delta }=\int\limits_{\tau_0}^\tau \Big((\delta +\frac{1}{2})(2-D)+(\delta +1)e^{\omega _0}(D-3)\Big)+\log \psi _{+\delta }(\tau _0,x_0) ~.
}{eq:appB8}
Without loss of generality we choose $\tau _0=0$ such that the integral is a function of $\tau$ satisfying periodic boundary conditions in $\tau\in[0,\Delta]$. This is so, because the zero mode of the integrand vanishes by the choice of $\delta$. In the vicinity of $x=1$ we express this function over the set of characteristic curves with $x_0=1-(1-x)\exp (\int_0^\tau F_1)$ as
\eq{
    \psi _{+\delta }(\tau ,x)=e^{g(\tau )}h\Big(\log(1-x)+\int\limits_0^\tau F_1\Big)
}{eq:appB9}
where 
\eq{
    g(\tau )=\int\limits_{\tau_0}^\tau \Big((\delta +\frac{1}{2})(2-D)+(\delta +1)e^{\omega _0}(D-3)\Big)
}{eq:appB10}
and 
\eq{
    h\Big(\log(1-x)+\int\limits_0^\tau F_1\Big)=h\Big(\log(1-x)+\tau \Bar{F}_1+\int\limits_0^\tau \Tilde{F}_1\Big)
}{eq:appB11}
has to be some function satisfying $h(y+\Delta \Bar{F}_1)=h(y)$. Therefore, this singular solution branch has one periodic function worth of free data. Schematically, it is of the form
\eq{
    \psi _{+\delta }=e^{g(\tau)}\Tilde{\psi }(\tau +\alpha \log(1-x))
}{eq:appB12}
with $\Tilde{\psi}$ a periodic function. 

\begin{figure}[htb]
    \centering
\includegraphics[width=0.9\linewidth]{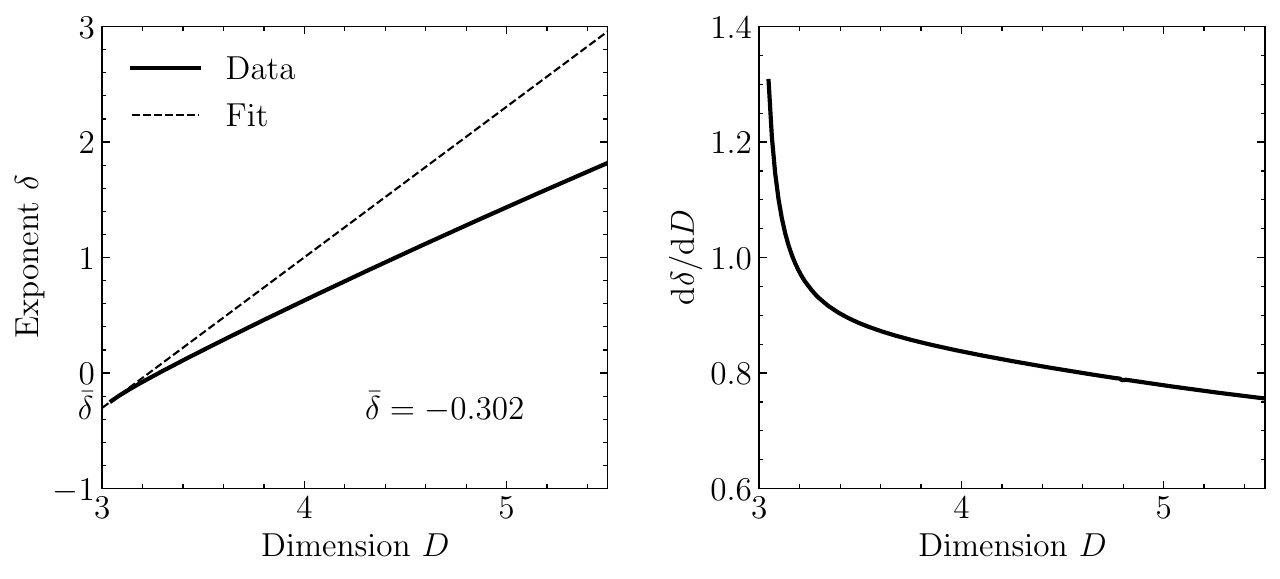}
    \caption[Exponent of singular solution branch]{\textbf{Left:} Value of $\delta $ for singular solution at SSH. \textbf{Right:} First derivative. Assuming a monotonous first derivative towards $D=3^+$ an upper bound at $D=3$ is $\bar{\delta } = -0.302$, consistent with the exact result $\delta=-\frac12$ in Section \ref{sec:6.2}.}
    \label{fig:deltaplot}
\end{figure}

The corresponding function $\delta (D)$ and its first derivative are plotted in Fig.~\ref{fig:deltaplot}. 
There are two points to be made here: First, it is evident that towards larger dimensions $\delta$ increases. Therefore, in an expansion close to the SSH the singular solution appears at increasingly higher orders in $(1-x)$. To still ensure convergence towards the critical solution, this singular contribution must be suppressed which in turn requires more and more Taylor orders at $\xr$ the higher $D$ gets. In the present case the Taylor series is truncated at $\mathcal{O}\big((1-x)^3\big)$ and we find that the Newton algorithm for minimizing the mismatch becomes unstable for $D\gtrsim 5.5$. Progressing to even larger dimensions would require more Taylor orders which becomes increasingly inefficient to implement. Second, as may be deduced from the monotonicity of $\dd \delta /\dd D$, the value of $\delta$ decreases further towards $D=3$ and one may find an upper bound by extrapolating the tangent at the lowest values for $D$ to $D=3$. (The dashed line labeled ``Fit''.) This gives $\bar{\delta}\approx -0.302$ and thus a divergent term in the expansion of $\psi_+$ around the SSH.\footnote{%
The value $\bar{\delta}\approx -0.302$ was deduced from the small-$D$ data near our lower value $D=3.05$. It is conceivable that the curve may get even steeper for smaller values of $D$ and it is possible that the correct value is $\bar{\delta}=-0.5$ as suggested by the exact solution \eqref{eq:smallD37}. 
} 
This behavior may be compared with the the solution \eqref{eq:smallD37} found in the limit $D\to 3$ suggesting that that solution corresponds to this singular branch. 


\section{Error budget and convergence}\label{app:Error}

In this Appendix, we present numerical tests and consistency checks of our results, with a focus on the numeric computation of CSCs. There are four main error sources entering the numerical code: The discretization errors in the $x$ and $\tau$ directions as well as the errors introduced by truncating the Taylor series around $x=0$ and $x=1$. While these errors are not completely independent we can check the convergence order with respect to their parameters. In order to do so, we assume that the numerical solution $Z(h)$ expands in a power series around the true solution, 
\eq{
    Z(h)=Z_{\textrm{\tiny true}}+Z_1h^{p+1}+\mathcal{O}(h^{p+2})
}{eq:Richardson_expansion}
where $p$ is the consistency order of the numerical procedure and $h$ stands for one of the small parameters we have in the problem. The error is then approximated by
\eq{
\text{err}(h)\approx Z_1h^{p+1}=\frac{Z(2h)-Z(h)}{2^{p+1}-1}+\mathcal{O}(h^{p+2})
}{eq:err_approx}
where $Z(2h)$ is the solution computed for the doubled value of the parameter. Moreover, we can crosscheck the value of $p$ by taking the difference with the error computed for twice the value of $h$, rescaled by the expected power of $2^{p+1}$, i.e.,
\eq{
    \text{err}(h)-\frac{1}{2^{p+1}}\text{err}(2h)\overset{!}{=}\mathcal{O}(h^{p+2}) ~.
}{eq:err_diff}
If the approximation by truncating \eqref{eq:Richardson_expansion} at (including) $\mathcal{O}(h^{p+1})$ is good, the right-hand side of \eqref{eq:err_diff} should be close to zero. 
In the following we focus on the two functions on the left-hand side of \eqref{eq:err_diff} and check how well they agree.

\subsection{Spatial grid-size convergence}

\begin{figure}[htb]
    \centering
    \includegraphics[width=\textwidth]{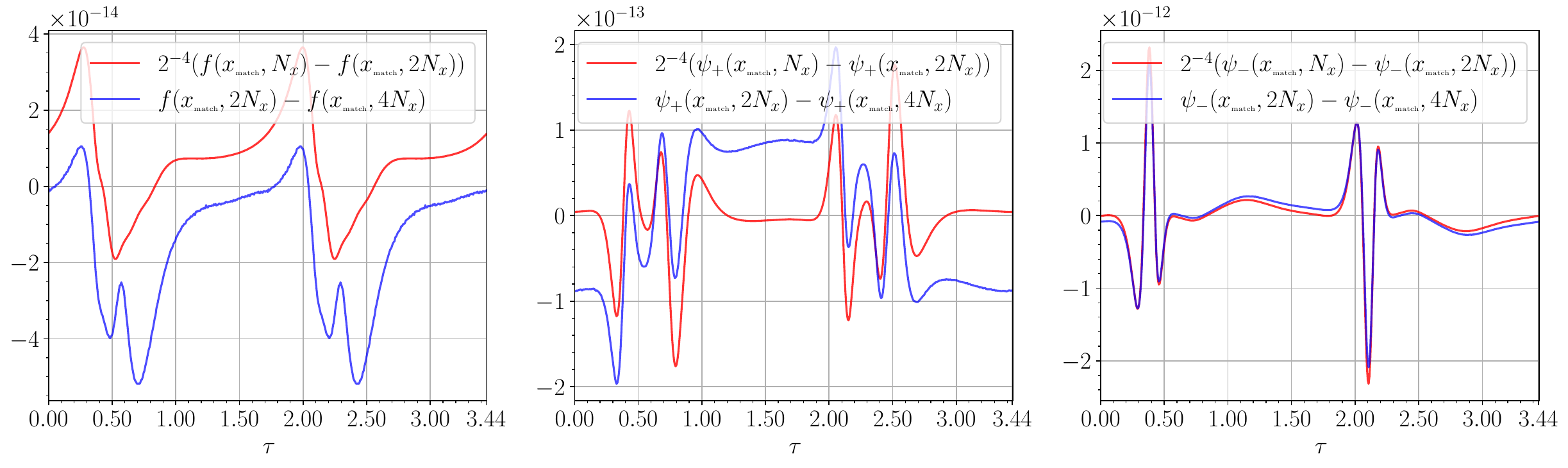}
    \includegraphics[width=\textwidth]{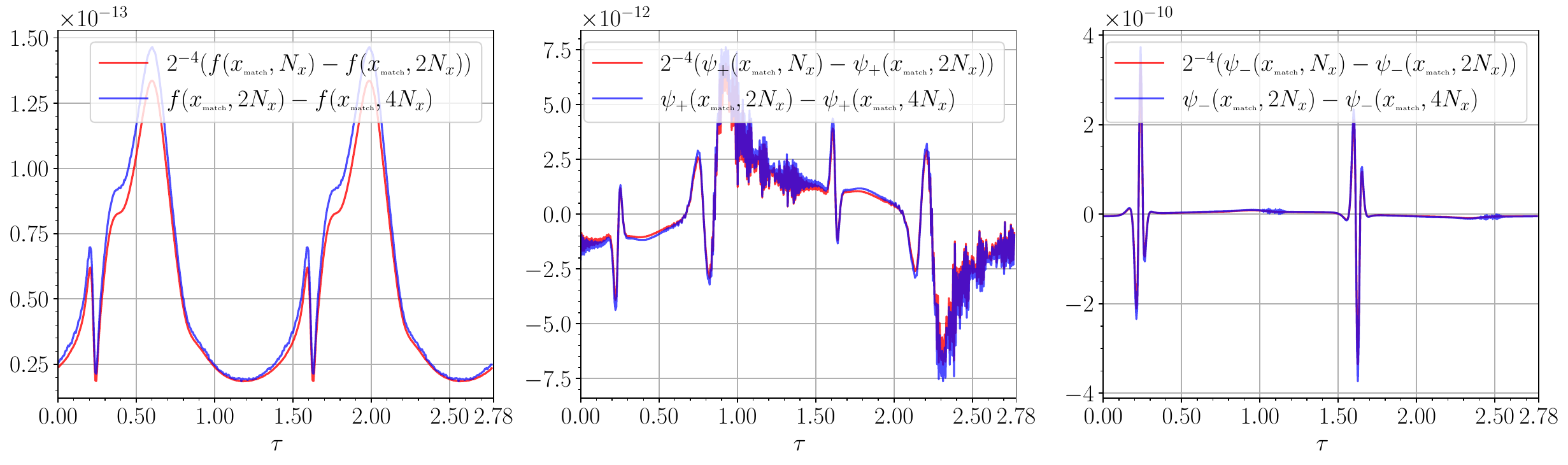}
    \caption[Global errors of evolution in spatial direction $x$]{Red: Global errors of $x$-evolution from $\xl $ to $x_{\textrm{\tiny match}}$ at $D=4$ (top) and $D=3.1$ (bottom).
    Blue: Errors for doubled number of gridpoints rescaled by $2^{-4}$. Curves agree reasonably, confirming convergence of $4^{\textrm{\tiny th}}$-order Runge--Kutta.}
    \label{fig:grid_convergence_L}
\end{figure}

As a first check we asses the convergence with respect to the discretization in $x$ and estimate how big the error is. Computing the differences of the functions in the shooting $\{\psi _+,\psi _-,f\}$ at the matching surface for three consecutively doubled numbers of gridpoints in $x$ leads to Fig.~\ref{fig:grid_convergence_L}, where we show results for $D=4$ and $D=3.1$ in the top and bottom row, respectively. The numbers of gridpoints chosen were $N_x=2000,\,4000,\,8000$ which is a range where the $x$-discretization error dominates the others. Besides the expected convergence for a fourth order Runge--Kutta method ($p=3$) one can see that the global error coming from the evolution is around $10^{-12}$ in the case of $\psi_-$ and even smaller for the other functions. Since this is the error bound for the lowest of the three numbers of gridpoints, we can expect that at $N_x=8000$, the discretization error will be subdominant.  Compared with the $D=4$ case, the errors in $D=3.1$ are slightly higher here which is due to the steep slopes developing at lower $D$. The high-frequency noise in the middle plot is not a problem in this case since the curves still lie on top of each other. It is, however, related to the Fourier approximation getting worse, see for example the slow convergence of the Fourier series of $\Psi _c$ in Fig.~\ref{fig:Nt_conv}.

\subsection{Cutoff dependence}

Once the shooting method has converged for a given dimension, the evolution data are determined in terms of the functions $\{f_c,\Psi_c, \psi_{-p}\}$. However, depending on the values of $\xl $ and $\xr$ chosen, they are still a more or less accurate approximation of the true critical solution. Since the Taylor series around $x=0$ is truncated at (including) $\mathcal{O}(x^5)$ we expect the converged free data to behave as
\eq{
    f_c(\xl ) = f_c^{\textrm{\tiny true}}+\delta f_c\xl ^6+\mathcal{O}(\xl ^7)
}{eq:appC1}
where the $\tau$-dependence in all functions is suppressed. This holds provided the cutoff at $\xl$ is the dominant source of error. Similar relations hold for the other functions. We can therefore again approximate the error like in \eqref{eq:err_approx} and plot the consecutive differences of the three functions if $\xl$ is doubled one or two times. The same steps are performed for $\xr$. 

\begin{figure}
    \centering
    \includegraphics[width=0.9\textwidth]{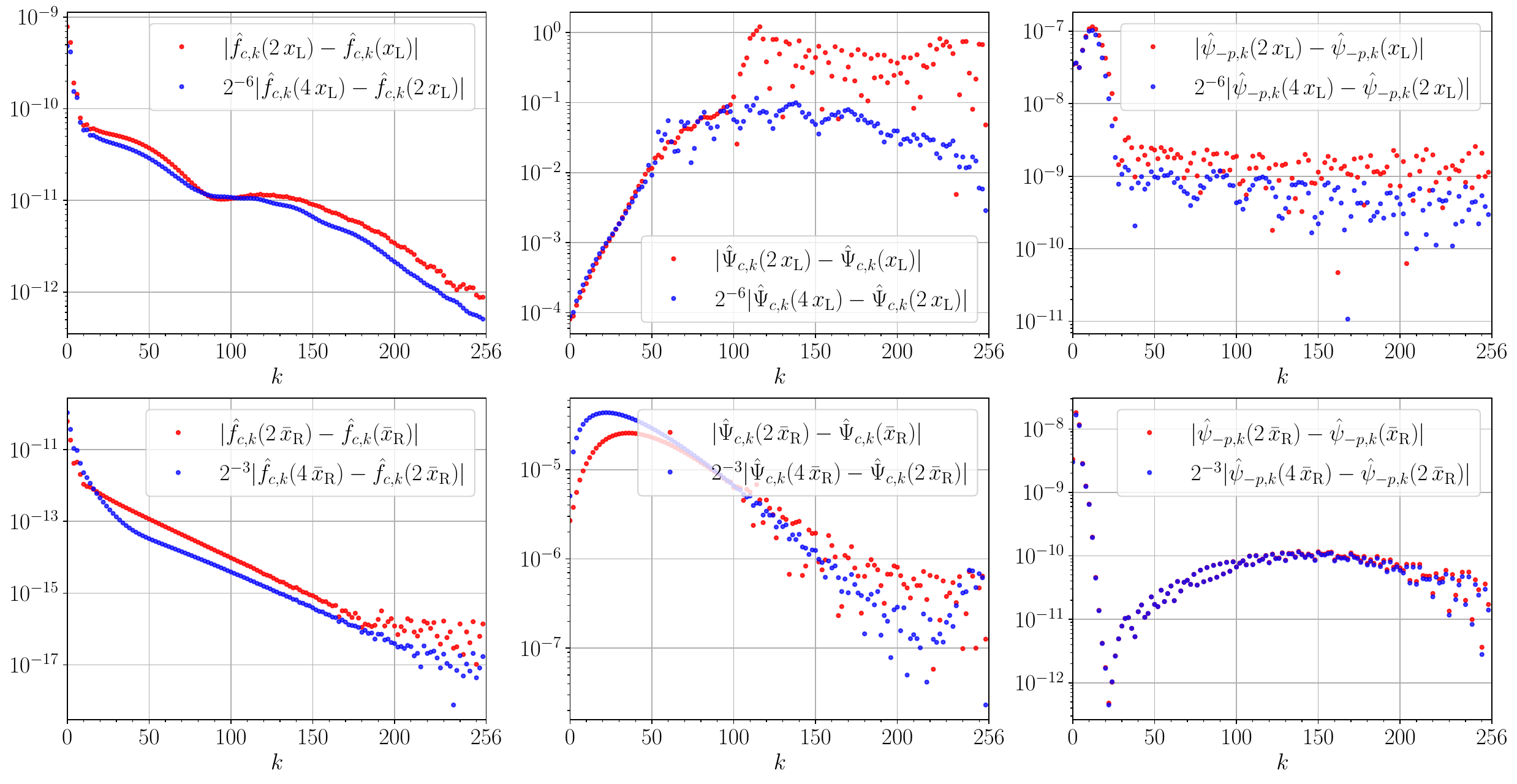}
    \includegraphics[width=0.9\textwidth]{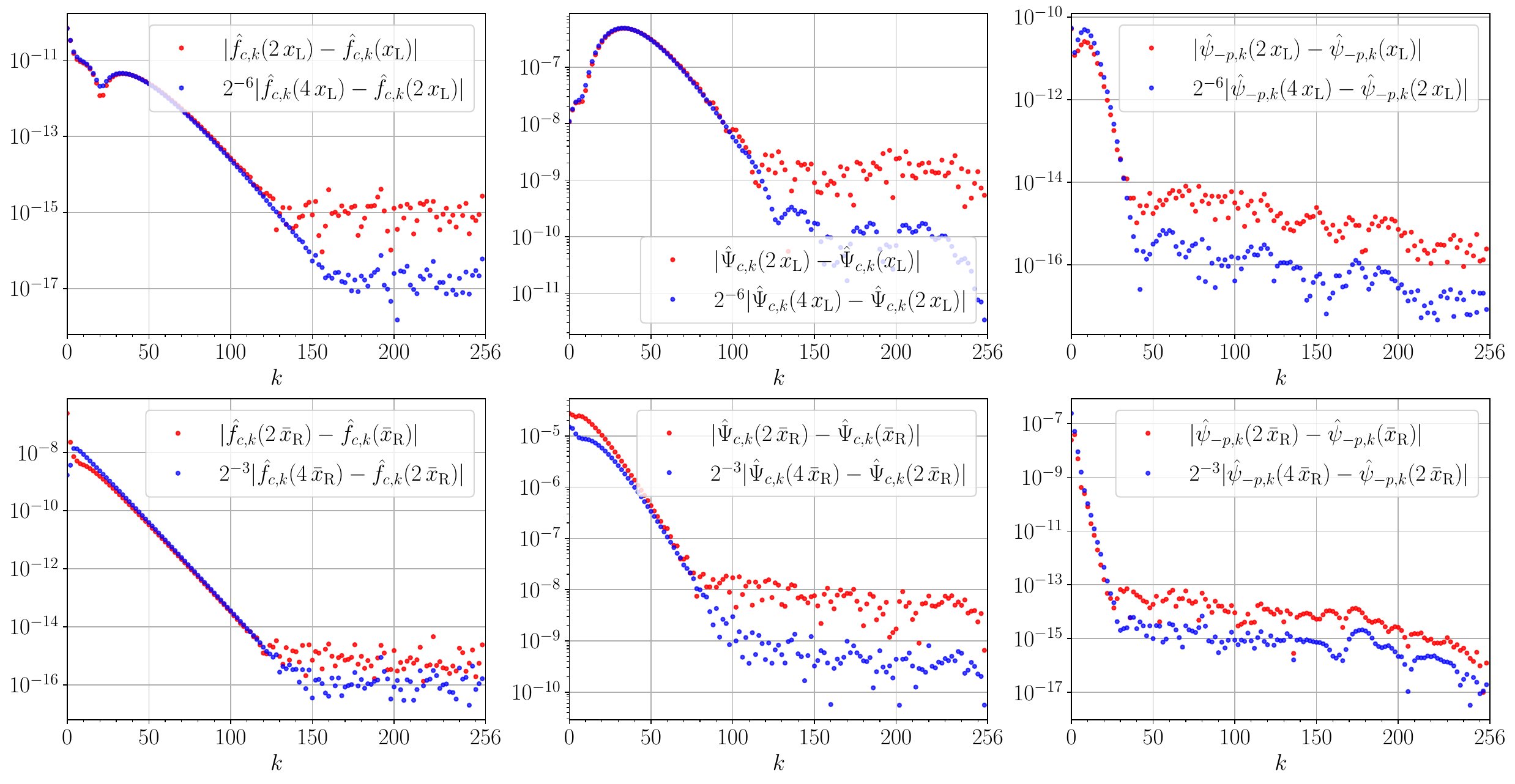}
    \includegraphics[width=0.9\textwidth]{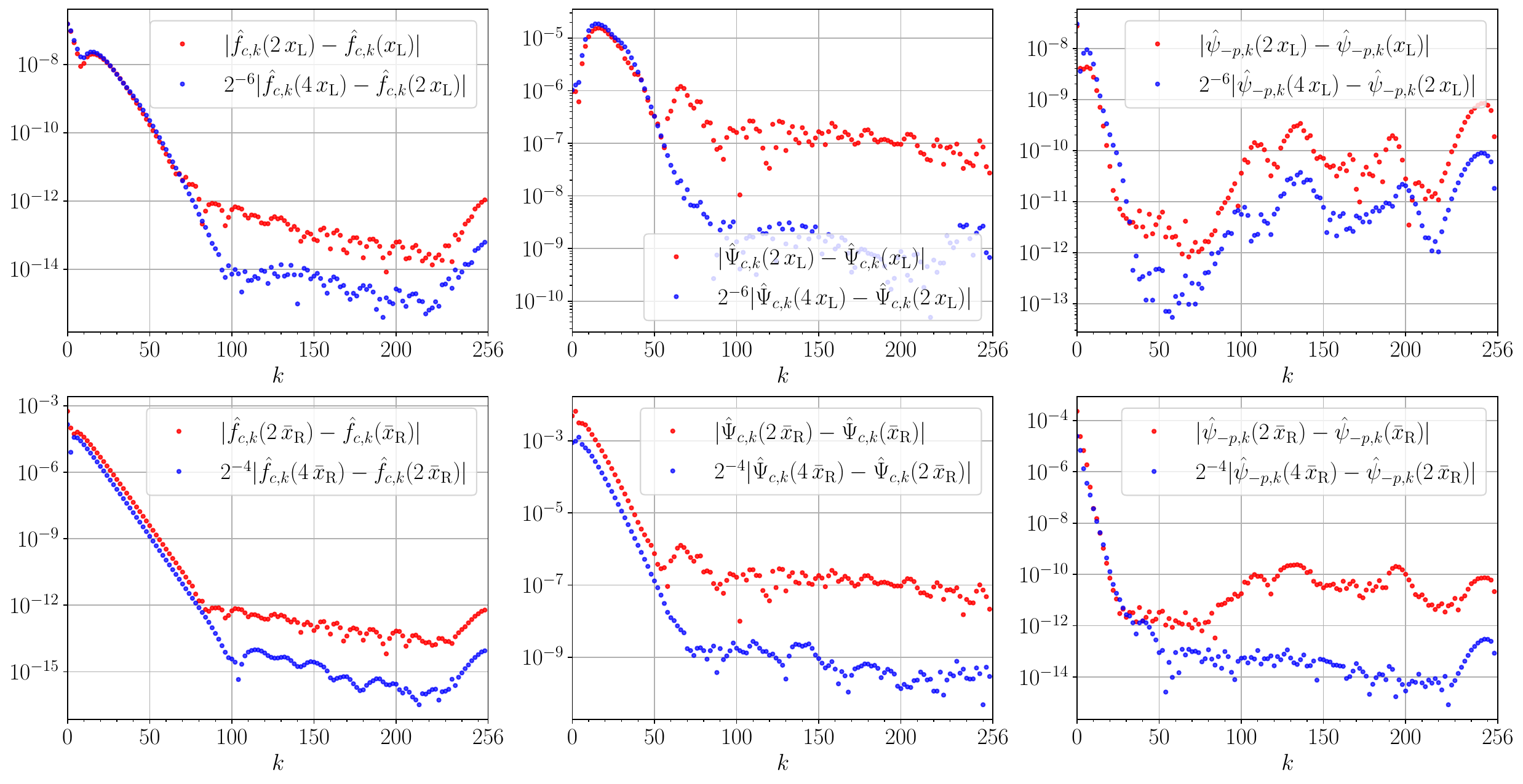}
    \caption[Convergence with respect to cutoff variations]{Convergence of data with respect to varying $\xl$ and $\xr$ at $D=3.1$ (first two rows), $D=4$ (second two rows) and $D=5.5$ (last two rows). First (second) row for each $D$ shows differences of power spectra if $\xl$ ($\xr$) is varied.}
    \label{fig:cut_conv_3.1D}
\end{figure}

Figure \ref{fig:cut_conv_3.1D} shows the corresponding power spectra at $D=3.1$, $D=4$ and $D=5.5$, respectively. 
The main point to observe in these plots is that in lower dimensions the numerical procedure is more error-prone which is most of all seen in the function $\Psi _c$. Not only is its error increasing, but its high frequency part does not converge any more. This is the main reason why we cannot get closer to $D=3$ with the current code. It should be pointed out that the large absolute error of $\Psi _c$ in Fig.~\ref{fig:cut_conv_3.1D} of order $10^{-1}$ is relatively seen still in the same region as it is for $D=4$ since the function $\Psi _c$ itself takes on values $\sim 10^5$ for $D=3.1$ while it is of the order $10^1$ for $D=4$. 
For $D=5.5$ one can also observe that the convergence gets worse even for the low frequency part in the spectrum. However, unlike for $D$ close to $3$, the functions themselves do not develop any steep gradients or sharp edges. Here, the decrease in convergence has its origin in the Taylor order at the SSH not being sufficient to resolve the singular solution any more, as discussed at the end of Appendix \ref{app:SingularBranch}. 

\subsection{Fourier approximation}\label{app:fourier}

A last source of error is the Fourier approximation of the various functions. Here we shall only focus on the converged initial data $\{f_c,\Psi_c,\psi_{-p}\}$ and show again how the error varies between the lowest and the highest value of $D$ we computed.
\begin{figure}
    \centering
    \includegraphics[width=\textwidth]{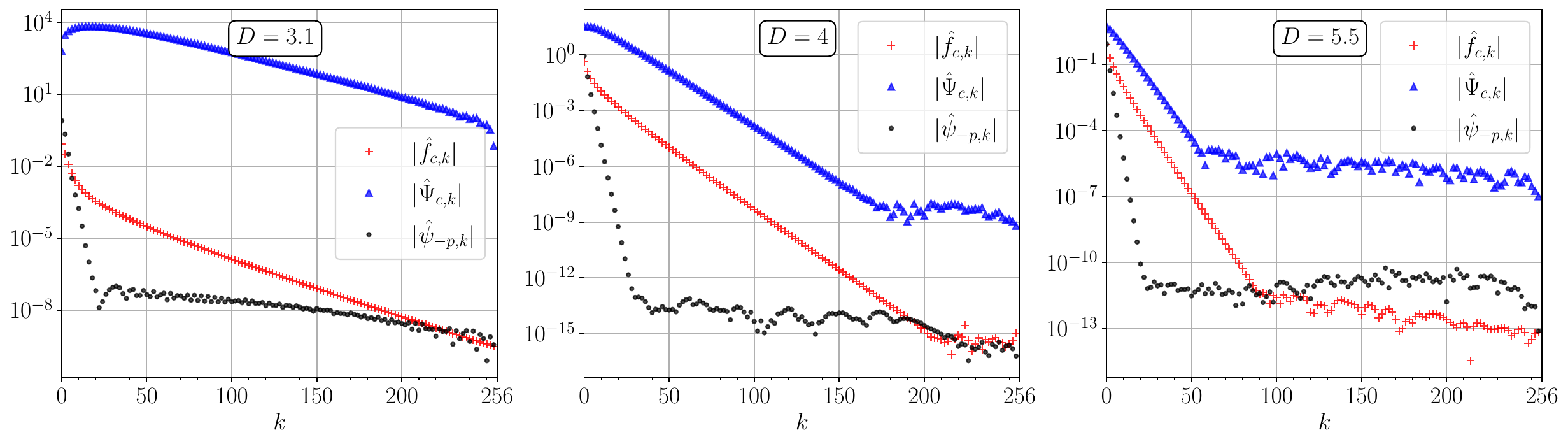}
    \caption[Power spectra of converged initial data]{Power spectra of converged initial data. Modes $\hat{\Psi }_{c,k}$ and $\hat{f}_{c,k}$ given in center ($x=0$) and $\hat{\psi }_{-p,k}$ at SSH ($x=1$).}
    \label{fig:Nt_conv}
\end{figure}
As can be seen in Fig.~\ref{fig:Nt_conv}, all the functions converge exponentially in the Fourier mode number $k$ until a noise plateau is reached. The height of the latter is ultimately a consequence of the finite precision the computation is done at, while the exponential convergence is a consequence of the functions being smooth. At the SSH, the modes $\hat{\psi}_{-p,k}$ always converge fast and one could thus use only the first few modes to approximate that function well. This is different for the modes in the center, especially for the function $\Psi_c$ and particularly at low dimensions.

At $D=3.1$, the peaky structure of that function makes the Fourier series converge rather slowly and the noise plateau is actually not even reached for $N_\tau =512$. The behavior of the blue power spectrum in Fig.~\ref{fig:Nt_conv} suggests that the limiting function $\lim_{D\to3^+}\Psi_c$ might not be smooth but rather a distribution, as the power spectrum no longer could exhibit exponential decay in $k$ in that limit but rather remain horizontal. If this indeed happens, the limiting distribution would be of $\delta$-type.

For $D=5.5$, the noise plateaus are quite high which is presumably again due to the loss of control over the singular solution for that range of $D$. This affects the stability of the minimum the Newton algorithm should find by introducing additional noise.

The errors in the initial data functions are summarized in Table \ref{tab:error_budget}. 
\setlength{\tabcolsep}{8pt} 
\renewcommand{\arraystretch}{1.5}
\begin{table}
    \let\mc\multicolumn
    \centering
    \begin{tabular}{c||c|c|c|c|c|c|c|c|c}
    & \multicolumn{3}{c|}{$f_c$} & \multicolumn{3}{c|}{$\Psi _c$} & \multicolumn{3}{c}{$\psi _{-p}$}\\
    \hline
    $D$ & $3.1$ & $4.0$ & $5.5$ & $3.1$ & $4.0$ & $5.5$ & $3.1$ & $4.0$ & $5.5$ \\
    \hline 
    \hline
         $\xl$ & $10^{-9}$ & $10^{-10}$ & $10^{-8}$ & $10^{-6}$ & $10^{-9}$ & $10^{-5}$ & $10^{-7}$ & $10^{-10}$ & $10^{-8}$   \\
         \hline 
         $\xr $ & $10^{-10}$ & $10^{-9}$ & $10^{-4}$ & $10^{-10}$ & $10^{-7}$ & $10^{-3}$ & $10^{-8}$ & $10^{-7}$ & $10^{-4}$  \\
         \hline
         $N_{\tau } $ & $10^{-9}$ & $10^{-15}$ & $10^{-11}$ & $10^{-6}$ & $10^{-10}$ & $10^{-5}$ & $10^{-8}$ & $10^{-13}$ & $10^{-10}$
    \end{tabular}
    \caption[Error estimates for initial data functions]{Error estimates for initial data functions. Rows labeled by error source.}
    \label{tab:error_budget}
\end{table}
In each case, we show relative errors, i.e., the error as determined by \eqref{eq:err_approx} divided by the absolute value of the function. At large $D$, one observes a clear dominance by the error associated with the truncation in $\xr $, which, as discussed above, has its origin in the contribution of the singular solution branch. At low $(D-3)$, on the other hand, the errors introduced by $\xl$ and the truncated Fourier series dominate. They are always of approximately the same size. The error coming from the discretization in $x$ is always negligible (around machine precision) and thus not included in Table \ref{tab:error_budget}. 


\section{Tabulated numerical results for \texorpdfstring{$\boldsymbol{\Delta}$}{Delta} and \texorpdfstring{$\boldsymbol{\gamma}$}{gamma}}\label{app:numbers}

The data in Table \ref{tab:DeltaGamma_2col_gray} below can be reconstructed using \cite{CSCdata:2026}. The values for the echoing period $\Delta$ were obtained from the C{}\verb!++! code described in Section \ref{sec:4.1} and the values for the Choptuik exponent $\gamma$ that are not in parentheses from the linearized code described in Section \ref{sec:chop_num}. The $\gamma$-values in parentheses are obtained by second-order interpolation for dimensions $3.1<D<5.5$ and by the best-fit extrapolation $\gamma\approx0.41\,(D-3)^{0.31}$ for $D<3.1$. To indicate that the extrapolated values become increasingly inaccurate we have successively reduced the number of digits in $\gamma$ that we display as we approach the lowest dimension, $D=3.05$. We have highlighted in bold the results in dimensions $D=3.05$, $D=3.1$, $D=3.2$, $D=3.5$, $D=D_{\textrm{\tiny crit}}$, $D=4$, $D=5$ and $D=5.5$.

{\renewcommand{\arraystretch}{1.02}
\begin{longtable}{llllll}
\caption[Tabulated data for $\Delta$ and $\gamma$]{
Tabulated data for $\Delta$ and $\gamma$, rounded to six digits after decimal point. 
}
\label{tab:DeltaGamma_2col_gray}\\

{$D$} & {$\Delta$} & {$\gamma$} & {$D$} & {$\Delta$} & {$\gamma$}\\
\cmidrule(lr){1-3}\cmidrule(lr){4-6}
\endfirsthead

{$D$} & {$\Delta$} & {$\gamma$} & {$D$} & {$\Delta$} & {$\gamma$}\\
\cmidrule(lr){1-3}\cmidrule(lr){4-6}
\endhead

\multicolumn{6}{r}{\emph{Continued on next page}}\\
\endfoot

\bottomrule
\endlastfoot

\textbf{3.05}   & \textbf{\num{2.4849271642909727}} & 
\textbf{(0.162)} &
3.0505 & \num{2.4886390007075363} & 
(0.163)\\

3.051  & \num{2.4923278133510474} & 
(0.163)&
3.0515 & \num{2.495993888885805}  & 
(0.164)\\

3.052  & \num{2.499637507897485}  & 
(0.164) &
3.0525 & \num{2.50325894523142}   & 
(0.165)\\

3.053  & \num{2.506858469975515}  & 
(0.165) &
3.054  & \num{2.513992851879058}  & 
(0.166) \\

3.055  & \num{2.5210426548910023} & 
(0.167) &
3.056  & \num{2.5280098672450704} & 
(0.168) \\

3.057  & \num{2.5348963773563375} & 
(0.169) &
3.058  & \num{2.5417040014463237} & 
(0.170) \\

3.059  & \num{2.548434487426444}  & 
(0.171) &
3.06   & \num{2.555089518474046}  & 
(0.172) \\

3.061  & \num{2.5616707164623898} & 
(0.1724) &
3.062  & \num{2.568179645020892}  & 
(0.1733) \\

3.063  & \num{2.5746178125033437} & 
(0.1741) &
3.064  & \num{2.5809866747359136} & 
(0.1750)\\

3.065  & \num{2.5872876375952587} & 
(0.1758) &
3.066  & \num{2.5935220594257897} & 
(0.1766) \\

3.067  & \num{2.5996912532181327} & 
(0.1774) &
3.068  & \num{2.605796488917223}  & 
(0.1782) \\

3.069  & \num{2.6118389952372785} & 
(0.1790) &
3.07   & \num{2.617819961522354}  & 
(0.1798) \\

3.071  & \num{2.623740540066316}  & 
(0.1806) &
3.072  & \num{2.629601846626925}  & 
(0.1814) \\

3.073  & \num{2.635404963165396}  & 
(0.1821) &
3.074  & \num{2.641150938744624}  & 
(0.1829) \\

3.075  & \num{2.6468407910488296} & 
(0.18364) &
3.076  & \num{2.6524755077121247} & 
(0.18439) \\

3.077  & \num{2.658056047566261}  & 
(0.18512) &
3.078  & \num{2.663583341825903}  & 
(0.18585) \\

3.079  & \num{2.6690582952539463} & 
(0.18658) &
3.08   & \num{2.6744817871748916} & 
(0.18730) \\

3.081  & \num{2.679854672382887}  & 
(0.18801)&
3.082  & \num{2.6851777824473313} & 
(0.18871)\\

3.083  & \num{2.690451926235144}  & 
(0.18941) &
3.084  & \num{2.695677890978999}  & 
(0.19011)\\

3.085  & \num{2.7008564430567654} & 
(0.19079)&
3.086  & \num{2.7059883287773343} & 
(0.19148)\\

3.087  & \num{2.711074275116743}  & 
(0.19215) &
3.088  & \num{2.71611499044527}   & 
(0.19284)\\

3.089  & \num{2.721111165176857}  & \Gnum{0.1934907891635717}{F} &
3.09   & \num{2.7260634724381343} & \Gnum{0.19415179959982193}{F} \\

3.091  & \num{2.730972568668603}  & \Gnum{0.1948077267533116}{F} &
3.092  & \num{2.7358390942120825} & \Gnum{0.19545866493769498}{F} \\

3.093  & \num{2.7406636738809484} & \Gnum{0.19610470571434732}{F} &
3.094  & \num{2.745446918458424}  & \Gnum{0.19674593800130774}{F} \\

3.095  & \num{2.7501894202318713} & \Gnum{0.1973824481768025}{F} &
3.096  & \num{2.7548917636827426} & \Gnum{0.1980143201776688}{F} \\

3.097  & \num{2.7595545153955023} & \Gnum{0.19864163559298512}{F} &
3.098  & \num{2.7641782301448297} & \Gnum{0.1992644737531832}{F} \\

3.099  & \num{2.7687634498990623} & \Gnum{0.19988291181490714}{F} &
\textbf{3.1}    & \textbf{\num{2.773310704599938}}  & \textbf{\Gnum{0.2005723277066773}{T}} \\

3.101  & \num{2.777820511942219}  & \Gnum{0.20101722540854147}{F} &
3.103  & \num{2.786729799543751}  & \Gnum{0.2023213146130108}{T} \\

3.104  & \num{2.79113025968113}   & \Gnum{0.2028938321689251}{T} &
3.105  & \num{2.795495233005331}  & \Gnum{0.2035148996102955}{T} \\

3.106  & \num{2.799825183503205}  & \Gnum{0.20411014716028383}{F} &
3.107  & \num{2.80412056540781}   & \Gnum{0.204710910869749}{T} \\

3.108  & \num{2.808381823330603}  & \Gnum{0.2053146125938287}{T} &
3.109  & \num{2.812609392728498}  & \Gnum{0.2058476133752614}{T} \\

3.11   & \num{2.816803700052979}  & \Gnum{0.20645043925174691}{T} &
3.111  & \num{2.820965163006331}  & \Gnum{0.2070449848510315}{T} \\

3.112  & \num{2.825094191147633}  & \Gnum{0.2076126172458098}{T} &
3.113  & \num{2.829191185378833}  & \Gnum{0.2081880872134084}{T} \\

3.114  & \num{2.833256538882469}  & \Gnum{0.20875200353156184}{F} &
3.115  & \num{2.837290637080947}  & \Gnum{0.2093231343615088}{T} \\

3.116  & \num{2.841293857839566}  & \Gnum{0.2098865430293362}{T} &
3.117  & \num{2.84526657170906}   & \Gnum{0.210451626451604}{T} \\

3.118  & \num{2.849209142117845}  & \Gnum{0.21100222854824485}{T} &
3.119  & \num{2.853121925572378}  & \Gnum{0.2115562464143425}{T} \\

3.12   & \num{2.857005271858138}  & \Gnum{0.212109581905406}{T} &
3.121  & \num{2.860859524140107}  & \Gnum{0.2126571349124793}{T} \\

3.122  & \num{2.864685019289756}  & \Gnum{0.21320286085310075}{F} &
3.123  & \num{2.868482087922093}  & \Gnum{0.213744365590189}{T} \\

3.124  & \num{2.872251054587792}  & \Gnum{0.2142860854551016}{T} &
3.125  & \num{2.875992238133347}  & \Gnum{0.2148211983300776}{T} \\

3.126  & \num{2.879705951013354}  & \Gnum{0.2153554948787637}{T} &
3.127  & \num{2.883392501043907}  & \Gnum{0.21588569387638648}{F} \\

3.128  & \num{2.887052189949732}  & \Gnum{0.2164141446466337}{T} &
3.129  & \num{2.890685314296197}  & \Gnum{0.2169395506404182}{T} \\

3.13   & \num{2.894292165497908}  & \Gnum{0.2174623725531978}{T} &
3.131  & \num{2.897873029870033}  & \Gnum{0.21798181970061659}{T} \\

3.132  & \num{2.901428188920919}  & \Gnum{0.2184997090517925}{T} &
3.133  & \num{2.904957918975463}  & \Gnum{0.2190138769261493}{T} \\

3.134  & \num{2.90846249221857}   & \Gnum{0.2195270043088933}{T} &
3.135  & \num{2.911942175888552}  & \Gnum{0.2200360412139153}{T} \\

3.14   & \num{2.928976300197405}  & \Gnum{0.2225463531392025}{T} &
3.145  & \num{2.945425834258389}  & \Gnum{0.22500456832142057}{F} \\

3.15   & \num{2.961319642404431}  & \Gnum{0.22741068676056964}{F} &
3.155  & \num{2.976684418495105}  & \Gnum{0.22976470845664979}{F} \\

3.16   & \num{2.991544911494664}  & \Gnum{0.23206663340966116}{F} &
3.165  & \num{3.00592412130987}   & \Gnum{0.2343164616196034}{F} \\

3.17   & \num{3.019843469604536}  & \Gnum{0.23651419308647667}{F} &
3.175  & \num{3.033322949418039}  & \Gnum{0.238659827810281}{F} \\

3.18   & \num{3.046381256744704}  & \Gnum{0.24075336579101653}{F} &
\textbf{3.2}    & \textbf{\num{3.094739902479072}}  & \textbf{\Gnum{0.2486065502832686}{T}} \\

3.25   & \num{3.193154171614816}  & \Gnum{0.2660305998340008}{T} &
3.3    & \num{3.267212633294251}  & \Gnum{0.2807136672413638}{T} \\

3.35   & \num{3.323462085040376}  & \Gnum{0.2933033346956138}{T} &
3.4    & \num{3.366251791567017}  & \Gnum{0.3042467195145317}{T} \\

3.45   & \num{3.398624651937266}  & \Gnum{0.313862203390977}{T} &
\textbf{3.5}    & \textbf{\num{3.422790934985493}}  & \textbf{\Gnum{0.3223857231847494}{T}} \\

3.55   & \num{3.440402350104208}  & \Gnum{0.3299980872602599}{T} &
3.6    & \num{3.452720982098932}  & \Gnum{0.3368411258952461}{T} \\

3.65   & \num{3.460728649532728}  & \Gnum{0.3430279960781047}{T} &
3.7    & \num{3.465200504469459}  & \Gnum{0.3486502642145055}{T} \\

3.71   & \num{3.465727392119557}  & \Gnum{0.3497138368747632}{T} &
3.72   & \num{3.466142336228941}  & \Gnum{0.3507583596296334}{T} \\

3.73   & \num{3.466449485694358}  & \Gnum{0.3517843769457151}{T} &
3.74   & \num{3.466652819561603}  & \Gnum{0.3527923877430348}{T} \\

3.75   & \num{3.46675615533853}   & \Gnum{0.3537829036932754}{T} &
3.755  & \num{3.466771474679724}  & \Gnum{0.3542716797683863}{T} \\

3.7554 & 3.466771673  & \Gnum{0.3543106554493394}{T} &
3.7555 & 3.466771699  & \Gnum{0.3543203668272787}{T} \\

3.7556 & 3.466771720  & \Gnum{0.3543300835869048}{T} &
\textbf{3.7557} & \textbf{3.466771723} & \textbf{\Gnum{0.3543397946180297}{T}} \\

3.7558 & 3.466771721 & \Gnum{0.3543495246984588}{T} &
3.7559 & 3.466771709 & \Gnum{0.3543592074648508}{T} \\

3.756  & 3.466771688 & \Gnum{0.3543689683739103}{T} &
3.76   & \num{3.466763156823134} & \Gnum{0.3547563451459015}{T} \\

3.77   & \num{3.466677341474497} & \Gnum{0.355713110255565}{T} &
3.78   & \num{3.466502087366576} & \Gnum{0.356653743842543}{T} \\

3.79   & \num{3.466240639745965} & \Gnum{0.3575785790819292}{T} &
3.8    & \num{3.465896117218149} & \Gnum{0.358488037718591}{T} \\

3.85   & \num{3.463028377852694} & \Gnum{0.3628176717375823}{T} &
3.9    & \num{3.458488125756194} & \Gnum{0.3668156914402242}{T} \\

3.95   & \num{3.45255263909535}  & \Gnum{0.3705193898685308}{T} &
\textbf{4.0}    & \textbf{\num{3.445452681947484}} & \textbf{\Gnum{0.3739606640604697}{T}} \\

4.05   & \num{3.437381336162633} & \Gnum{0.3771668865164782}{T} &
4.1    & \num{3.428500921857038} & \Gnum{0.3801619538082258}{T} \\

4.15   & \num{3.418948473997138} & \Gnum{0.3829663793824839}{T} &
4.2    & \num{3.408840115326819} & \Gnum{0.3855983474750595}{T} \\

4.25   & \num{3.398274576491131} & \Gnum{0.3880737108135106}{T} &
4.3    & \num{3.3873360512191}   & \Gnum{0.3904064459326017}{T} \\

4.35   & \num{3.376096528363015} & \Gnum{0.3926089445750959}{T} &
4.39   & \num{3.366929958936248} & \Gnum{0.3942845969101932}{T} \\

4.4    & \num{3.364617709481664} & \Gnum{0.3946921784743028}{T} &
4.41   & \num{3.362298031723455} & \Gnum{0.3950954021718596}{T} \\

4.42   & \num{3.359971312100125} & \Gnum{0.3954943344576772}{T} &
4.43   & \num{3.357637925420562} & \Gnum{0.3958890274902691}{T} \\

4.44   & \num{3.355298235711826} & \Gnum{0.3962795736170897}{T} &
4.45   & \num{3.352952596185309} & \Gnum{0.3966660109765917}{T} \\

4.46   & \num{3.35060134986588}  & \Gnum{0.3970484144871775}{T} &
4.47   & \num{3.348244829671562} & \Gnum{0.3974268908212756}{T} \\

4.48   & \num{3.34588335902438}  & \Gnum{0.3978014190548826}{T} &
4.49   & \num{3.343517251799103} & \Gnum{0.3981721784785263}{T} \\

4.5    & \num{3.341146812904805} & \Gnum{0.3985391454850785}{T} &
4.51   & \num{3.338772338491041} & \Gnum{0.3989024013197133}{T} \\

4.52   & \num{3.336394116214719} & \Gnum{0.3992620164470315}{T} &
4.53   & \num{3.334012425515197} & \Gnum{0.399618023532967}{T} \\

4.54   & \num{3.331627537798366} & \Gnum{0.399970510802947}{T} &
4.55   & \num{3.329239716896443} & \Gnum{0.400319520703915}{T} \\

4.56   & \num{3.326849219061752} & \Gnum{0.4006650635704677}{T} &
4.57   & \num{3.324456293541339} & \Gnum{0.4010072588244785}{T} \\

4.58   & \num{3.322061182232963} & \Gnum{0.4013461777514364}{T} &
4.59   & \num{3.319664120742258} & \Gnum{0.4016817688813438}{T} \\

4.6    & \num{3.317265337760083} & \Gnum{0.4020141534626752}{T} &
4.61   & \num{3.314865055872529} & \Gnum{0.4023433705582878}{T} \\

4.62   & \num{3.312463491421084} & \Gnum{0.4026694580386362}{T} &
4.63   & \num{3.310060855030621} & \Gnum{0.4029924786293551}{T} \\

4.64   & \num{3.30765735128719}  & \Gnum{0.4033124241920351}{T} &
4.65   & \num{3.305253179814654} & \Gnum{0.403629464112792}{T} \\

4.66   & \num{3.302848534346191} & \Gnum{0.4039435357653387}{T} &
4.67   & \num{3.300443603652178} & \Gnum{0.4042546558937389}{T} \\

4.68   & \num{3.298038571962922} & \Gnum{0.4045630063955002}{T} &
4.69   & \num{3.295633618067134} & \Gnum{0.4048684550415698}{T} \\

4.7    & \num{3.293228916470628} & \Gnum{0.4051711679861106}{T} &
4.71   & \num{3.290824637235167} & \Gnum{0.405471214902443}{T} \\

4.72   & \num{3.288420946544505} & \Gnum{0.405768516316612}{T} &
4.73   & \num{3.286018005448856} & \Gnum{0.4060630841229684}{T} \\

4.74   & \num{3.283615971987003} & \Gnum{0.406355112474207}{T} &
4.75   & \num{3.281215000421347} & \Gnum{0.4066445425312444}{T} \\

4.76   & \num{3.278815240573692} & \Gnum{0.4069313641400514}{T} &
4.77   & \num{3.276416839676711} & \Gnum{0.4072156525224468}{T} \\

4.78   & \num{3.274019941220516} & \Gnum{0.4074974032803471}{T} &
4.79   & \num{3.271624685669399} & \Gnum{0.4077764872447603}{T} \\

4.8    & \num{3.269188154125189} & \Gnum{0.4080537568467425}{F} &
4.805  & \num{3.267990614265432} & \Gnum{0.408191711261822}{F} \\

4.81   & \num{3.266793518538213} & \Gnum{0.4083292120862938}{F} &
4.815  & \num{3.26559688169071}  & \Gnum{0.40846625932015784}{F} \\

4.82   & \num{3.264400718227686} & \Gnum{0.40860285296341414}{F} &
4.825  & \num{3.263205042453722} & \Gnum{0.4087389930160627}{F} \\

4.83   & \num{3.262009868501074} & \Gnum{0.4088746794781035}{F} &
4.835  & \num{3.26081521025738}  & \Gnum{0.4090099123495366}{F} \\

4.84   & \num{3.259621081378713} & \Gnum{0.40914469163036193}{F} &
4.845  & \num{3.258427495457739} & \Gnum{0.40927901732057953}{F} \\

4.85   & \num{3.257234465713027} & \Gnum{0.4094128894201894}{F} &
4.855  & \num{3.256042005329914} & \Gnum{0.40954630792919156}{F} \\

4.86   & \num{3.254850127216899} & \Gnum{0.40967927284758593}{F} &
4.865  & \num{3.253658844137121} & \Gnum{0.4098117841753726}{F} \\

4.87   & \num{3.252468168642458} & \Gnum{0.4099438419125515}{F} &
4.875  & \num{3.251278113102167} & \Gnum{0.4100754460591227}{F} \\

4.88   & \num{3.250088689717131} & \Gnum{0.4102065966150861}{F} &
4.885  & \num{3.248899910585311} & \Gnum{0.41033729358044185}{F} \\

4.89   & \num{3.247711787538231} & \Gnum{0.41046753695518984}{F} &
4.895  & \num{3.24652433223664}  & \Gnum{0.41059732673933}{F} \\

4.9    & \num{3.245337556235431} & \Gnum{0.4107266629328626}{F} &
4.905  & \num{3.244151470867072} & \Gnum{0.41085554553578735}{F} \\

4.91   & \num{3.242966087385067} & \Gnum{0.4109839745481043}{F} &
4.915  & \num{3.241781416820728} & \Gnum{0.4111119499698136}{F} \\

4.92   & \num{3.240597469993392} & \Gnum{0.41123947180091514}{F} &
4.925  & \num{3.239414257696911} & \Gnum{0.411366540041409}{F} \\

4.93   & \num{3.238231790518284} & \Gnum{0.41149315469129505}{F} &
4.935  & \num{3.237050078812466} & \Gnum{0.41161931575057337}{F} \\

4.94   & \num{3.235869132938858} & \Gnum{0.411745023219244}{F} &
4.945  & \num{3.234688962989888} & \Gnum{0.41187027709730684}{F} \\

4.95   & \num{3.233509578943125} & \Gnum{0.41199507738476193}{F} &
4.955  & \num{3.232330990708248} & \Gnum{0.41211942408160934}{F} \\

4.96   & \num{3.23115320796679}  & \Gnum{0.41224331718784896}{F} &
4.965  & \num{3.229976240258481} & \Gnum{0.4123667567034809}{F} \\

4.97   & \num{3.228800097077066} & \Gnum{0.4124897426285051}{F} &
4.975  & \num{3.227624787732192} & \Gnum{0.41261227496292147}{F} \\

4.98   & \num{3.226450321373179} & \Gnum{0.4127343537067302}{F} &
4.985  & \num{3.22527670708291}  & \Gnum{0.41285597885993114}{F} \\

4.99   & \num{3.224103953708376} & \Gnum{0.41297715042252436}{F} &
4.995  & \num{3.222932070158106} & \Gnum{0.41309786839450985}{F} \\

\textbf{5.0}    & \textbf{\num{3.22176106504378}}  & \textbf{\Gnum{0.4132181327758876}{F}} &
5.005  & \num{3.220590946978639} & \Gnum{0.4133379435666576}{F} \\

5.025  & \num{3.21591951177496}  & \Gnum{0.41381265082366026}{F} &
5.045  & \num{3.211262923470132} & \Gnum{0.4142801006309391}{F} \\

5.065  & \num{3.206621675989017} & \Gnum{0.4147402929884942}{F} &
5.085  & \num{3.201996237254468} & \Gnum{0.4151932278963254}{F} \\

5.105  & \num{3.197387050630366} & \Gnum{0.4156389053544329}{F} &
5.125  & \num{3.192794536659431} & \Gnum{0.4160773253628165}{F} \\

5.135  & \num{3.190504657451362} & \Gnum{0.4162938138233619}{F} &
5.145  & \num{3.188219094641818} & \Gnum{0.4165084879214764}{F} \\

5.149  & \num{3.18730608858678}  & \Gnum{0.4165938495392415}{F} &
5.151  & \num{3.186849848314896} & \Gnum{0.4166364214863782}{F} \\

5.153  & \num{3.186393783635426} & \Gnum{0.41667892085901764}{F} &
5.155  & \num{3.18593789495624}  & \Gnum{0.4167213476571599}{F} \\

5.157  & \num{3.185482182655294} & \Gnum{0.41676370188080486}{F} &
5.159  & \num{3.185026647072316} & \Gnum{0.4168059835299526}{F} \\

5.169  & \num{3.182751632408463} & \Gnum{0.4170163031582328}{F} &
5.179  & \num{3.180481088763941} & \Gnum{0.417224808424082}{F} \\

5.189  & \num{3.178215059246646} & \Gnum{0.4174314993275003}{F} &
5.199  & \num{3.175953586091871} & \Gnum{0.41763637586848756}{F} \\

5.209  & \num{3.173696710323502} & \Gnum{0.41783943804704393}{F} &
5.219  & \num{3.171444472369104} & \Gnum{0.4180406858631694}{F} \\

5.229  & \num{3.169196911453381} & \Gnum{0.41824011931686383}{F} &
5.239  & \num{3.16695406622353}  & \Gnum{0.41843773840812737}{F} \\

5.249  & \num{3.164715974363608} & \Gnum{0.41863354313695994}{F} &
5.259  & \num{3.162482672751605} & \Gnum{0.41882753350336155}{F} \\

5.269  & \num{3.160254197898846} & \Gnum{0.41901970950733225}{F} &
5.279  & \num{3.158030585075303} & \Gnum{0.41921007114887193}{F} \\

5.319  & \num{3.149185444321158} & \Gnum{0.4199533740907213}{F} &
5.349  & \num{3.142604292883155} & \Gnum{0.4204918004915834}{F} \\

5.359  & \num{3.140420785498169} & \Gnum{0.4206676472336755}{F} &
5.369  & \num{3.138242434853689} & \Gnum{0.4208416796133367}{F} \\

5.379  & \num{3.13606927133676}  & \Gnum{0.4210138976305669}{F} &
5.389  & \num{3.133901325068746} & \Gnum{0.42118430128536616}{F} \\

5.409  & \num{3.129581203182176} & \Gnum{0.42151966550767184}{F} &
5.419  & \num{3.127429086652913} & \Gnum{0.42168462607517826}{F} \\

5.459  & \num{3.118874260364288} & \Gnum{0.42232632472089443}{F} &
5.489  & \num{3.112515264534111} & \Gnum{0.4227885478996566}{T} \\

5.499  & \num{3.110406631554393} & \Gnum{0.4229552182324366}{F} &
\textbf{5.5}    & \textbf{\num{3.110196073833881}} & \textbf{\Gnum{0.4229718106480393}{T}} \\

\end{longtable}

}


\bibliography{review}

\end{document}